\documentclass[12pt,letterpaper]{article}
\usepackage{amsmath,amssymb,array,calc,rotating,epsfig,psfrag,amscd, cite}

\makeatletter
\renewcommand\section{\@startsection {section}{1}{\z@}%
                                   {-3.5ex \@plus -1ex \@minus -.2ex}
                                   {2.3ex \@plus.2ex}%
                                   {\normalfont\large\bfseries}}
\renewcommand\subsection{\@startsection{subsection}{2}{\z@}%
                                     {-3.25ex\@plus -1ex \@minus -.2ex}%
                                     {1.5ex \@plus .2ex}%
                                     {\normalfont\bfseries}}
\makeatother

\let\non\nonumber

\let\a=\alpha
\let\l=\lambda
\let\r=\rho
\let\s=\sigma

\let\S=\Sigma

\newcommand{\bea}{\begin{eqnarray}}
\newcommand{\eea}{\end{eqnarray}}
\newcommand{\be}{\begin{equation}}
\newcommand{\ee}{\end{equation}}

\newcommand{\G}{\Gamma}

\newcommand{\m}{\mu}
\newcommand{\n}{\nu}
\newcommand{\p}{\partial}

\newcommand{\C}[1]{$(\ref{#1})$}

\def\IZ{\relax\ifmmode\mathchoice
{\hbox{\cmss Z\kern-.4em Z}}{\hbox{\cmss Z\kern-.4em Z}}
{\lower.9pt\hbox{\cmsss Z\kern-.4em Z}} {\lower1.2pt\hbox{\cmsss
Z\kern-.4em Z}}\else{\cmss Z\kern-.4em Z}\fi}
\def\IR{\relax{\rm I\kern-.18em R}}

\def\one{{\hbox{ 1\kern-.8mm l}}}

\newlength{\bredde}
\def\slash#1{\settowidth{\bredde}{$#1$}\ifmmode\,\raisebox{.15ex}{/}
\hspace*{-\bredde} #1\else$\,\raisebox{.15ex}{/}\hspace*{-\bredde}
#1$\fi}

\newsavebox{\zzzbar}
\sbox{\zzzbar}
  {\setlength{\unitlength}{0.9em}
  \begin{picture}(0.6,0.7)
  \thinlines
  \put(0,0){\line(1,0){0.6}}
  \put(0,0.75){\line(1,0){0.575}}
  \multiput(0,0)(0.0125,0.025){30}{\rule{0.3pt}{0.3pt}}
  \multiput(0.2,0)(0.0125,0.025){30}{\rule{0.3pt}{0.3pt}}
  \put(0,0.75){\line(0,-1){0.15}}
  \put(0.015,0.75){\line(0,-1){0.1}}
  \put(0.03,0.75){\line(0,-1){0.075}}
  \put(0.045,0.75){\line(0,-1){0.05}}
  \put(0.05,0.75){\line(0,-1){0.025}}
  \put(0.6,0){\line(0,1){0.15}}
  \put(0.585,0){\line(0,1){0.1}}
  \put(0.57,0){\line(0,1){0.075}}
  \put(0.555,0){\line(0,1){0.05}}
  \put(0.55,0){\line(0,1){0.025}}
  \end{picture}}

\newcommand{\ena}{\end{eqnarray}}
\newcommand{\beqa}{\begin{eqnarray}}
\newcommand{\eeqa}{\end{eqnarray}}

\def\G{\Gamma}

\newcommand{\g}{\gamma}

\def\a{\alpha}

\def\g{\gamma}

\def\l{\lambda}
\def\m{\mu}
\def\n{\nu}

\def\r{\rho}
\def\s{\sigma}

\def\G{\Gamma}

\def\S{\Sigma}

\begin{document}
\begin{titlepage}

\begin{center}



\vskip 2 cm
{\Large \bf Low momentum expansion of one loop amplitudes in heterotic string theory }\\
\vskip 1.25 cm { Anirban Basu\footnote{email address:
    anirbanbasu@hri.res.in} } \\
{\vskip 0.5cm  Harish--Chandra Research Institute, HBNI, Chhatnag Road, Jhusi,\\
Allahabad 211019, India}

\end{center}

\vskip 2 cm

\begin{abstract}
\baselineskip=18pt

We consider the low momentum expansion of the four graviton and the two graviton--two gluon amplitudes in heterotic string theory at one loop in ten dimensions, and analyze contributions upto the $D^2\mathcal{R}^4$ interaction from the four graviton amplitude, and the $D^4 \mathcal{R}^2\mathcal{F}^2$ interaction from the two graviton--two gluon amplitude. The calculations are performed by obtaining equations for the relevant modular graph functions that arise in the modular invariant integrals, and involve amalgamating techniques used in the type II theory and the calculation of the elliptic genus in the heterotic theory.

\end{abstract}

\end{titlepage}


\section{Introduction}

The effective action of string theory defined on various backgrounds provides information about perturbative as well as non--perturbative aspects of the theory. Certain terms in the effective action have been obtained in various cases based on spacetime as well as worldsheet techniques. This primarily includes maximally supersymmetric string theories in various dimensions obtained by toroidally compactifying type II string theory. In particular, BPS interactions in the effective action have been obtained in~\cite{Gross:1986iv,Green:1997tv,Kiritsis:1997em,Green:1998by,Obers:1998fb,Green:1999pu,Green:2005ba,Basu:2007ru,Basu:2007ck,Green:2008bf,Basu:2008cf,Green:2010kv,Basu:2011he,Basu:2014hsa,Pioline:2015yea,Bossard:2015uga,Fleig:2015vky,Basu:2016kon} primarily based on considerations of supersymmetry and U--duality. The weak coupling expansion of the moduli dependent coefficient functions of the BPS interactions matches results in string perturbation theory at one loop~\cite{Green:1999pv,Green:2008uj,D'Hoker:2015foa,D'Hoker:2015zfa,Basu:2015ayg,D'Hoker:2015qmf,Basu:2016fpd,D'Hoker:2016jac,Basu:2016xrt,Basu:2016kli}~\footnote{Also see~\cite{Green:2013bza,Florakis:2016boz,Basu:2016mmk,DHoker:2016quv,Kleinschmidt:2017ege} for relevant details regarding calculating various one loop amplitudes.} as well as at higher loops~\cite{D'Hoker:2005vch,D'Hoker:2005jhf,Gomez:2013sla,D'Hoker:2014gfa,Basu:2015dqa,Pioline:2015nfa}, while the one loop results for non--BPS interactions lead to predictions for U--duality. Thus an intricate interplay between spacetime and worldsheet techniques in maximally supersymmetric compacifications allows us constrain various terms in the effective action.    

However, the effective action in theories with lesser supersymmetry has not been as well analyzed, which is essentially because of the reduced supersymmetry leading to a more involved analysis. Theories which preserve sixteen supersymmetries are the simplest to analyze in this class, and result from various compactifications as well as arising as the world volume theories of D--branes. Certain terms in the effective action of such theories have been analyzed, for example, in~\cite{Bergshoeff:1989de,deRoo:1992zp,Tseytlin:1995bi,Garousi:1996ad,Harvey:1996ir,Bachas:1997mc,Gregori:1997hi,Bachas:1999um,Basu:2008gt,Bossard:2013rza,Lin:2015ixa,Green:2016tfs,Garousi:2017fbe,Bossard:2017wum} based on diverse spacetime and worldsheet techniques.

Since the low momentum expansion of the superstring amplitudes strongly constrain the effective action, we shall consider this expansion at one loop in heterotic string theory~\cite{Gross:1984dd} in ten dimensions. This is among the simplest examples among closed string theories with sixteen supersymmetries where one can perform such an analysis. In particular, we consider the low momentum expansion of the four graviton, and the two graviton-two gluon amplitudes in heterotic string theory at one loop in ten dimensions. These amplitudes are the same  in the ${\rm Spin}(32)/\mathbb{Z}_2$ and $E_8 \times E_8$ theories.  

For these massless four point scattering amplitudes, the external momenta $k_i^{\m}$ ($i=1, \ldots, 4$) satisfy $\sum_{i=1}^4 k_i^\mu =0$, and the Mandelstam variables are given by 
\be s= -(k_1 +k_2)^2, \quad t= - (k_1 + k_4)^2,\quad u= -(k_1 +k_3)^2\ee
satisfying $s+t+u=0$. We shall neglect overall numerical factors for the various amplitudes which are irrelevant for our purposes. 

The leading interactions in the effective action obtained from the low momentum expansion of the four graviton and the two graviton--two gluon amplitude are schematically denoted by the $\mathcal{R}^4$ and $\mathcal{R}^2 \mathcal{F}^2$ terms respectively, where $\mathcal{R}$ and $\mathcal{F}$ denote the Riemann curvature and the Yang--Mills field strength respectively.
For the four graviton amplitude, we consider the first subleading contribution given by the $D^2\mathcal{R}^4$ term, while for the two graviton--two gluon amplitude we consider the $D^2\mathcal{R}^2 \mathcal{F}^2$ and the $D^4\mathcal{R}^2 \mathcal{F}^2$ terms, where each factor of $D$ schematically stands for an external momenta. Thus we consider all the interactions upto ten derivatives which result from these string amplitudes.

We first consider the four graviton and the two graviton--two gluon amplitudes at tree level and perform the low momentum expansion. This yields the various spacetime structures that arise and also fixes the coefficients at tree level.  We next consider these four point amplitudes at one loop, and perform the low momentum expansion. To obtain the coefficients of the various terms in the effective action, it amounts to evaluating modular invariant integrals involving $SL(2,\mathbb{Z})$ covariant modular graph functions over the truncated fundamental domain of $SL(2,\mathbb{Z})$. While some of the graphs are quite simple for the others we obtain second order differential equations satisfied by these graphs which allow us to solve for them, and hence perform these integrals over moduli space. We get these differential equations by varying the graphs with respect to the Beltrami differential on the toroidal worldsheet, leading to compact expressions for eigenvalue equations these graphs satisfy. These Poisson equations have source terms that are determined by the structure of the graphs. We then extract the coefficients of these interactions at one loop by integrating modular invariant combinations of these graphs over the truncated fundamental domain of $SL(2,\mathbb{Z})$. This makes crucial use of the asymptotic expansions of the various graphs which is performed separately. This generalizes similar analysis done for the one loop amplitude in type II string theory which only involves $SL(2,\mathbb{Z})$ invariant graphs~\cite{Basu:2015ayg,Basu:2016xrt,Basu:2016kli,Kleinschmidt:2017ege}, and the calculation of the $\mathcal{R}^4$, $\mathcal{R}^2 \mathcal{F}^2$ coefficients in the heterotic theory, as well as the anomaly cancelling terms~\cite{Lerche:1987qk,Ellis:1987dc,Abe:1988cq}. These coefficients in the heterotic theory involved integrating almost anti--holomorphic integrands with mild holomorphicity over the truncated fundamental domain of $SL(2,\mathbb{Z})$. This has been done by directly calculating the amplitude~\cite{Ellis:1987dc,Abe:1988cq}, as well as by calculating the elliptic genus~\cite{Lerche:1987qk}. We shall see that the analysis we perform is considerably more involved, given the nature of the integrands.  Hence we determine these interactions at one loop which contribute to the effective action of heterotic string theory. The various technical details are given in the appendices.      

Compared to the analysis of the interactions in the type II string theory which has been considered to higher orders in the derivative expansion than what we do for the heterotic theory, we see that the analysis is considerably more involved. This is a consequence of the reduced supersymmetry the heterotic theory enjoys compared to the type II  theory. While we perform the analysis only upto the $D^2\mathcal{R}^4$ and $D^4\mathcal{R}^2\mathcal{F}^2$ interactions in the low momentum expansion of the four graviton and the two graviton--two gluon amplitude respectively, our method can be extended to higher orders in the momentum expansion. It will be interesting to understand the structure of the interactions that follow at higher orders in the momentum expansion, as well as in cases with lesser supersymmetry. It will also be very interesting to understand the structure of the low momentum expansion of these amplitudes at higher loops.

\section{The tree level amplitudes in heterotic string theory}

To begin with, we consider the low momentum expansion of the tree level amplitudes in heterotic string theory\cite{Gross:1985rr,Kawai:1985xq,Cai:1986sa,Gross:1986mw,Kikuchi:1986cz}. At tree level, the amplitudes are the same in the ${\rm Spin(32)}/\mathbb{Z}_2$ and $E_8 \times E_8$ theories.

\subsection{The four graviton amplitude}

At tree level, the four graviton amplitude is given by
\bea \label{t1}&&A^{tree}_{4g} (k_i, \epsilon^{(i)})= -e^{-2\phi}\frac{\Gamma(-\alpha' s/4)\Gamma(-\alpha' t/4)\Gamma(-\alpha' u/4)}{\Gamma(1+\alpha' s/4)\Gamma(1+\alpha' t/4)\Gamma(1+\alpha' u/4)}\prod_{i=1}^4 \epsilon^{(i)}_{\mu_i\nu_i} K^{\mu_1 \mu_2 \mu_3 \mu_4} \non \\ &&\times \Big[ K^{\nu_1\nu_2\nu_3\nu_4} -\frac{\alpha'stu}{16}\Big( \frac{\eta^{\nu_1\nu_2} \eta^{\nu_3\nu_4}}{1+\alpha's/4}+ \frac{\eta^{\nu_1\nu_4} \eta^{\nu_2\nu_3}}{1+\alpha't/4}+ \frac{\eta^{\nu_1\nu_3} \eta^{\nu_2\nu_4}}{1+\alpha'u/4}\Big)\non \\ &&+\frac{\alpha' ut}{8(1+\alpha's/4)}\Big(\frac{\alpha'}{2} k_1^{\n_2} k_2^{\n_1}k_3^{\n_4} k_4^{\n_3}- \eta^{\n_1\nu_2} k_3^{\n_4} k_4^{\n_3}- \eta^{\n_3\n_4} k_1^{\n_2} k_2^{\n_1}\Big) \non \\ &&+\frac{\alpha' us}{8(1+\alpha't/4)}\Big(\frac{\alpha'}{2} k_1^{\n_4} k_4^{\n_1}k_2^{\n_3} k_3^{\n_2}- \eta^{\n_1\nu_4} k_2^{\n_3} k_3^{\n_2}- \eta^{\n_2\n_3} k_1^{\n_4} k_4^{\n_1}\Big)\non \\ &&+\frac{\alpha' st}{8(1+\alpha'u/4)}\Big(\frac{\alpha'}{2} k_1^{\n_3} k_3^{\n_1}k_2^{\n_4} k_4^{\n_2}- \eta^{\n_1\nu_3} k_2^{\n_4} k_4^{\n_2}- \eta^{\n_2\n_4} k_1^{\n_3} k_3^{\n_1}\Big)\non \\ &&+\frac{\alpha's}{12}\Big(k_1^{\n_3} k_2^{\n_4} k_4^{\n_1} k_3^{\n_2} + k_1^{\n_4} k_2^{\n_3} k_3^{\n_1} k_4^{\n_2} +(k_3^{\n_1} k_3^{\n_2} + k_4^{\n_1} k_4^{\n_2})(k_1^{\n_3}k_1^{\n_4} + k_2^{\n_3} k_2^{\n_4})\Big)\non \\ && +\frac{\alpha't}{12}\Big(k_1^{\n_3} k_4^{\n_2} k_2^{\n_1} k_3^{\n_4} + k_1^{\n_2} k_4^{\n_3} k_3^{\n_1} k_2^{\n_4} +(k_2^{\n_1} k_2^{\n_4} + k_3^{\n_1} k_3^{\n_4})(k_1^{\n_2}k_1^{\n_3} + k_4^{\n_2} k_4^{\n_3})\Big)\non \\ &&+\frac{\alpha'u}{12}\Big(k_1^{\n_2} k_3^{\n_4} k_4^{\n_1} k_2^{\n_3} + k_1^{\n_4} k_3^{\n_2} k_2^{\n_1} k_4^{\n_3} +(k_1^{\n_2} k_1^{\n_4} + k_3^{\n_2} k_3^{\n_4})(k_2^{\n_1}k_2^{\n_3} + k_4^{\n_1} k_4^{\n_3})\Big) \Big],\eea
where $\epsilon^{(i)}_{\mu\nu}$ ($i=1,\cdots, 4$) is the polarization tensor for the graviton which carries momentum $k_i$ respectively, and $\phi$ is the dilaton~\footnote{This will be our convention at one loop as well.}. Here we have that
\bea K^{\m_1\m_2\m_3\m_4}=\frac{1}{4} (ut \eta^{\m_1\m_2} \eta^{\m_3\m_4} + st\eta^{\m_1\m_3}\eta^{\m_2\m_4} + su \eta^{\m_1\m_4}\eta^{\m_2\m_3})\non \\ -\frac{s}{2} (\eta^{\m_2\m_4} k_4^{\m_1}k_2^{\m_3} +\eta^{\m_1\m_3} k_1^{\m_4} k_3^{\m_2} +\eta^{\m_2\m_3} k_2^{\m_4} k_3^{\m_1} + \eta^{\m_1\m_4} k_4^{\m_2} k_1^{\m_3})\non \\ -\frac{t}{2} (\eta^{\m_2\m_4} k_2^{\m_1}k_4^{\m_3} +\eta^{\m_1\m_3} k_3^{\m_4} k_1^{\m_2} +\eta^{\m_1\m_2} k_2^{\m_4} k_1^{\m_3} + \eta^{\m_3\m_4} k_3^{\m_1} k_4^{\m_2})\non \\ -\frac{u}{2} (\eta^{\m_1\m_2} k_1^{\m_4}k_2^{\m_3} +\eta^{\m_2\m_3} k_3^{\m_4} k_2^{\m_1} +\eta^{\m_1\m_4} k_1^{\m_2} k_4^{\m_3} + \eta^{\m_3\m_4} k_3^{\m_2} k_4^{\m_1})\eea
leading to a manifestly crossing symmetric expression for the amplitude. In fact, 
\be K^{\m_1\m_2\m_3\m_4}= t_8^{\m_1\n_1\m_2\n_2\m_3\n_3\m_4\n_4} \prod_{i=1}^4 k_{i\n_i}\ee
where $t_8$ is given by the standard expression. Thus the first term in \C{t1} which has two powers of $K$ is the type II four graviton amplitude, while the rest arise from reduced supersymmetry.

\subsection{The low momentum expansion of the four graviton amplitude}

We now perform the low momentum expansion of the four graviton amplitude. We shall be interested in contact terms in the effective action resulting from terms in the amplitude having at least eight powers of momenta. Terms with lesser powers of momenta in the amplitude arise from contributions that result from the supergravity action involving two derivative terms, as well as the four derivative $\mathcal{R}^2$ contact term. These terms do not receive contributions beyond tree level and hence they are tree level exact. At one loop $n$--point amplitudes of massless particles vanish unless $n \geq 4$ which follows from a zero mode analysis and we consider only those interactions that arise from amplitudes which receive contributions beyond tree level. For the four graviton amplitude at tree level, these are contact term contributions which involve at least eight powers of momenta, and they lead to terms in the effective action which we schematically denote as $D^{2k} \mathcal{R}^4$ for $k \geq 0$. Along with the $\mathcal{R}^4$ contribution which has already been analyzed in the literature, we shall analyze the $D^2\mathcal{R}^4$ term at tree level and at one loop. The interactions at higher orders in the derivative expansion can be analyzed in a similar way.          

The various manifestly crossing symmetric spacetime tensors that arise in our analysis are given in appendix A. 

From \C{t1}, we see that the $\mathcal{R}^4$ term is given by
\bea \label{tR4}A^{tree}_{\mathcal{R}^4} = e^{-2\phi} \prod_{i=1}^4 \epsilon^{(i)}_{\mu_i\nu_i} K^{\mu_1 \mu_2 \mu_3 \mu_4} \Big[2\zeta(3) K -\Big( I_{4,0} +\frac{2}{\alpha'} I_{2;1,0} +\frac{4}{\alpha'^2} I_{1;2,0}\Big)  \Big]^{\n_1\n_2\n_3\n_4}, \eea
while the $D^2 \mathcal{R}^4$ term is given by
\bea \label{tD2R4}A^{tree}_{D^2\mathcal{R}^4} &=& e^{-2\phi} \prod_{i=1}^4 \epsilon^{(i)}_{\mu_i\nu_i} K^{\mu_1 \mu_2 \mu_3 \mu_4} \Big[ \Big( I_{4,1} +\frac{2}{\alpha'} I_{2;2,0}+\frac{4}{\alpha'^2} I_{1;3,0}\Big)  \non \\ &&-\frac{4\zeta(3)}{\alpha'} \Big(I_{2;0,1} +\frac{2}{\alpha'}I_{1;1,1} -\frac{\alpha'}{6} (I_{5,1} + I_{6,1})\Big)\Big]^{\n_1\n_2\n_3\n_4}. \eea

The various expressions in \C{tR4} and \C{tD2R4} lead to terms in the effective action that can be written in terms of covariant expressions involving curvature tensors. Here and elsewhere, we refrain from giving the expressions as they are not relevant for our purposes.

\subsection{The two graviton--two gluon amplitude}

The two graviton--two gluon amplitude is given by
\bea \label{t2}A^{tree}_{2g,2a} (k_i, \epsilon^{(i)},e^{(i)})=-e^{-2\phi}\frac{\Gamma(-\alpha' s/4)\Gamma(-\alpha' t/4)\Gamma(-\alpha' u/4)}{\Gamma(1+\alpha' s/4)\Gamma(1+\alpha' t/4)\Gamma(1+\alpha' u/4)} \epsilon^{(1)}_{\mu_1\nu_1} \epsilon^{(2)}_{\m_2\n_2} e^{(3)}_{a\n_3} e^{(4)}_{b\n_4} \non \\ \times Tr(T^aT^b) K^{\n_1 \n_2 \n_3 \n_4}\Big[ \frac{ut}{2(1+\alpha's/4)}\Big(\frac{\alpha'}{2} k_1^{\m_2} k_2^{\m_1} - \eta^{\m_1\m_2}\Big) +t k_3^{\m_1} k_4^{\m_2} +u k_4^{\m_1} k_3^{\m_2}\Big]\eea
where $\epsilon^{(i)}_{\m_i\n_i}$ ($i=1,2$) is the polarization tensor for the graviton which carries momentum $k_i$ respectively, while $e^{(i)}_{a\m_i}$ ($i=3,4$) is the polarization vector for the gluon which carries momentum $k_i$ respectively. This will also be our convention at one loop as well. The color trace is in the vector representation of $SO(32)$ in the ${\rm Spin}(32)/\mathbb{Z}_2$ theory and is defined by ${\rm Tr}_V(T^a T^b) = 2\delta^{ab}$, and in the adjoint representation of $E_8\times E_8$ and is defined by ${\rm Tr}_A (T^a T^b) = 2\delta^{ab}/30$ in the $E_8 \times E_8$ theory.  

\subsection{The low momentum expansion of the two graviton--two gluon amplitude}

We now perform the low momentum expansion of the two graviton--two gluon amplitude, which leads to contact terms in the effective action with at least six powers of momenta. Terms with lesser powers of momenta arise from the Einstein--Hilbert, $\mathcal{F}^2$ and $\mathcal{R}^2$ contributions which do not receive contributions beyond tree level. We denote these terms which also receive contributions at one loop as $D^{2k} \mathcal{R}^2 \mathcal{F}^2$.  

We find it useful to define the tensors
\bea J_1^{\mu_1\mu_2} &=& k_1^{\m_2} k_2^{\m_1} +\frac{s}{2} \eta^{\m_1\m_2}, \non \\ J_2^{\mu_1\mu_2} &=& t k_3^{\m_1}k_4^{\m_2} + u k_4^{\m_1} k_3^{\m_2} -\frac{ut}{2} \eta^{\m_1\m_2}. \eea

Thus from \C{t2}, we see that the $\mathcal{R}^2 \mathcal{F}^2$ term is given by
\bea A^{tree}_{\mathcal{R}^2\mathcal{F}^2} = \frac{4e^{-2\phi}}{\alpha'}\epsilon^{(1)}_{\mu_1\nu_1} \epsilon^{(2)}_{\m_2\n_2} e^{(3)}_{a\n_3} e^{(4)}_{b\n_4} Tr(T^aT^b) K^{\n_1 \n_2 \n_3 \n_4}  J_1^{\m_1\m_2} , \eea
while the $D^2\mathcal{R}^2 \mathcal{F}^2$ term is given by
\bea A^{tree}_{D^2\mathcal{R}^2\mathcal{F}^2} = -e^{-2\phi}\epsilon^{(1)}_{\mu_1\nu_1} \epsilon^{(2)}_{\m_2\n_2} e^{(3)}_{a\n_3} e^{(4)}_{b\n_4} Tr(T^aT^b) K^{\n_1 \n_2 \n_3 \n_4} \Big[ s J_1^{\m_1\m_2} +2\zeta(3) J_2^{\m_1\m_2}\Big] , \eea
and finally the $D^4\mathcal{R}^2 \mathcal{F}^2$ term is given by
\bea A^{tree}_{D^4\mathcal{R}^2\mathcal{F}^2} = \frac{\alpha'e^{-2\phi}}{4}\epsilon^{(1)}_{\mu_1\nu_1} \epsilon^{(2)}_{\m_2\n_2} e^{(3)}_{a\n_3} e^{(4)}_{b\n_4} Tr(T^aT^b) K^{\n_1 \n_2 \n_3 \n_4} \Big(s^2- 2\zeta(3)ut\Big)J_1^{\m_1\m_2}. \non \\ \eea

While the $\mathcal{R}^2\mathcal{F}^2$ term has been analyzed in the literature, we shall also analyze the $D^2 \mathcal{R}^2\mathcal{F}^2$ and $D^4 \mathcal{R}^2\mathcal{F}^2$  terms upto one loop, which is the order in the momentum expansion to which we analyze the four graviton amplitude. 

\section{The one loop amplitudes in heterotic string theory}

We now consider the four graviton and the two graviton--two gluon amplitudes at one loop~\cite{Gross:1985rr,Sakai:1986bi,Ellis:1987dc,Abe:1988cq} in the heterotic theory. Before considering the explicit expressions for these amplitudes, we briefly go through some generalities relevant for our purposes.

For the one loop amplitudes, we have to integrate the modular invariant integrand over the fundamental domain of $SL(2,\mathbb{Z})$ parametrized by the complex structure $\tau$ of the torus. This integrand involves an integral over the insertion points of the vertex operators on the toroidal worldsheet $\S$. At one loop, $z_i$ $(i=1,\cdots,4)$ are the positions of insertions of the four vertex operators on the worldsheet with complex structure $\tau$. Thus $d^2 z_i = d({\rm Re} z_i) d({\rm Im}z_i)$, where
\be -\frac{1}{2} \leq {\rm Re} z_i \leq \frac{1}{2}, \quad 0 \leq {\rm Im} z_i\leq \tau_2 \ee
for all $i$. The various amplitudes involve the scalar Green function $\hat{G}_{ij}$ (as well as its derivatives) between points $z_i$ and $z_j$, and so\footnote{We suppress the $\bar{z}, \bar\tau$ dependence in $\hat{G}(z,\bar{z};\tau,\bar\tau)$ and simply write it as $\hat{G}(z;\tau)$ for brevity, which is standard. }
\be \hat{G}_{ij} \equiv \hat{G}(z_i - z_j;\tau).\ee
It is given by~\cite{Lerche:1987qk,Green:1999pv}
\bea \label{Prop}\hat{G}(z;\tau) &=& -{\rm ln} \Big\vert \frac{\theta_1 (z\vert\tau)}{\theta_1'(0\vert\tau)} \Big\vert^2 + \frac{2\pi({\rm Im}z)^2}{\tau_2} \non \\ &=& \frac{1}{\pi} \sum_{(m,n)\neq(0,0)} \frac{\tau_2}{\vert m\tau+n\vert^2} e^{\pi[\bar{z}(m\tau+n)-z(m\bar\tau+n)]/\tau_2} + 2{\rm ln} \vert \sqrt{2\pi} \eta(\tau)\vert^2.\eea
The Green function itself enters the amplitude only in the Koba--Nielsen factor and thus the position independent second term in the second expression in \C{Prop} cancels using $s+t+u=0$. Only the worldsheet derivatives of the Green function arise elsewhere in the amplitudes which follows from the structure of the vertex operators, and thus the second term in the second expression in \C{Prop} can be ignored. Hence we consider the Green function to be simply given by
\be \label{Green}G(z;\tau) = \frac{1}{\pi} \sum_{(m,n)\neq(0,0)} \frac{\tau_2}{\vert m\tau+n\vert^2} e^{\pi[\bar{z}(m\tau+n)-z(m\bar\tau+n)]/\tau_2}.\ee
Note that $G(z;\tau)$ is modular invariant, and single valued as it is doubly periodic. Thus
\be \label{sv}G(z;\tau) = G(z+1;\tau) = G(z+\tau;\tau).\ee
We denote $G(z_{ij};\tau) \equiv G_{ij}$ for brevity. This structure of the Green function is very useful for us in the various manipulations. Using its single valuedness, we can integrate $\p_z G(z,w)$ by parts while integrating over $z$ without picking up boundary terms. Also from the structure of \C{Green} it follows that all one particle reducible diagrams vanish. It also follows that
\be \int_\S d^2 z \p_z G(z,w) = \int_\S d^2 z \p_z^2 G(z,w)=0\ee    
which proves to be useful.

Finally, we have that
\bea \label{eigen}\bar{\p}_w\p_z G(z,w) = \pi \delta^2 (z-w) - \frac{\pi}{\tau_2}, \non \\ \bar{\p}_z\p_z G(z,w) = -\pi \delta^2 (z-w) + \frac{\pi}{\tau_2} \eea
which we shall repeatedly use in our analysis.

\subsection{The four graviton amplitude}

The one loop four graviton amplitude is given by
\bea \label{1loop4g}A^{1-loop}_{4g} (k_i, \epsilon^{(i)}) = \prod_{i=1}^4 \epsilon^{(i)}_{\mu_i\nu_i} K^{\mu_1 \mu_2 \mu_3 \mu_4} \int_{\mathcal{F}} \frac{d^2\tau}{\tau_2^2} \frac{{\bar{E}}_4^2}{{\bar{\eta}}^{24}}\prod_{i=1}^4 \int_\S  \frac{d^2 z^i}{\tau_2} e^{\mathcal{D}} \mathcal{T}^{\n_1\n_2\n_3\n_4}, \eea
where we have integrated over $\mathcal{F}$, the fundamental domain of $SL(2,\mathbb{Z})$ and $d^2 \tau = d\tau_1 d\tau_2$. Here we have defined
\be 2\zeta(2k) E_{2k}(\tau) = G_{2k} (\tau)\ee
for $k \geq 2$, where $G_{2k}(\tau)$ is the holomorphic Eisenstein series of modular weight $2k$ defined by\footnote{A modular form $\Phi^{(m,n)}(\tau,\bar\tau)$ of weight $(m,n)$ transforms under $SL(2,\mathbb{Z})$ transformations $\tau \rightarrow (a\tau+b)/(c\tau+d)$ where $a,b,c,d \in \mathbb{Z}, ad-bc=1$ as
\be \Phi^{(m,n)}(\tau,\bar\tau) \rightarrow (c\tau+d)^m (c\bar\tau +d)^n \Phi^{(m,n)} (\tau,\bar\tau).\ee}
\be \label{defG}G_{2k} (\tau) = \sum_{(m,n)\neq (0,0)} \frac{1}{(m+n\tau)^{2k}}.\ee
Also in the Koba--Nielsen factor, $\mathcal{D}$ is defined by
\be 4 \alpha'^{-1}\mathcal{D} = s (G_{12} + G_{34}) + t (G_{14} + G_{23}) +u (G_{13} +G_{24}). \ee
Finally, we consider the manifestly crossing symmetric tensor $\mathcal{T}^{\m_1\m_2\m_3\m_4}$. It depends on the locations of the insertions $z_i$ on the worldsheet, and the complex structure $\tau$, and is defined by
\bea \label{T}&&\mathcal{T}^{\m_1\m_2\m_3\m_4} = A^{\m_1} A^{\m_2} A^{\m_3} A^{\m_4} +\frac{1}{2\alpha'} \Big(\eta^{\m_1\m_2} R_{12} A^{\m_3} A^{\m_4} + \eta^{\m_1\m_3} R_{13} A^{\m_2} A^{\m_4} \non \\ &&+\eta^{\m_1\m_4} R_{14} A^{\m_2}A^{\m_3} + \eta^{\m_2\m_3} R_{23} A^{\m_1} A^{\m_4} +\eta^{\m_2\m_4} R_{24} A^{\m_1} A^{\m_3} +\eta^{\m_3\m_4} R_{34} A^{\m_1}A^{\m_2}\Big)\non \\ &&+\frac{1}{(2\alpha')^2} \Big(\eta^{\m_1\m_2}\eta^{\m_3\m_4}R_{12}R_{34} + \eta^{\m_1\m_3}\eta^{\m_2\m_4}R_{13}R_{24} +\eta^{\m_1\m_4}\eta^{\m_2\m_3}R_{14}R_{23}\Big),\eea
where\footnote{Thus, for example,
\be A^{\m_1} = \frac{1}{4\pi i} (k_2^{\m_1} \bar\p_2 G_{21}+ k_3^{\m_1} \bar\p_3 G_{31}+k_4^{\m_1} \bar\p_4 G_{41})\ee
where the term involving $k_1^{\m_1}$ vanishes in \C{defAR} using transversality on contracting with the polarization tensor. Hence there is no potential divergence from the coincident Green function. This is, in fact, how $A^{\m_i}$ is defined.}
\bea \label{defAR}A^{\mu_i} &=& \frac{1}{4\pi i}\sum_{j=1}^4 k_j^{\m_i} \bar\p_j G_{ji}, \non \\ R_{ij} &=& -\frac{1}{4\pi^2} \bar\p^2_i G_{ij},\eea
which involves derivatives of the Green functions.

In the expression for the four graviton amplitude \C{1loop4g}, the factor of $\bar\eta^{24}$ in the denominator arises from the left moving oscillator modes, while $\bar{E}_4^2$ in the numerator arises from the lattice sum while summing over the spin structures after performing the GSO projection. In the ${\rm Spin}(32)/\mathbb{Z}_2$ theory the sum yields\footnote{We denote $\theta_\alpha (0,\tau) \equiv \theta_\alpha$ for brevity.}
\be \label{s1}\frac{1}{2} \sum_{\alpha=2}^4 \bar\theta_\alpha^{16} = \bar{E}_8,\ee
while in the $E_8\times E_8$ theory it yields
\be \label{s2}\frac{1}{4} (\sum_{\alpha=2}^4\bar\theta_\alpha^8  )^2 = \bar{E}_4^2.\ee
Both these sums are equal using $\bar{E}_4^2 = \bar{E}_8$, hence the four graviton amplitude is the same in both the theories. 

\subsection{The two graviton--two gluon  amplitude}

We next consider the two graviton--two gluon amplitude, which is given by
\bea A^{1-loop}_{2g,2a}(k_i, \epsilon^{(i)}, e^{(i)}) = \epsilon^{(1)}_{\m_1\n_1} \epsilon^{(2)}_{\m_2\n_2} e^{(3)}_{a\n_3} e^{(4)}_{b\n_4}Tr(T^aT^b) K^{\nu_1 \nu_2 \nu_3 \nu_4} \non \\ \times \int_{\mathcal{F}} \frac{d^2\tau}{\tau_2^2} \frac{1}{{\bar{\eta}}^{24}}\prod_{i=1}^4 \int_\S  \frac{d^2 z^i}{\tau_2} e^{\mathcal{D}}  \Big(\frac{1}{2\alpha'} R_{12}\eta^{\m_1\m_2} + A^{\m_1} A^{\m_2}\Big)\bar{L}.\eea
The factor of $L$ in the above expression involving the lattice sum is given by complex conjugating 
\be \label{L1}L_{{\rm Spin}(32)/\mathbb{Z}_2} = \sum_{\alpha=2}^4 \theta_\alpha^{16} \Big[ \frac{\theta_\alpha (z_{34};\tau)\theta_1'(0;\tau)}{\theta_\alpha (0;\tau) \theta_1 (z_{34};\tau)}\Big]^2 \ee
in the ${\rm Spin}(32)/\mathbb{Z}_2$ theory, and
\be \label{L2}L_{E_8\times E_8} = E_4 \sum_{\alpha=2}^4 \theta_\alpha^{8} \Big[ \frac{\theta_\alpha (z_{34};\tau)\theta_1'(0;\tau)}{\theta_\alpha (0;\tau) \theta_1 (z_{34};\tau)}\Big]^2\ee
in the $E_8\times E_8$ theory. These somewhat involved expressions can be expressed differently as we now discuss, which is very useful for our purposes.   

The sums involving theta functions in \C{L1} and \C{L2} can be simplified~\cite{Ellis:1987dc} using the relation
\be \label{P}\mathcal{P} (z;\tau) = \p^2_z G(z;\tau) - \hat{G}_2 (\tau) = e_{\alpha -1}(\tau) +\Big[ \frac{\theta_\alpha (z;\tau)\theta_1'(0;\tau)}{\theta_\alpha (0;\tau) \theta_1 (z;\tau)}\Big]^2,\ee
which holds for $\alpha =2,3,4$. In \C{P}, $\mathcal{P}$ is the Weierstrass p--function, and $e_{\alpha-1}$ are defined in terms of the theta functions by the equations
\bea e_1 (\tau) &=& 2\zeta(2)(\theta_3^4 + \theta_4^4), \non \\ e_2 (\tau) &=& 2\zeta(2) (\theta_2^4 -\theta_4^4), \non \\ e_3(\tau) &=& -2\zeta(2)(\theta_2^4 +\theta_3^4).\eea
Finally, $\hat{G}_2$ is defined by\footnote{We denote $\hat{G}_2 (\tau,\bar\tau)$ as simply $\hat{G}_2(\tau)$ for brevity, as is standard.}
\be \hat{G}_2 (\tau) = G_2 (\tau) - \frac{\pi}{\tau_2},\ee
which has holomorphic modular weight 2, where $G_2$ is defined by \C{defG}. Thus
\be \label{simpleL}L_{{\rm Spin}(32)/\mathbb{Z}_2} = 2E_4 \Big(2\zeta(2) E_6 + (\p_3^2 G(z_{34},\tau) -\hat{G}_2 )E_4\Big)\ee  
on using \C{P}, $\sum_{\alpha= 2}^4 \theta_\alpha^{16} = 2 E_4^2$, and~\cite{Ellis:1987dc}
\be \sum_{\alpha=2}^4 \theta_\alpha^{16} e_{\alpha -1} = -4\zeta(2) E_4 E_6.\ee
This is equal to $L_{E_8\times E_8}$ on using $\sum_{\alpha=2}^4 \theta_\alpha^8 = 2 E_4$ and~\cite{Abe:1988cq}
\be \sum_{\alpha=2}^4 \theta_\alpha^8 e_{\alpha -1} = -4\zeta(2) E_6.\ee
Thus the two graviton--two gluon amplitude is the same in both the heterotic string theories~\cite{Abe:1988cq}. Note that $L$ as expressed in \C{simpleL}, involves Green functions rather than theta functions, which will be very useful for us.  

\section{The low momentum expansion of the one loop four graviton amplitude}

We want to obtain local interactions from the low momentum expansion of the one loop amplitudes. The one loop amplitude is generically of the form
\be \label{whole}\int_{\mathcal{F}}\frac{d^2\tau}{\tau_2^2} f(\tau,\bar\tau; k_i)\ee
where $f(\tau,\bar\tau;k_i)$ is modular invariant. Now \C{whole} contains contributions which are analytic as well as non--analytic in the external momenta. While the former yield local terms in the effective action, the latter yield non--local terms. The non--local terms are obtained from the boundary of moduli space as $\tau_2 \rightarrow \infty$. Thus to obtain the local contributions, we expand the integrand in powers of $\alpha'$ and integrate over the truncated fundamental domain given by 
~\cite{Green:1999pv,Green:2008uj}
\be \mathcal{F}_L = \{ -\frac{1}{2} \leq \tau_1 \leq \frac{1}{2}, \quad \vert \tau \vert \geq 1, \quad \tau_2 \leq L\},\ee
and take $L\rightarrow \infty$. On the other hand, the non--analytic contribution involves an integral over $\mathcal{R}_L$ defined by
\be \label{RL}\mathcal{R}_L = \{ -\frac{1}{2} \leq \tau_1 \leq \frac{1}{2}, \quad \vert \tau \vert \geq 1, \quad \tau_2 > L\},\ee
with appropriate integrands depending on the amplitude, leading to contributions which are non--perturbative in the external momenta. Note that $\mathcal{F} = \mathcal{F}_L \oplus \mathcal{R}_L$ is the fundamental domain of $SL(2,\mathbb{Z})$.

The analytic contributions are obtained by integrating over the truncated fundamental domain of $SL(2,\mathbb{Z})$, and the integral yields terms which are finite as well as divergent in this limit. While the finite contributions yield the desired local terms in the effective action, the divergences cancel those coming from the integral over $\mathcal{R}_L$.

We now perform the low momentum expansion of the four graviton amplitude given in \C{1loop4g}. The first non--vanishing term in this expansion has eight powers of external momenta, and terms involving lesser powers of external momenta vanish using elementary properties of the Green function on performing the integrals over the insertion points of the vertex operators. The leading $\mathcal{R}^4$ contribution has been calculated using the expression \C{1loop4g} in \cite{Ellis:1987dc}. 

At every order in the derivative expansion, from the expression \C{T} we obtain contributions from the terms of the form $A^4$, $A^2 R/\alpha'$ and $R^2/\alpha'^2$ schematically from $\mathcal{T}$. Each of these contributions must be multiplied by a term involving appropriate powers of the external momenta that result from expanding the Koba--Nielsen factor $e^{\mathcal{D}}$ as a polynomial in $\mathcal{D}$. We list the details that yield the various contributions below.

Thus on performing the $\alpha'$ expansion we get that
\be \label{1G}A^{1-loop}_{4g} (k_i, \epsilon^{(i)}) = \prod_{i=1}^4 \epsilon^{(i)}_{\mu_i\nu_i} K^{\mu_1 \mu_2 \mu_3 \mu_4} \int_{\mathcal{F}_L} \frac{d^2\tau}{\tau_2^2} \frac{{\bar{E}}_4^2}{{\bar{\eta}}^{24}} \mathcal{X}^{\n_1\n_2\n_3\n_4}\ee
where $\mathcal{X}^{\n_1\n_2\n_3\n_4}$ is given below for the various interactions, and we have integrated over the truncated fundamental domain.

We now give the various contributions in terms of various modular graph functions which are summarized in appendix B. 

\subsection{The $\mathcal{R}^4$ term}

From the $A^4 e^{\mathcal{D}}$ term, in the integrand we obtain
\bea \frac{1}{(4\pi)^4} (Q_1 + Q_2^2)I_{4,0}^{\nu_1\nu_2\nu_3\nu_4},\eea
while from the $A^2 R e^{\mathcal{D}}/\alpha'$ term, in the integrand we obtain
\bea \frac{2}{(4\pi)^4\alpha'} \Big[(Q_1 +  Q_2^2) I_{2;1,0} + \frac{\alpha'}{2} Q_1 I_{3,0}\Big]^{\nu_1\nu_2\nu_3\nu_4}.\eea
Finally from the $R^2 e^{\mathcal{D}}/\alpha'^2$ term, in the integrand we obtain
\bea \frac{1}{(8\pi^2)^2 \alpha'^2}\Big[ (Q_1 + Q_2^2) I_{1;2,0} - 2 Q_1 I_{1;0,1}\Big]^{\nu_1\nu_2\nu_3\nu_4}.\eea
Thus, for the $\mathcal{R}^4$ term, the total contribution is given by
\bea \label{c7}\mathcal{X}^{\n_1\n_2\n_3\n_4} = \frac{1}{(4\pi)^4} \Big[(Q_1 + Q_2^2) \Big( I_{4,0} +\frac{2}{\alpha'} I_{2;1,0} +\frac{4}{\alpha'^2} I_{1;2,0}\Big)- 2 Q_1 K\Big]^{\nu_1\nu_2\nu_3\nu_4}.\eea
Hence from \C{tR4} we see that the spacetime structures that arise at tree level and at one loop are the same.

\subsection{The $D^2\mathcal{R}^4$ term}

From the $A^4 e^{\mathcal{D}}$ term, in the integrand we obtain
\bea \frac{1}{(4\pi)^4} \Big[2(Q_2 Q_3 - Q_5) I_{4,1} +2(2 Q_7 + Q_8 - Q_6) I_{5,1} - 2(Q_2 Q_4 + Q_5 + Q_7) I_{6,1}\Big]^{\nu_1\nu_2\nu_3\nu_4}, \eea
while from the $A^2 R e^{\mathcal{D}}/\alpha'$ term, in the integrand we obtain
\bea &&\frac{1}{(8\pi^2)^2 \alpha'} \Big[ (Q_2 Q_3 - Q_5) I_{2;2,0} + (Q_2 Q_4 + Q_5 + 2 Q_6 -3 Q_7 -2 Q_8)I_{2;0,1}\non \\ && -\frac{\alpha'}{4} (Q_2 Q_4 + Q_5 - Q_6+ 3 Q_7 + Q_8)I_{3,1}\Big]^{\nu_1\nu_2\nu_3\nu_4}.\eea
Finally from the $R^2 e^{\mathcal{D}}/\alpha'^2$ term, in the integrand we obtain
\bea \frac{2}{(8\pi^2)^2 \alpha'^2}\Big[ (Q_2 Q_3 - Q_5) I_{1;3,0} + (2 Q_2 Q_4 + 2 Q_5 +  Q_6 - Q_8) I_{1;1,1}\Big]^{\nu_1\nu_2\nu_3\nu_4}.\eea

Thus for the $D^2\mathcal{R}^4$ term, the total contribution is given by
\bea \label{c6}&&\mathcal{X}^{\n_1\n_2\n_3\n_4} = \frac{2}{(4\pi)^4} (Q_2 Q_3 - Q_5) \Big(I_{4,1} + \frac{2}{\alpha'} I_{2;2,0} +\frac{4}{\alpha'^2} I_{1;3,0}\Big)^{\n_1\n_2\n_3\n_4} \non \\ &&+\frac{1}{(8\pi^2)^2\alpha'}(Q_2 Q_4 + Q_5 + 2 Q_6 - 3 Q_7 - 2 Q_8) \Big( I_{2;0,1} +\frac{2}{\alpha'} I_{1;1,1} -\frac{\alpha'}{6}(I_{5,1} + I_{6,1})\Big)^{\n_1\n_2\n_3\n_4}\non \\ &&+\frac{2}{(4\pi)^4} (Q_2 Q_4 + Q_5 - Q_6 + 3 Q_7 + Q_8)\Big(\widetilde{K}+\frac{1}{3} (I_{5,1} -2 I_{6,1})\Big)^{\n_1\n_2\n_3\n_4}, \eea
where we have defined
\be \widetilde{K}^{\m_1\m_2\m_3\m_4} = \frac{4}{\alpha'^2} I_{1;1,1}^{\m_1\m_2\m_3\m_4} - \frac{1}{2} I_{3,1}^{\m_1\m_2\m_3\m_4}.\ee
Hence from \C{tD2R4} we see that a new spacetime structure appears at one loop.

\section{The low momentum expansion of the one loop two graviton--two gluon amplitude}

We next consider the low momentum expansion of the two graviton--two gluon amplitude.
The first non--vanishing term in this expansion  has six powers of external momenta, since terms involving lesser powers of external momenta vanish using elementary properties of the Green function. This amplitude is given by
\be \label{2C}A^{1-loop}_{2g,2a}(k_i, \epsilon^{(i)}, e^{(i)}) = \epsilon^{(1)}_{\m_1\n_1} \epsilon^{(2)}_{\m_2\n_2} e^{(3)}_{a\n_3} e^{(4)}_{b\n_4}Tr(T^aT^b) K^{\nu_1 \nu_2 \nu_3 \nu_4} \int_{\mathcal{F}_L} \frac{d^2\tau}{\tau_2^2} \frac{\mathcal{Y}^{\m_1\m_2}}{{\bar{\eta}}^{24}} .\ee
In \C{2C}, the tensor $\mathcal{Y}^{\m_1\m_2}$ is given by 
\be\mathcal{Y}^{\m_1\m_2} = \mathcal{Y}_1^{\m_1\m_2} + \mathcal{Y}_2^{\m_1\m_2},\ee
where
\bea \label{two}\mathcal{Y}_1^{\m_1\m_2} &=& \frac{2\pi^2}{3} \bar{E}_4 (\bar{E}_6 - \bar{\hat{E}}_2 \bar{E}_4) \prod_{i=1}^4 \int_\S  \frac{d^2 z^i}{\tau_2} e^{\mathcal{D}} \Big(\frac{R_{12}\eta^{\m_1\m_2}}{2\alpha'} + A^{\m_1} A^{\m_2}\Big)\non \\ \mathcal{Y}_2^{\m_1\m_2} &=& - 8\pi^2 \bar{E}_4^2 \prod_{i=1}^4 \int_\S  \frac{d^2 z^i}{\tau_2} e^{\mathcal{D}} R_{34}\Big(\frac{R_{12}\eta^{\m_1\m_2}}{2\alpha'} + A^{\m_1}A^{\m_2}\Big).\eea 
We split the total contribution into a sum of two terms as this is useful for our purposes. To obtain the answer at various orders in the derivative expansion, we have to expand the Koba--Nielsen factor upto the appropriate order.  We now list the contributions to the terms in the effective action that arise from the two terms in \C{two}. The $\mathcal{R}^2 \mathcal{F}^2$ term has been obtained from \C{2C} in \cite{Ellis:1987dc}.

\subsection{Contribution from $\mathcal{Y}_1^{\m_1\m_2}$}

This contributes 
\be \label{c1}\mathcal{Y}_1^{\m_1\m_2} = -\frac{Q_2}{24}\bar{E}_4 (\bar{E}_6 - \bar{\hat{E}}_2 \bar{E}_4) J_1^{\mu_1\mu_2}\ee
to the $\mathcal{R}^2 \mathcal{F}^2$ term,
and
\be \label{c2}\mathcal{Y}_1^{\m_1\m_2} = -\frac{\alpha'}{96}\bar{E}_4 (\bar{E}_6 - \bar{\hat{E}}_2 \bar{E}_4)(Q_3 s J_1^{\mu_1\mu_2} - 2 Q_4 J_2^{\mu_1\mu_2})\ee
to the $D^2 \mathcal{R}^2 \mathcal{F}^2$ term. 

This also contributes
\bea \label{c4}\mathcal{Y}_1^{\m_1\m_2} &=& -\frac{\alpha'^2}{3\cdot 256} \bar{E}_4 (\bar{E}_6 - \bar{\hat{E}}_2 \bar{E}_4)\Big[ \Big(Q_9 s^2 + (s^2 + 2t^2 + 2u^2)Q_2 \mathcal{E}_2 \non \\ &&+  4utQ_{10} + 2(s^2-ut)Q_{11}\Big) J_1^{\m_1\m_2} + 2 Q_{11}s J_2^{\m_1\m_2}\Big]\eea
to the $D^4 \mathcal{R}^2 \mathcal{F}^2$ term.

\subsection{Contribution from $\mathcal{Y}_2^{\m_1\m_2}$}

While this does not contribute to the $\mathcal{R}^2 \mathcal{F}^2$ term, it contributes
\be \label{c3}\mathcal{Y}_2^{\m_1\m_2} = -\frac{\alpha'}{32\pi^2}\bar{E}_4^2 \Big( (Q_1 + Q_2^2) s J_1^{\mu_1\mu_2} + 2 Q_1 J_2^{\mu_1\mu_2}\Big)\ee
to the $D^2 \mathcal{R}^2 \mathcal{F}^2$ term, and
\bea \label{c5}\mathcal{Y}_2^{\m_1\m_2}&=&-\frac{\alpha'^2}{128\pi^2}\bar{E}_4^2\Big[ 2 Q_2 Q_3 s^2 J_1^{\m_1\m_2} + 2Q_2 Q_4 (tu J_1^{\m_1\m_2} - s J_2^{\m_1\m_2})\non \\ &&-Q_5 \Big((s^2 + t^2 + u^2)J_1^{\m_1\m_2} +2 s J_2^{\m_1\m_2}\Big) - 6 Q_7 (tu J_1^{\m_1\m_2} + s J_2^{\m_1\m_2})\non \\ &&+2 (Q_6 -Q_8)(2ut J_1^{\m_1\m_2} + s J_2^{\m_1\m_2})\Big   ] \non \\ && = -\frac{\alpha'^2}{64\pi^2}\bar{E}_4^2\Big[ \Big(s^2(Q_2 Q_3 - Q_5)+ut(Q_2 Q_4 +Q_5 +2Q_6 - 3 Q_7 -2 Q_8)\Big)J_1^{\m_1\m_2} \non \\ &&-s(Q_2 Q_4 + Q_5 - Q_6 + 3 Q_7 + Q_8)J_2^{\m_1\m_2}\Big]\eea
to the $D^4\mathcal{R}^2\mathcal{F}^2$ term. Note the striking equality of the structure of the integrands between \C{c7} and \C{c3}, and between \C{c6} and \C{c5} as well.

Thus the total contribution to the $\mathcal{R}^2 \mathcal{F}^2$ term is given by \C{c1}, and to the $D^2\mathcal{R}^2 \mathcal{F}^2$ term is given by the sum of \C{c2} and \C{c3}. Finally, the total contribution to the $D^4\mathcal{R}^2 \mathcal{F}^2$ term is given by the sum of \C{c4} and \C{c5}. Note that the two tensors $J_1^{\m_1\m_2}$ and $J_2^{\m_1\m_2}$ which appeared in the amplitudes at tree level appear at one loop as well, and there are no new tensors that arise. 

\section{Various equations for the modular graph functions}

From the previous analysis, we see that the integrands we need involve modular graph functions where the links are either Green functions $G_{ij}$ or their antiholomorphic derivatives $\bar\p_i G_{ij}$. To evaluate these integrands, we now analyze them in detail.   

\subsection{Graphs obtained by direct evaluation}

It is easy to evaluate graphs where every vertex has coordination number 2 and every link is given by a Green function. For a graph with $s$ links each given by a Green function, this gives us
\be \mathcal{G}_s = \prod_{i=1}^s \int_{\S} \frac{d^2 z^i}{\tau_2} G_{12} G_{23} \ldots G_{s1} =  \sum_{(m,n) \neq (0,0)} \frac{\tau_2^s}{\pi^s \vert m\tau+ n\vert^{2s}} = \mathcal{E}_s (\tau,\bar\tau),\ee  
where $\mathcal{E}_s$ is the $SL(2,\mathbb{Z})$ invariant non--holomorphic Eisenstein series, as depicted for $\mathcal{E}_6$ in figure 1. For later purposes, we note that $\mathcal{E}_s$ satisfies the Laplace equation
\be \label{eigenE}4\tau_2^2 \frac{\p^2 \mathcal{E}_s}{\p \tau \p\bar\tau} = s(s-1) \mathcal{E}_s ,\ee
and it has the large $\tau_2$ expansion given by
\bea \label{expE}\mathcal{E}_s (\tau,\bar\tau) &=& \frac{2\zeta(2s)}{\pi^s} \tau_2^s + \frac{2\Gamma(s-1/2)\zeta(2s-1)}{\Gamma(s) \pi^{s-1/2}} \tau_2^{1-s} \non \\ &&+ \frac{4\sqrt{\tau_2}}{\Gamma(s)} \sum_{m\neq 0} \vert m \vert^{s-1/2} \s_{1-2s}(m)K_{s-1/2} (2\pi\vert m\vert \tau_2) e^{2\pi i m\tau_1},\eea
where $\s_{\alpha} (m)$ is the divisor function defined by
\be \s_{\alpha} (m) = \sum_{d|m, d>0}d^\a.\ee

\begin{figure}[ht]
\begin{center}
\[
\mbox{\begin{picture}(190,100)(0,0)
\includegraphics[scale=.6]{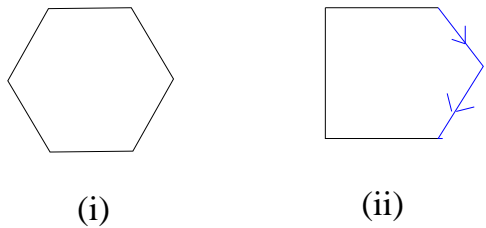}
\end{picture}}
\]
\caption{ The graphs (i) $\mathcal{E}_6$ and (ii) $R_{3,2}$}
\end{center}
\end{figure}

Next consider the case where every vertex has coordination number 2, but some of the links are given by the Green function $G_{ij}$, while the others are given by its antiholomorphic derivative $\bar\p_i G_{ij}$. For a graph with a total of $s$ links, where $p$ links involve an antiholomorphic  derivative and the remaining  $s-p$ links do not, we have that~\footnote{This graph is non--vanishing only if $p$ is an even integer.}
\bea R_{s,p} &=& \prod_{i=1}^s \int_{\S} \frac{d^2 z^i}{\tau_2} (\bar\p_2 G_{12} \bar\p_3 G_{23} \ldots \bar\p_{p+1} G_{p,p+1}) G_{p+1,p+2} \ldots G_{s1} \non \\ &=& \sum_{(m,n) \neq (0,0)} \frac{1}{(m\bar\tau + n)^p} \times  \frac{\tau_2^{s-p}}{\pi^{s-p} \vert m\tau+ n\vert^{2(s-p)}}\eea
as depicted for $R_{3,2}$ in figure 1. We have taken the derivatives to be on consecutive links without loss of generality as all other graphs where the derivatives are not on consecutive links are related to it by trivial integration by parts. Note that $R_{s,0} = \mathcal{G}_s$.  

When $s=p$, we have that 
\be R_{s,s} = \sum_{(m,n) \neq (0,0)} \frac{1}{(m\bar\tau + n)^s}.\ee
For $s > 2$, $R_{s,s}$ is antiholomorphic and $R_{s,s} = \overline{G_s(\tau)}$. For $s=2$, this has to be regularized, and $R_{2,2} = \overline{\hat{G}_2} (\tau,\bar\tau) = \overline{G_2 (\tau)} - \pi/\tau_2$. Note that $R_{s,p}$ is a modular form of weight $(0,p)$.  

In fact, $R_{s,p}$ can be related to $\mathcal{E}_{s-p/2}$ by the action of appropriate number of modular covariant derivatives. They are defined by
\be \label{modcov}D_m = i\Big( \frac{\p}{\p\tau} - \frac{im}{2\tau_2}\Big), \quad \bar{D}_n = - i\Big(\frac{\p}{\p\bar\tau} +\frac{in}{2\tau_2}\Big).\ee  
Now $D_m$ maps a modular form $\Phi^{(m,n)}$ of weight $(m,n)$ to a modular form of weight $(m+2,n)$, while $\bar{D}_n$ maps $\Phi^{(m,n)}$ to a modular form of weight $(m,n+2)$. We get that
\be \label{defR}R_{s,p} = (2\pi)^{p/2} \frac{\G(s-p/2)}{\G(s)} \bar{D}_{p-2} \ldots \bar{D}_4 \bar{D}_2 \bar{D}_0 \mathcal{E}_{s-p/2}.\ee 

It is very useful for our purposes to note that $R_{s,p}$ satisfies the eigenvalue equation
\be 4\tau_2^2 \bar{D}_p D_0 R_{s,p} = (s-1)(s-p) R_{s,p}.\ee
Now using \C{modcov}, we also define the modular covariant derivatives
\be \mathcal{D}_m = \tau_2 D_m, \quad \bar{\mathcal{D}}_n =\tau_2 \bar{D}_n.\ee
It follows that $\mathcal{D}_m$ maps a modular form $\Phi^{(m,n)}$ of weight $(m,n)$ to a modular form of weight $(m+1,n-1)$, while $\bar{\mathcal{D}}_n$ maps $\Phi^{(m,n)}$ to a modular form of weight $(m-1,n+1)$.

\subsubsection{The equations for $Q_1$, $Q_2$, $Q_4$ and $Q_{11}$}

Thus based on the discussion above, we have that
\bea Q_1 = 2 \zeta(4) \bar{E}_4, \quad Q_2 = 2\zeta(2) \overline{\hat{E}_2}, \quad Q_4 = \pi\bar{D}_0 \mathcal{E}_2, \quad Q_{11} = \frac{2\pi}{3} \bar{D}_0 \mathcal{E}_3. \eea 
Thus these graphs are very easily related to simple covariant modular forms.  

\subsection{Graphs obtained by solving differential equations}

The remaining modular graph functions do not belong to the class of graphs we have discussed above, and we next deduce differential equations satisfied by them. In order to deduce these equations, rather than analyzing the variations of these graphs with respect to the complex structure $\tau$, we find it convenient to analyze the variations with respect to the Beltrami differential. In this way of analyzing deformations of the complex structure, only the worldsheet metric is deformed, and hence in the modular graph functions we need to vary only the Green functions and their worldsheet derivatives.  

To obtain these variations, we note that for the Beltrami differential $\mu$, the holomorphic variation $\delta_\mu \Psi$ for any $\Psi$ is given by
\be \label{Beltrami}\delta_\mu \Psi = \frac{1}{\pi} \int_\S d^2 z \mu_{\bar{z}}^{~z} \delta_{zz} \Psi.\ee 
On the torus, the Beltrami differential $\m_{\bar{z}}^{~z} =1$, but it leads to non--trivial variations. We denote the holomorphic and anti--holomorphic infinitesimal variations as $\delta_{\mu}$ and $\delta_{\bar\m}$ respectively.

On using the known formulae for the variations of the prime form, period matrix and the Abelian differential in \C{Beltrami}~\cite{Verlinde:1986kw,DHoker:1988pdl}, the  relevant variation of $G_{ij}$ is given by
\bea  \label{onevar} \delta_\mu G(z_1,z_2) &=& -\frac{1}{\pi} \int_\S d^2 z \p_z G(z,z_1) \p_z G(z,z_2),
\eea
while the relevant variations of $\bar\p_i G_{ij}$ are given by~\footnote{To deduce the first equation in \C{morevar}, we also use $\delta_{\bar{w}\bar{w}} \p_z = \pi \delta^2(z-w) \bar\p_z$.}
\bea \label{morevar}\delta_{\m} \bar\p_{z_1} G(z_1,z_2) &=& 0, \non \\
\delta_{\bar{\m}} \bar\p_{z_1} G(z_1,z_2) &=& -\frac{1}{\pi} \bar\p_{z_1}\int_\S d^2 z \bar\p_z G(z,z_1) \bar\p_z G(z,z_2).\eea
We note that $\delta_{\m} \bar\p_{z_1} G(z_1,z_2) =0$ is true only in the bulk of moduli space. It can receive contributions from the boundary of moduli space which we shall discuss separately~\footnote{In fact, this is also true of $\delta_{\bar{\m}} \delta_\mu \bar\p_{z_1}G(z_1,z_2) = 0$.}.  

We now explain why we find it very convenient to analyze the variation of the complex structure in terms of variations of the Beltrami differential rather than directly analyzing variations of $\tau$. This is because of the structure of the modular graph functions that arise in our analysis. Since the vertices of the graphs are integrated over, every graph only depends on $\tau$ and $\bar\tau$. However, in the explicit calculations to determine the variations, we have to perform the variations of the Green functions and their derivatives that are the links of the graphs. These variations are simpler for the Beltrami differentials rather than the complex structure itself. In fact, we have that
\bea \label{imp}-\frac{i}{2} \delta_{\m} G(z,0) &=& -i \tau_2 D_0 G(z,0) + z_2 \frac{\p G (z,0)}{\p z} , \non \\ -\frac{i}{2} \delta_{\m} \frac{\p G(z,0)}{\p z} &=&-i \tau_2 D_1 \frac{\p G(z,0)}{\p z}+ z_2 \frac{\p^2 G (z,0)}{\p z^2}\eea   
leading to involved calculations if variations of $\tau$ are directly studied.  Note that the variation with respect to $\tau$ arises in a covariant way since it involves $D_m$, where $m$ depends on the modular weight of the object on which $\delta_{\mu}$ acts.

Note that graphs in \C{graphs} that remain to be analyzed are $(0,p)$ modular forms, where $p=2,4$. Thus since $\delta_{\m}\bar\p_z G (z,0)=0$ upto contributions that arise from the boundary of moduli space,
our primary strategy will be to first act with $\delta_{\m}$ on them, and then with $\delta_{\bar\mu}$, which will yield an eigenvalue equation we shall shortly discuss. On manipulating the equations we obtain, this will yield modular covariant second order differential equations the various graphs satisfy on also including the contributions that arise from the boundary of moduli space. 

We now analyze the action of $\delta_{\mu}$ and $\delta_{\bar\mu}$ on the various modular graph functions that are relevant for our purposes, where the vertices are integrated over. We first act with $\delta_{\m}$ on $(0,p)$ modular forms, which are graphs with $p$ links involving $\bar\p G$, while the remaining links involve $G$. For links involving  $G$, we use the identity
\be \label{L}-i \delta_{\m} G(z_{ij}) = -2i \tau_2 D_0 G(z_{ij}) + 2\Big({\rm Im} z_i \frac{\p}{\p z_i} + {\rm Im}z_j \frac{\p}{\p z_j}\Big) G(z_{ij}), \ee   
while for links involving $\bar\p G$ we use the identity
\be i\tau_2 D_0 \bar\p_i G(z_{ij}) = \Big({\rm Im} z_i \frac{\p}{\p z_i} + {\rm Im}z_j \frac{\p}{\p z_j}\Big) \bar\p_i G(z_{ij}),\ee
which trivially follows from $\delta_{\m} \bar\p_i G(z_{ij})=0$ and hence is true upto boundary terms. 
Including all the contributions and using momentum conservation at each vertex, all the terms involving derivatives of $z_i$ cancel leading to
\be \delta_{\m} = 2 \mathcal{D}_0\ee 
on the $(0,p)$ graph. 

After acting with $\delta_{\m}$ on the $(0,p)$ graph, in all the cases we shall encounter, we are left with graphs having either 2 or 0 $\bar\p G$ links, while the remaining links are all given by $G$~\footnote{There is also a case where the the resulting graph has two links with $\p G$ on them. The action of $\delta_{\bar\mu}$ on it can be trivially obtained by complex conjugation of what we have described above.}. We now have to act on them with $\delta_{\bar\m}$. If the graph has no links given by $\bar\p G$ and hence is $SL(2,\mathbb{Z})$ invariant, then simply using the complex conjugate of \C{L} on every link and momentum conservation at the vertices, we see that
\be \delta_{\bar\m} = 2 \bar{\mathcal{D}}_0.\ee 
Finally if there are $q$ links given by $\bar\p G$ as well, let us consider the action of $\delta_{\bar\m}$ on the graph. We use the complex conjugate of \C{L} for the links involving $G$, and the complex conjugate of
\be -i \delta_{\m} \p_i G(z_{ij}) = -2i \tau_2 D_1 \p_i G(z_{ij}) + 2 \Big({\rm Im} z_i \frac{\p}{\p z_i} + {\rm Im}z_j \frac{\p}{\p z_j}\Big)\p_i G(z_{ij})\ee
for those involving $\bar\p G$. Adding the contributions and using momentum conservation at each vertex, we see that
\be\delta_{\bar\m} = 2\bar{\mathcal{D}}_q. \ee  
Thus, based on the cases we consider and modular covariance, we obtain the general result that acting on a modular form of weight $(m,n)$, we have that
\be \delta_{\mu} = 2 \mathcal{D}_m, \quad \delta_{\bar\m} = 2\bar{\mathcal{D}}_n,\ee 
simply yielding modular covariant derivatives. 

Thus first acting with $\delta_{\m}$ and then with $\delta_{\bar\m}$ on the various graphs we get second order differential equations, with the operator acting on the graphs given by    
\be \delta_{\bar\m}\delta_{\mu} = 4 \bar{\mathcal{D}}_{n-1}  \mathcal{D}_m = 4 \tau_2^2 \bar{D}_n D_m \ee  
which maps a modular form of weight $(m,n)$ to a modular form of weight $(m,n)$. Hence this is a modular covariant Laplacian of $SL(2,\mathbb{Z})$. Note that acting on the modular form of weight $(m,n)$ in the reverse order, one can also define another modular covariant Laplacian 
\be \delta_{\m} \delta_{\bar\m} = 4 \mathcal{D}_{m-1} \bar{\mathcal{D}}_n = 4\tau_2^2 D_m \bar{D}_n.\ee
They satisfy the simple relation
\be 4(\bar{\mathcal{D}}_{n-1}  \mathcal{D}_m - \mathcal{D}_{m-1} \bar{\mathcal{D}}_n) = n-m,\ee
and hence they commute only when $m=n$.

This generalizes the analysis for $SL(2,\mathbb{Z})$ invariant graphs. Note that the $SL(2,\mathbb{Z})$ invariant Laplacian acting on them is given by
\be \Delta  = 4 \bar{\mathcal{D}}_{-1}  \mathcal{D}_0 = 4\mathcal{D}_{-1} \bar{\mathcal{D}}_0= 4\tau_2^2\frac{\p^2}{\p\tau \bar\p\tau} .\ee

Before proceeding, we consider the issue of possible contributions to the differential equations that arise from factors of $\delta_{\m} \bar\p_z G (z,w)$ in the graphs. As we have mentioned above, such contributions vanish in the bulk of moduli space on simply using \C{morevar}. However, this manipulation involves using the relation
\be \delta_{{u} {u}} \Big(\bar\p_z G(z,w)\Big) = \frac{\pi}{2} \p_u \delta^2 (z-u) +\frac{\pi}{\tau_2} \p_u \Big(G(u,w) - G(u,z)\Big)\ee
which has a contact term contribution, which can give a non--vanishing answer even though it leads to a total derivative. These are interpreted as potential contributions coming from the boundary of moduli space when $z$ and $u$ are coincident\footnote{Such boundary contributions, along with the boundary contributions that arise from $\tau_2 \rightarrow \infty$, are part of the definition of the Deligne--Mumford compactification of the moduli space of Riemann surfaces with punctures. }. In fact, indeed in some of the cases we consider, there are non--vanishing contributions.    

Let us first consider the simplest example where this happens, which captures the main issue that is central to all the other cases. Consider the graph $Q_2$ which is a weight $(0,2)$ modular form given by
\be \label{G2def}{Q}_2 = \overline{\hat{G}}_2 = \prod_{i=1}^2 \int_{\S}\frac{d^2 z^i}{\tau_2} \bar\p_1 G_{12} \bar\p_2 G_{12} =  {\rm lim}_{s\rightarrow 0} \sum_{(m,n)\neq (0,0)} \frac{1}{(m+n\bar\tau)^2 \vert m+n\tau\vert^{2s}}\ee
which arises in our analysis. On acting with $\delta_{\m}$ we see that the entire contribution comes from the factors of $\delta_{\m} \bar\p G$ in the integrand. However, instead of evaluating it directly this way, let us evaluate it using the expression given by the lattice sum.  
Naively on setting $s=0$, it is antiholomorphic and hence $\delta_{\m} \overline{\hat{G}}_2$ vanishes, leading to an incorrect answer simply because this is not modular covariant, and one needs to regularize it as in \C{G2def} to obtain a modular covariant expression. It is this regularization that breaks antiholomorphy leading to
\be \delta_{\m} \overline{\hat{G}}_2 = \frac{\pi}{\tau_2}\ee   
which trivially follows from the asymptotic expansion of $\overline{\hat{G}}_2$ given by
\be \label{term}\overline{\hat{G}}_2 (\tau,\bar\tau)= \overline{{G}_2 (\tau)} - \frac{\pi}{\tau_2 } = 2\zeta(2) - \frac{\pi}{\tau_2} +\ldots,\ee
where we have ignored terms that are exponentially suppressed as $\tau_2\rightarrow \infty$ which are functions only of $\bar{\tau}$. Thus we see that $\delta_{\m} \bar\p G$  yields non--vanishing contributions for $Q_2$. In fact, we see that $\delta_{\m} \overline{\hat{G}}_2$  receives contribution only from the second term in the last equation in \C{term} which has less transcendentality than the first (we assign zero transcendentality to $\tau_2$ and $k$ to $\zeta(k)$). In fact, in this case, this is the only contribution. 

This is the route we take in determining possible contributions to a graph that arise from factors of $\delta_{\m} \bar\p G$ in the integrand, rather than calculating them directly. We start with the regularized expression for the graph expressed as a lattice sum and perform the asymptotic expansion and keep only the terms that are power behaved as $\tau_2 \rightarrow \infty$. These terms separates into two distinct kinds of contributions: 

(i) A set of terms that precisely match with those in the asymptotic expansion obtained by acting with $\delta_{\m}$ on the graph, and  simply setting  $\delta_{\m} \bar\p G=0$ for all the links. This is done by directly obtaining the variation of the graph and performing the matching.

(ii) The rest, which gives the non--vanishing contributions that must result from the variations $\delta_{\m} \bar\p G$ along the various links.  
   
The contribution (i) from $\delta_{\m}$ acting on the graph yields an $SL(2,\mathbb{Z})$ covariant expression. Thus we then extend the power behaved terms in (ii) that give additional contributions to $\delta_{\m}$ acting on the same graph, to its $SL(2,\mathbb{Z})$ covariant completion to get the complete answer. For the cases we are interested in, (ii) yields simple expressions which have obvious $SL(2,\mathbb{Z})$ covariant completions.  It will be interesting to obtain these results by directly computing the $\delta_{\m}$ variations acting on $\bar\p G$ in the graphs for more general cases. Perhaps generalizations of the techniques in~\cite{DHoker:2013sqy,Berkovits:2014rpa,Sen:2015uoa} will be useful in evaluating these boundary contributions.    

Thus this leads to the complete equation for $\delta_{\m}$ acting on any graph $Q$. Further acting on it with $\delta_{\bar\m}$ gives the desired $SL(2,\mathbb{Z})$ covariant Poisson equation of the form
\be \delta_{\bar\m}  \delta_{\m}Q = \ldots,\ee
where the terms on the right hand side are easily determined\footnote{For the cases we consider, $\delta_{\m}Q$ does not yield any graph with $\p G$ as a link on which the subsequent action of $\delta_{\bar\m}$  gives a non--vanishing contribution. For cases where there are such such contributions, we have to evaluate them by taking complex conjugate of the results that follow from the discussion above.}. 
   
For the cases we consider, there are some common features that arise in our analysis for the contributions from $\delta_{\m} \bar\p G$ to a graph as mentioned above:

{\bf{(i)}} These contributions to a graph $Q$  that arise from the boundary of moduli space are given by very simple $SL(2,\mathbb{Z})$ covariant modular forms. On the other hand, the remaining contributions to $Q$ satisfy involved $SL(2,\mathbb{Z})$ covariant Poisson equations. Heuristically this is expected as boundary terms result from regularized expressions and they receive very few power behaved contributions in the large $\tau_2$ expansion, which have simple $SL(2,\mathbb{Z})$ covariant completions for the cases we consider. 

{\bf{(ii)}} The asymptotic expansions of these contributions that arise from the boundary of moduli space have less transcendentality  than the terms with maximal transcendentality in the asymptotic expansion of the remaining terms which arise from the bulk of moduli space\footnote{Note that there also some contributions with reduced transcendentality that arise in the asymptotic expansion of the remaining terms. These boundary terms are directly determined in our analysis and do not have to be dealt with separately.}. 

{\bf{(iii)}} Finally, there is a very simple diagnostic to ascertain which graphs receive contributions from the boundary of moduli space resulting from non--vanishing $\delta_{\m} \bar\p G$: these are the only ones which have a factor of $(\bar\p_1 G_{12})^2$ in the integrand. Heuristically this is expected, because it is only $\overline{\hat{G}}_2$ which needs regularization, and has  $(\bar\p_1 G_{12})^2$ as the integrand. Note that $R_{s,p}$ for $s \geq 3$ do not receive such contributions. Thus for example, $\delta_{\m} R_{3,2} \sim \mathcal{E}_2/\tau_2$. It would be interesting to check if this diagnostic works in all cases.     

Hence among the various graphs we consider only $Q_2$, $Q_3$, $Q_5$, $Q_9$ and $Q_{10}$ receive such contributions. Thus only for these graphs, we have to determine these boundary terms separately. 

We now obtain modular covariant eigenvalue equations of the Poisson type having source terms for the various graphs, based on the discussions above. In the equations below, the vertices in the various graphs are always integrated with measure $d^2 z/\tau_2$. The relevant asymptotic expansions of the various graphs are described in detail in appendix C.    

Let us now summarize the strategy we follow in obtaining the eigenvalue equations in the analysis below. We first consider $\delta_{\mu} Q$ for any graph $Q$, ignoring contributions of the form $\delta_{\mu} \bar\p G $, leading to
\be \label{act}\delta_{\mu} Q = \tilde{Q}\ee 
where $\tilde{Q}$ is modular covariant. This leads to a modular covariant Poisson equation for $Q$ on further acting on \C{act} with $\delta_{\bar\m}$ given by
\be \label{incomp}\delta_{\bar\m} \delta_{\m} Q = \delta_{\bar\m} \tilde{Q}.\ee
We next match the power behaved terms in the large $\tau_2$ expansion of $Q$ obtained from \C{incomp} with the exact asymptotic expansion of $Q$ analyzed in appendix C. Note that this also includes the zero mode of the modular covariant Laplacian. This can lead to extra contributions which modify \C{act} to the exact equation
\be \label{act2}\delta_{\mu} Q = \tilde{Q} + f,\ee
where $f$ is modular covariant and includes the contributions from $\delta_{\mu} \bar\p G $ acting on $Q$. This finally leads to the exact Poisson equation 
\be \delta_{\bar\m} \delta_{\m} Q = \delta_{\bar\m} \tilde{Q} +\delta_{\bar\m} f\ee
for the graph $Q$.  

For the various cases we consider, in the process of matching the power behaved terms in the asymptotic expansions obtained from \C{incomp} with the exact analysis in the appendix, the extra terms in $Q$ that lead to $f$ in \C{act2} (this includes the zero mode of the Laplacian) can be completed to $SL(2,\mathbb{Z})$ covariant modular forms\footnote{This completion is unique, given the power behaved terms in the large $\tau_2$ expansion, modular covariance and the boundary condition that all other subleading contributions are exponentially suppressed as $\tau_2 \rightarrow \infty$. Note that we are neglecting contributions from cusp forms in our analysis.}, hence leading to additional contributions to $Q$. In the analysis below, we present the exact equations.

\subsubsection{The equation for $Q_3$}

We first consider the equation for the graph $Q_3$. Varying with respect to the Beltrami differential $\mu$, we get that 
\be \delta_{\m} Q_3 = \frac{2\pi}{\tau_2} \mathcal{E}_2 -\frac{\pi}{ \tau_2},\ee
where the second term on the right hand side is a boundary contribution we have fixed using the asymptotic expansion. Note that each of the contributions is modular covariant.

This further leads to
\be \delta_{\bar\m}  \delta_{\m}  Q_3 = 4 Q_4,\ee
yielding the differential equation
\be \label{E1}\tau_2^2 \bar{D}_2 D_0 Q_3 = Q_4 = \pi \bar{D}_0 \mathcal{E}_2.\ee
Now to solve for $Q_3$, we consider the eigenvalue equation satisfied by $\bar{D}_0 \mathcal{E}_s$ given by
\be \label{vaL}\tau_2^2 \bar{D}_2 D_0 (\bar{D}_0 \mathcal{E}_s) = \frac{s(s-1)}{4} (\bar{D}_0 \mathcal{E}_s).\ee
Thus \C{E1} yields the solution  
\be \label{f1}Q_3 = 2\pi \bar{D}_0 \mathcal{E}_2 -2\zeta(2) \overline{\hat{E}}_2,\ee
since $\bar{E}_2 \sim  \bar{D}_0 \mathcal{E}_1$ is a covariant solution of the homogeneous equation \C{E1} having modular weight $(0,2)$, and the normalization is fixed by the asymptotic expansion. Thus we have that
\be Q_3 = 2 Q_4 -Q_2 .\ee
Hence we see that the boundary contribution to $Q_3$ simply involves $\overline{\hat{E}}_2$ which is a zero mode of the covariant Laplacian.

\subsubsection{The equation for $Q_{10}$}

We next consider the graph $Q_{10}$. Proceeding as above, we have that 
\be \delta_{\m} Q_{10} = \frac{4\pi}{\tau_2} \mathcal{E}_3 -\frac{\pi}{\tau_2}\mathcal{E}_2 ,\ee
where the second term is a boundary term whose structure is determined by the asymptotic expansion.  
This further leads to
\be \delta_{\bar\m}  \delta_{\m}  Q_{10} = 12 Q_{11} - 2 Q_4,\ee
which we rewrite as
\be \label{E2}\tau_2^2 \bar{D}_2 D_0 Q_{10} = 3 Q_{11} - \frac{1}{2} Q_4 = 2\pi \bar{D}_0 \mathcal{E}_{3} -\frac{\pi}{2} \bar{D}_0 \mathcal{E}_2.\ee
On using \C{vaL} for $s=2$ and $s=3$, we get the covariant solution
\be \label{f2}Q_{10} = \frac{4\pi}{3} \bar{D}_0 \mathcal{E}_3 -\pi  \bar{D}_0 \mathcal{E}_2.\ee
Thus we obtain the relation
\be Q_{10} = 2 Q_{11} - Q_4.\ee
Note that a contribution proportional to $\overline{\hat{E}}_2$, the zero mode of the covariant Laplacian  is ruled out based on the asymptotic expansion. Also the boundary term contribution involves $\bar{D}_0 \mathcal{E}_2$, which is not a zero mode of the Laplacian.

\subsubsection{The equation for $Q_5$}

We next consider the equation for $Q_5$. We have that 
\be \delta_{\m} Q_5 = -\frac{\pi}{\tau_2} Q_{10} -\frac{\pi^2}{\tau_2} \bar{D}_0 \mathcal{E}_2 = -\frac{4\pi^2}{3\tau_2} \bar{D}_0 \mathcal{E}_3,\ee
where the boundary contribution arising from $\delta_{\m} \bar\p G$ is given by $-\pi^2\bar{D}_0 \mathcal{E}_2/\tau_2$. This has been chosen such that it cancels a contribution from $Q_{10}$ based on the asymptotic expansion. 
This further  leads to
\be \delta_{\bar\m}  \delta_{\m}  Q_5 =  -8 R_{5,4}= -\frac{8\pi^2}{3} \bar{D}_2 \bar{D}_0 \mathcal{E}_3,\ee
where we have used
\be \label{use}\delta_{\bar\m} \bar{D}_0 \mathcal{E}_3 = \frac{3}{2\pi} \delta_{\bar\m} Q_{11} = \frac{6\tau_2}{\pi^2} R_{5,4}.\ee
Thus we obtain the equation
\be \label{hg}\tau_2^2 \bar{D}_4 D_0 Q_5 = -\frac{2\pi^2}{3} \bar{D}_2 \bar{D}_0 \mathcal{E}_3.\ee
On using the eigenvalue equation
\be \label{e}\tau_2^2 \bar{D}_4 D_0 (\bar{D}_2 \bar{D}_0 \mathcal{E}_s) = \frac{(s+1)(s-2)}{4} (\bar{D}_2 \bar{D}_0 \mathcal{E}_s),\ee
we get the solution
\be \label{f3}Q_5 = -\frac{2\pi^2}{3} \bar{D}_2 \bar{D}_0 \mathcal{E}_3 +\frac{2\pi^2}{3} \bar{D}_2 \bar{D}_0 \mathcal{E}_2,\ee
since $\bar{E}_4 \sim \bar{D}_2 \bar{D}_0 \mathcal{E}_2$ is a covariant solution of the homogeneous equation \C{hg} having modular weight $(0,4)$, and the constant is fixed based on the asymptotic expansion.  Thus
\be Q_5 = -2 R_{5,4} + Q_1.\ee
Hence the boundary contribution to $Q_5$ arising from $\delta_{\m} \bar\p G$ is proportional to $\bar{D}_2 \bar{D}_0 \mathcal{E}_2$ which is a zero mode of the covariant Laplacian. 

Thus we see that the graphs $Q_3$, $Q_{10}$ and $Q_5$ are given by very simple expressions which solve the eigenvalue equation. The remaining graphs require more involved analysis, to which we now turn. Of these graphs, $Q_6$, $Q_7$ and $Q_8$ do not receive any boundary contributions arising from $\delta_{\m} \bar\p G$ contributions. 

\subsubsection{The equation for $Q_6$}

We next consider the equation for $Q_6$, which gives us\footnote{Using the various expressions for the asymptotic expansions given in appendix C, it is straightforward to check \C{consis1} for the power behaved terms. This provides a non--trivial check of the boundary term contribution to $Q_{10}$.}
\be \label{consis1}\delta_{\mu} Q_6 = \frac{2\pi}{\tau_2} Q_{11} +\frac{\pi}{\tau_2} Q_{10}- \frac{\pi}{\tau_2} \mathcal{E}_2 Q_2.\ee

Now using \C{use}, the relations
\be \delta_{\bar\m} Q_4 = \frac{3\tau_2}{\pi}Q_1, \quad \delta_{\bar\m} \mathcal{E}_2 = \frac{2\tau_2}{\pi}Q_4, \ee
and~\footnote{We have used the regularized expression
\be \label{reG}\mathcal{E}_1 = -{\rm ln} (\tau_2 \vert\eta (\tau)\vert^4),\ee
which leads to
\be \label{needlat}\bar{D}_0 \mathcal{E}_1 = \frac{\pi}{6} \overline{\hat{E}}_2.\ee
We also use the identity
\be \bar{D}_2 \overline{\hat{E}}_2 = \frac{\pi}{6}(\bar{E}_4 -  \overline{\hat{E}}_2^2).\ee}
\be \delta_{\bar\m} Q_2 = 4\pi\tau_2 \bar{D}_2 \bar{D}_0 \mathcal{E}_1 = \frac{\tau_2}{\pi}(5Q_1 -  Q_2^2),\ee
we get the eigenvalue equation
\bea \label{Q6}\tau_2^2 \bar{D}_4 D_0 Q_6 &=& 4 R_{5,4} - \frac{3}{4} Q_1 - \frac{1}{2}Q_2 Q_4 -\frac{1}{4} (5 Q_1 -Q_2^2) \mathcal{E}_2\non \\ &=& 
\frac{4\pi^2}{3}\bar{D}_2 \bar{D}_0 \mathcal{E}_3 -\frac{3\zeta(4)}{2} \bar{E}_4- \frac{\pi^3}{6} \overline{\hat{E}}_2 \bar{D}_0 \mathcal{E}_2- \frac{\pi^4}{36}(\bar{E}_4 - \overline{\hat{E}}_2^2)\mathcal{E}_2 \non \\ &=& \frac{4\pi^2}{3}\bar{D}_2 \bar{D}_0 \mathcal{E}_3 -\frac{3\zeta(4)}{2} \bar{E}_4 -\frac{\pi^3}{6} \bar{D}_2 (\mathcal{E}_2\overline{\hat{E}}_2)\eea
satisfied by $Q_6$ which is considerably more involved than the earlier ones. However, this can be simplified using the relation 
\be \frac{\pi}{2} {D}_2 (\mathcal{E}_2\overline{\hat{E}}_2) = \tau_2^2 \bar{D}_4 D_0 (R_{3,2}\overline{\hat{E}}_2) - \frac{9\zeta(4)}{2\pi^2} \bar{E}_4\ee
and also \C{e}, and we see that \C{Q6} reduces to
\be \tau_2^2 \bar{D}_4 D_0 \Big(Q_6 + \frac{\pi^2}{3} R_{3,2}\overline{\hat{E}}_2 - \frac{4\pi^2}{3} \bar{D}_2 \bar{D}_0 \mathcal{E}_3\Big) =0,\ee
leading to the covariant solution
\be Q_6 = -\frac{\pi^2}{3} R_{3,2}\overline{\hat{E}}_2 + \frac{4\pi^2}{3} \bar{D}_2 \bar{D}_0 \mathcal{E}_3 + \m\bar{E}_4,\ee
where $\m$ is a constant, which is fixed using the asymptotic expansion \C{asympQ6} giving us
\be \m  = -2\zeta(4).\ee
Thus we have that
\bea \label{finQ6}Q_6 &=& -\frac{\pi^2}{3} R_{3,2}\overline{\hat{E}}_2 + \frac{4\pi^2}{3} \bar{D}_2 \bar{D}_0 \mathcal{E}_3 -2\zeta(4)\bar{E}_4 \non \\ &=& -Q_2 Q_4 + 4 R_{5,4}-Q_1,\eea
leading to a simple equation at the end of the analysis.

\subsubsection{The equation for $Q_7$}

We next consider the equation for $Q_7$, where we obtain
\be \label{consis2}\delta_{\m} Q_7 = \frac{\pi}{\tau_2} Q_{11} +\frac{\pi}{\tau_2} Q_{10}-\frac{\pi}{\tau_2}\mathcal{E}_2 Q_2.\ee
Now from \C{consis1}, we have that
\be \delta_{\mu} (Q_6 - 2 R_{5,4}) = \frac{\pi}{\tau_2}Q_{10} - \frac{\pi}{\tau_2} \mathcal{E}_2 Q_2,\ee
while from \C{consis2}, we get that
\be \delta_{\m} (Q_7 - R_{5,4}) = \frac{\pi}{\tau_2}Q_{10} - \frac{\pi}{\tau_2} \mathcal{E}_2 Q_2.\ee
Thus modular covariance yields the relation
\be Q_6 = Q_7 +R_{5,4}\ee
between the various  graph functions. Given the various asymptotic expansions, it is straightforward to check this equality for the power behaved terms.

\subsubsection{The equation for $Q_8$}

Proceeding as above, for $Q_8$ we have that
\be \label{consis3}\delta_{\m} Q_8 = - \frac{2\pi}{\tau_2} Q_{11} -\frac{2\pi}{\tau_2} Q_{10}+ \frac{2\pi}{\tau_2}\mathcal{E}_2 Q_2.\ee
Thus from \C{consis2} and \C{consis3} we obtain
\be \delta_{\m}(2 Q_7 + Q_8) =0,\ee
leading to the simple relation between the graphs 
\be 2 Q_7 + Q_8 = 0\ee
based on modular covariance. Again, this relation is easily seen to hold for the power behaved terms in the asymptotic expansion mentioned in the appendix. 

\subsubsection{The equation for $Q_9$}

Finally, we consider the equation for $Q_9$ which gives us
\be \label{Q9}\delta_{\m} Q_9 =\frac{2\tau_2}{\pi} P_1 + \frac{2\pi}{\tau_2} P_2 - \frac{3\pi}{\tau_2}\mathcal{E}_2 +\frac{2\pi}{\tau_2},\ee
where the graphs $P_1$ and $P_2$ are defined by
\be P_1 =  \prod_{i=1}^3 \int_\S  \frac{d^2 z^i}{\tau_2} (\bar\p_1 G_{12})^2 \p_1 G_{12} \p_1 G_{13} G_{23}, \quad P_2 =  \prod_{i=1}^2 \int_\S  \frac{d^2 z^i}{\tau_2} G_{12}^3 = \mathcal{E}_3 +\zeta(3) \ee
as depicted in figure 2.
\begin{figure}[ht]
\begin{center}
\[
\mbox{\begin{picture}(170,120)(0,0)
\includegraphics[scale=.75]{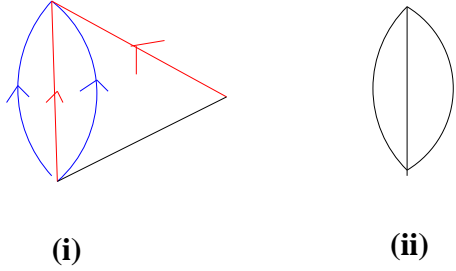}
\end{picture}}
\]
\caption{ (i) $P_1$ and (ii) $P_2$}
\end{center}
\end{figure}

The last two terms in \C{Q9} have been fixed using the asymptotic expansion. While the expression for $P_2$ is known~\cite{D'Hoker:2015foa,Basu:2016kli}, we need to express $P_1$ in a way such that it is useful for our analysis.

To do so, it is very useful for our purposes to introduce an appropriate auxiliary diagram~\cite{Basu:2016xrt}. This is a diagram which trivially reduces to the diagram we want to evaluate using \C{eigen} along with other easily tractable terms, however, on the other hand it can be evaluated independently leading to a simplified expression. Equating the two results leads to the desired expression.   

\begin{figure}[ht]
\begin{center}
\[
\mbox{\begin{picture}(160,110)(0,0)
\includegraphics[scale=.75]{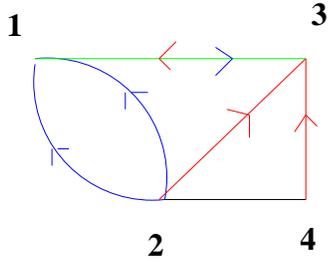}
\end{picture}}
\]
\caption{The auxiliary diagram A}
\end{center}
\end{figure}

For this case, we introduce the auxiliary diagram $A$ defined by
\be A =  \prod_{i=1}^4 \int_\S  \frac{d^2 z^i}{\tau_2} (\bar\p_1 G_{12})^2 \p_1 \bar\p_3 G_{13} \p_3 G_{23} \p_3 G_{34} G_{24}\ee
as depicted in figure 3. 
Evaluating it trivially using \C{eigen} for the $1-3$ link gives us
\be A = \frac{\pi}{\tau_2} P_1 - \frac{\pi}{\tau_2} Q_2 \bar{Q}_4.\ee
On the other hand, moving the $\p$ in the $1-3$ link to the left gives us that
\be A = -\frac{\pi^3}{\tau_2^3} P_2 + \frac{2\pi^3}{\tau_2^3}\mathcal{E}_3.\ee
This gives us that
\be \label{MatcH}\delta_{\m} Q_9 = \frac{4\pi}{\tau_2} \mathcal{E}_3 +\frac{2\tau_2}{\pi} Q_2 \bar{Q}_4 - \frac{3\pi}{\tau_2}\mathcal{E}_2 +\frac{2\pi}{\tau_2},\ee
leading to the eigenvalue equation
\bea \label{VE}\tau_2^2 \bar{D}_2 D_0 Q_9 &=& 3 Q_{11} +\frac{\tau_2^2}{2\pi^2} (5 Q_1 - Q_2^2)\bar{Q}_4 +\frac{1}{2} Q_2\mathcal{E}_2 - \frac{3}{2} Q_4 \non \\ &=&2\pi\bar{D}_0 \mathcal{E}_3 +\frac{\pi^3\tau_2^2}{18} (\bar{E}_4 - \overline{\hat{E}}_2^2)D_0\mathcal{E}_2 +\zeta(2) \overline{\hat{E}}_2 \mathcal{E}_2 - \frac{3\pi}{2} \bar{D}_0 \mathcal{E}_2.\eea

Hence manipulating the various terms in \C{VE} as before, we get that
\be \tau_2^2 \bar{D}_2 D_0 \Big(Q_9 -\frac{4\pi}{3}\bar{D}_0 \mathcal{E}_3 +4\pi\bar{D}_0 \mathcal{E}_2 -\frac{\pi^2}{3} \overline{\hat{E}}_2 \mathcal{E}_2\Big) =0,\ee
leading to the relation
\bea Q_9 &=& \frac{4\pi}{3}\bar{D}_0 \mathcal{E}_3 -4\pi\bar{D}_0 \mathcal{E}_2 +\frac{\pi^2}{3} \overline{\hat{E}}_2 \mathcal{E}_2  +4\zeta(2) \overline{\hat{E}}_2\non \\ &=& 2 Q_{11} - 4 Q_4 + Q_2\mathcal{E}_2 + 2 Q_2 \non \\ &=& 2 Q_{11} -2 Q_3 + Q_2\mathcal{E}_2\eea
on adding a contribution to $Q_9$ given by $4\zeta(2)\overline{\hat{E}}_2$ on using the asymptotic expansion for $Q_9$.

\section{Relations between various modular graph functions}

From the above discussion, we  see that there are several non--trivial relations between the different modular graph functions, given by 
\bea \label{rel}Q_3 &=& 2 Q_4 - Q_2, \non \\ Q_{10} &=& 2 Q_{11} - Q_4, \non \\ Q_5 &=& -2 R_{5,4} + Q_1, \non \\ Q_6 &=& -Q_2 Q_4 + 4 R_{5,4}-Q_1 , \non \\ Q_7 &=& Q_6 -R_{5,4}, \non \\  Q_8 &=& -2 Q_7 , \non \\ Q_9 &=& 2 Q_{11} - 2 Q_3 + Q_2\mathcal{E}_2 . \eea
Thus topologically distinct graphs can be related, which also reduces the number of elements in the basis of graphs with a fixed modular weight. This is analogous to similar relations which exist among several $SL(2,\mathbb{Z})$ invariant graphs that arise in the type II theory. However, unlike that analysis, the relations between graphs in heterotic string theory involve modular covariant, rather than modular invariant graphs. We expect a rich structure of similar relations between topologically distinct graphs for all modular weights that arise in the low momentum expansion of the string amplitude at all orders in the derivative expansion. It will be interesting to analyze such modular covariant relations in general.    

Note that for the cases we have considered, from \C{rel} it follows that all the graphs can be expressed in terms of a few elementary ones involving $R_{s,p}$ leading to an enormous simplification. Clearly this is not true in general, and at higher orders in the derivative expansion we shall encounter graphs which cannot be expressed in this way. In order to obtain their contribution to the string amplitude, one has to directly use the Poisson equations they satisfy to reduce their contribution to boundary integrals over moduli space upto source terms that have to be separately considered, as in the type II analysis. In fact, expressing the various integrands as total derivatives is a powerful technique in calculating the integrals, which we shall heavily use in the analysis below.       

\begin{figure}[ht]
\begin{center}
\[
\mbox{\begin{picture}(140,80)(0,0)
\includegraphics[scale=.65]{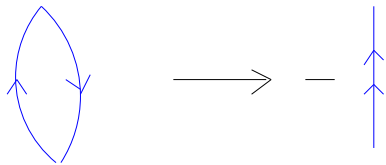}
\end{picture}}
\]
\caption{A pattern for the boundary terms}
\end{center}
\end{figure} 

Before proceeding, let us consider the equations the graphs $Q_3, Q_{10}, Q_5$ and $Q_9$ satisfy in \C{rel}, which are the only graphs in the list in \C{rel} which have a factor of $(\bar\p_1 G_{12})^2$ in the integrand. Apart from the contributions on the right hand side of \C{rel} to these graphs that directly arise in our analysis using the variation of the Beltrami differentials, there are also boundary contributions to each of these graphs that we fixed using the asymptotic analysis as discussed earlier, which naively vanish using $\delta_{\m}(\bar\p  G)=0$. For these graphs, we see these contributions are given by\footnote{For $Q_9$, this actually arises from the sum of \C{last} and \C{verylast}.} 
\be Q_3 \rightarrow - Q_2, \quad Q_{10} \rightarrow -Q_4, \quad Q_5 \rightarrow Q_1, \quad Q_9 \rightarrow -2 Q_3. \ee             
These contributions suggest a striking pattern that can be summarized by figure 4. That is, in each of these graphs, to obtain these non--trivial contributions, simply erase the factor of $(\bar\p_1 G_{12})^2$ in the integrand, and replace it by $- \bar\p_1^2  G_{12}$ keeping the rest of the graph intact. Whether this pattern is more general or not and deriving it if so, will be interesting. It suggests that these boundary contributions should be determined in general by simple properties of how the factor of $(\bar\p_1 G_{12})^2$ is embedded in the integrand of the graph, and is independent of various details of the rest of the graph.   

\section{Evaluating the various four point amplitudes}

Using the equations satisfied by the various modular graph functions, we shall now perform the integrals over the truncated fundamental domain of $SL(2,\mathbb{Z})$ and evaluate the coefficients of the various interactions that arise from the low momentum expansion of the amplitudes upto ten derivatives. 
We first consider the four graviton amplitude, and then the two graviton--two gluon amplitude.  

\subsection{The four graviton amplitude} 

To simplify the notation, we define the tensors
\bea \label{tensor}Z_1^{\m_1\m_2\m_3\m_4} &=& \Big( I_{4,0} +\frac{2}{\alpha'} I_{2;1,0} +\frac{4}{\alpha'^2} I_{1;2,0}\Big)^{\m_1\m_2\m_3\m_4}, \non \\ Z_2^{\m_1\m_2\m_3\m_4} &=& \Big(I_{4,1} + \frac{2}{\alpha'} I_{2;2,0} +\frac{4}{\alpha'^2} I_{1;3,0}\Big)^{\m_1\m_2\m_3\m_4} , \non \\ Z_3^{\m_1\m_2\m_3\m_4} &=& \frac{1}{\alpha'}\Big( I_{2;0,1} +\frac{2}{\alpha'} I_{1;1,1} -\frac{\alpha'}{6}(I_{5,1} + I_{6,1})\Big)^{\m_1\m_2\m_3\m_4}, \non \\ Z_4^{\m_1\m_2\m_3\m_4} &=& \Big(\widetilde{K}+\frac{1}{3} (I_{5,1} -2 I_{6,1})\Big)^{\m_1\m_2\m_3\m_4}.\eea
Using the the tensors in \C{tensor} and various relations between the modular graph functions in \C{rel}, from \C{1G} we obtain the expressions for the $\mathcal{R}^4$ term given by
\be A^{1-loop}_{\mathcal{R}^4}  = \frac{1}{(4\pi)^4}\prod_{i=1}^4 \epsilon^{(i)}_{\mu_i\nu_i} K^{\mu_1 \mu_2 \mu_3 \mu_4} \int_{\mathcal{F}_L} \frac{d^2\tau}{\tau_2^2} \frac{{\bar{E}}_4^2}{{\bar{\eta}}^{24}} \Big[ (Q_1 + Q_2^2) Z_1 - 2 Q_1 K\Big]^{\n_1\n_2\n_3\n_4},\ee
and the $D^2\mathcal{R}^4$ term given by
\bea &&A^{1-loop}_{D^2 \mathcal{R}^4} = \frac{2}{(4\pi)^4}\prod_{i=1}^4 \epsilon^{(i)}_{\mu_i\nu_i} K^{\mu_1 \mu_2 \mu_3 \mu_4} \int_{\mathcal{F}_L} \frac{d^2\tau}{\tau_2^2} \frac{{\bar{E}}_4^2}{{\bar{\eta}}^{24}} \Big[ \Big(2 Q_2 Q_4 - Q_1 -Q_2^2 + 2 R_{5,4}\Big)Z_2 \non \\ &&- 2 \Big(2 Q_2 Q_4 +2 Q_1 -9 R_{5,4}\Big) Z_3 + \Big(Q_2 Q_4 + Q_1 - 3 R_{5,4}\Big)Z_4\Big]^{\n_1\n_2\n_3\n_4}.\eea

Let us first consider the terms involving $Q_1$ and $Q_2^2$ in the integrand. These are particular cases of the general integral
\be \mathcal{I}_1 =\int_{\mathcal{F}_L} \frac{d^2\tau}{\tau_2^2} \overline{F(\tau)} \overline{\hat{E}}_2^m \ee 
where $F(\tau)$ is a purely holomorphic modular form, which we evaluate in a somewhat different way from~\cite{Lerche:1987qk}, which we very briefly outline\footnote{For our purposes, $F(\tau)$  depends on $E_{2k}(\tau)$ for $k \geq 2$ and $\eta(\tau)$.}. From \C{reG}, we have that
\be 4\tau_2^2 D_0 \bar{D}_0 \mathcal{E}_1 = 1,\ee
leading to
\bea \label{I1}\mathcal{I}_1 &=& 4\int_{\mathcal{F}_L} d^2\tau\overline{F(\tau)} \overline{\hat{E}}_2^m D_0 \bar{D}_0 \mathcal{E}_1  = \frac{2\pi i}{3(m+1)} \int_{\mathcal{F}_L} d^2\tau \frac{\p}{\p\tau} \Big(  \overline{F(\tau)} \overline{\hat{E}}_2^{m+1}\Big) \non \\ &=& \frac{\pi}{3(m+1)} \Big( \overline{F(\tau)} \overline{\hat{E}}_2^{m+1}\Big)\Big\vert_{\bar{q}^0}\eea
where we have used \C{needlat} and the antiholomorphicity of $\overline{F(\tau)}$. In the last term, we have that $q = e^{2\pi i\tau}$, where the subscript means that we have to keep the coefficient of the $\bar{q}^0$ term only while expanding the resulting modular form around the cusp $\tau_2\rightarrow \infty$ in non--negative integer powers of $\bar{q}$. Here $E_{2k}(\tau)$ is given by
\be E_{2k} (\tau) = 1+ c_{2k} \sum_{n=1}^\infty \frac{n^{2k-1} q^n}{1-q^n} \ee
for $k \geq 1$, where
\be c_{2k} = \frac{(-1)^k (2\pi)^{2k}}{\Gamma(2k) \zeta(2k)}.\ee
We shall need the values
\be \label{cval}c_2 = -24, \quad c_4 = 240, \quad c_6 = -504, \quad c_8 = 480, \quad c_{10}= -264. \ee
in our analysis. 
We also need the expression for the Dedekind eta function given by
\be \eta(\tau) = q^{1/24} \prod_{n=1}^\infty (1-q^n).\ee

Thus using \C{I1} we have that
\be \int_{\mathcal{F}_L}\frac{d^2\tau}{\tau_2^2} \frac{\bar{E}_4^2 Q_1}{\bar{\eta}^{24}} =480\pi \zeta(4), \quad \int_{\mathcal{F}_L}\frac{d^2\tau}{\tau_2^2} \frac{\bar{E}_4^2 Q_2^2}{\bar{\eta}^{24}} = 192\pi \zeta(2)^2. \ee
leading to the expression for the $\mathcal{R}^4$ term given by~\cite{Ellis:1987dc}
\be A^{1-loop}_{\mathcal{R}^4}  = \frac{\pi}{24}\prod_{i=1}^4 \epsilon^{(i)}_{\mu_i\nu_i} K^{\mu_1 \mu_2 \mu_3 \mu_4}   (Z_1 -  K)^{\n_1\n_2\n_3\n_4}.\ee

We now focus on the other integrals that are needed in evaluating the $D^2\mathcal{R}^4$ term. 
To start with, we consider the integral
\be \mathcal{I}_2 =  \int_{\mathcal{F}_L}\frac{d^2\tau}{\tau_2^2}\frac{\bar{E}_4^2}{\bar{\eta}^{24}} R_{5,4}.\ee  

Given the non--holomorphic factor $\bar{E}_4^2/\bar{\eta}^{24}$ in the integrand, we would like to act with the Laplacian $4\tau_2^2 D_0\bar{D}_p$ on the other factor in the integrand using its eigenvalue equation, to easily perform the calculation\footnote{This will also be the case, for example, in \C{later} where the integrand has an extra factor of $\overline{\hat{E}}_2$ which has a mildly holomorphic part as well. We shall evaluate those integrals based on similar ideas. } on integrating by parts. However, this does not prove to be directly useful as
\be 4\tau_2^2 D_0 \bar{D}_p R_{p+1,p} =0\ee  
leading to zero eigenvalue, and ours is a particular case with $p=4$. So we proceed to evaluate the integral somewhat differently. 

Consider the integral
\be \tilde{\mathcal{I}}_{s,p} = \int_{\mathcal{F}_L}\frac{d^2\tau}{\tau_2^2}\frac{\bar{E}_{12-p}}{\bar{\eta}^{24}} R_{s,p}\ee 
where $s \neq p+1$. Using
\be\label{LaP} 4\tau_2^2 D_0\bar{D}_p R_{s,p} = s(s-p-1) R_{s,p},\ee 
we have that
\be \label{epsilon1}s(s-p-1) \tilde{\mathcal{I}}_{s,p} = 2\int_{-1/2}^{1/2} d\tau_1 \frac{\bar{E}_{12-p}}{\bar{\eta}^{24}} \bar{D}_p R_{s,p}\big\vert_{\tau_2 = L \rightarrow \infty}.\ee
Defining $\epsilon = s-(p+1)$ and letting $\epsilon \rightarrow 0$, we want to analyze \C{epsilon1}. The left hand side yields $(p+1) \epsilon \tilde{\mathcal{I}}_p$, where
\be \label{tilde}\tilde{\mathcal{I}}_p = \int_{\mathcal{F}_L}\frac{d^2\tau}{\tau_2^2}\frac{\bar{E}_{12-p}}{\bar{\eta}^{24}} R_{p+1,p}\ee
and hence $\tilde{\mathcal{I}}_4 = \mathcal{I}_2$, which is the integral we want to evaluate. 

Now first let us evaluate the right hand side of \C{epsilon1} when $s =p+1$, which yields a contribution proportional to
\be \int_{-1/2}^{1/2} d\tau_1 \frac{\bar{E}_{12-p} \bar{E}_{p+2}}{\bar{\eta}^{24}}\big\vert_{\tau_2 = L \rightarrow \infty} = 24 + c_{12-p} + c_{p+2}.\ee   
Now the only possibilities are $p=2,4,6,8$, for which $24+c_{12-p} +c_{p+2}=0$ on using \C{cval}, and hence the right hand side of \C{epsilon1} is actually $O(\epsilon)$. Thus taking $\epsilon$ to be very small, we now evaluate the right hand side of \C{epsilon1} to $O(\epsilon)$, noting that the $O(1)$ term cancels.   

To do so, we first consider the terms in $\bar{D}_p R_{s,p}$ that are independent of $\tau_1$. Using \C{defR} and \C{expE}, we see there are two such contributions given by
\be R_{s,p} = \frac{2\zeta(2s-p)}{\pi^{s-p}} \tau_2^{s-p} +\frac{2(-1)^{p/2}\zeta(2s-p-1)\G(s-p/2)\G(s-p/2-1/2)}{\pi^{s-p-1/2}\G(s)\G(s-p)}\tau_2^{1-s},\ee
leading to
\bea \label{DR}\bar{D}_p R_{s,p} &=& \frac{s\zeta(2s-p)}{\pi^{s-p}} \tau_2^{s-(p+1)}\non \\ &&+ \frac{(s-p-1)(-1)^{1+ p/2}\zeta(2s-p-1)\G(s-p/2)\G(s-p/2-1/2)}{\pi^{s-p-1/2}\G(s)\G(s-p)\tau_2^s}. \eea
In the small $\epsilon$ expansion, ignoring the $O(1)$ contribution, the contribution of the first term on the right hand side of \C{DR} to $\tilde{\mathcal{I}}_p$ is given by
\be \label{1C}\tilde{\mathcal{I}}_p = \frac{2}{\pi} (24 + c_{12-p}) \zeta(p+2) \Big[ {\rm ln}\frac{L}{\pi} +\frac{1}{p+1} + \frac{2\zeta'(p+2)}{\zeta(p+2)}\Big],\ee
hence leading to a contribution that diverges logarithmically as $L \rightarrow \infty$. Such contributions signal the possible presence of a term that is non--analytic in the external momenta arising from the integral over $\mathcal{R}_L$ defined in \C{RL}~\cite{Green:2008uj}. Schematically it is of the form $a_0{\rm ln} (-\alpha' L s)D^2\mathcal{R}^4$ such that the ${\rm ln} L$ term cancels between the contributions from $\mathcal{F}_L$ and $\mathcal{R}_L$. Hence the coefficient $a_0$ is fixed in the process. Thus this leads to a non--analytic term in the effective action of the form ${\rm ln} (-\alpha' \mu s)  D^2\mathcal{R}^4$ where the scale $\m$ involves contributions both from $\mathcal{F}_L$ and $\mathcal{R}_L$. However, since
\be {\rm ln} (-\alpha' \mu s)  D^2\mathcal{R}^4 = {\rm ln} (-\alpha' \mu' s)  D^2\mathcal{R}^4 + {\rm ln} (\mu/\mu')D^2\mathcal{R}^4,\ee      
we see that changing the scale amounts to mixing the analytic and non--analytic contributions, and hence it is the total contribution which is invariant. Thus our analysis yields the total contribution from $\mathcal{F}_L$. Note that such a logarithmic contribution also arises in the type II theory, however the scale of the logarithm is irrelevant as $D^2\mathcal{R}^4 = (s+t+u)\mathcal{R}^4 =0$ in the type II theory, but not in the heterotic theory due to the different tensor structure.

Next note that the second term on the right hand side of \C{DR} has an overall factor of $\epsilon$ and its contribution to $\tilde{\mathcal{I}}_p$ is given by
\be \tilde{\mathcal{I}}_p = \frac{2(-1)^{1+p/2} (c_{12-p} +24)\zeta(p+1)\G(p/2+1/2)\G(1+p/2)}{\sqrt{\pi} \G(2+p) L^{p+1}} \rightarrow 0\ee
as $L\rightarrow \infty$. Hence this does not contribute to $\mathcal{I}_3$. 

Finally, consider the contribution to \C{epsilon1} from terms in $\bar{D}_p R_{s,p}$ which depend on $\tau_1$. In this analysis, we keep only those terms that are non--vanishing as $\tau_2 \rightarrow \infty$. These are obtained from the last term on the right hand side of \C{expE} on setting
\be K_s (x) \rightarrow \sqrt{\frac{\pi}{2x}} e^{-x},\ee
as subleading contributions vanish, leading to
\be \label{np}\mathcal{E}_{s-p/2} (\tau,\bar\tau)\rightarrow \frac{2}{\G(s-p/2)} \sum_{m=1}^\infty m^{s-p/2 -1} \s_{1-2s +p}(m)\bar{q}^m,\ee 
which gives us the relevant terms
\be \bar{D}_p R_{s,p} = 2(-1)^{1+p/2} \frac{(2\pi)^{p+1}}{\G(s)} \sum_{m=1}^\infty m^s \s_{1-2s +p}(m)\bar{q}^m\ee
in the expression for $\bar{D}_p R_{s,p}$. Thus its contribution to \C{epsilon1} is given by
\be s\epsilon \tilde{\mathcal{I}}_p =4(-1)^{1+p/2}\frac{(2\pi)^{p+1}}{\G(s)}\sum_{m=1}^\infty m^s \s_{1-2s +p}(m) \frac{\bar{E}_{12-p}\bar{q}^m}{{\bar\eta}^{24}}\Big\vert_{\bar{q}^0} .\ee
On using 
\be \frac{\bar{E}_{12-p}\bar{q}^m}{{\bar\eta}^{24}}\Big\vert_{\bar{q}^0} = \delta_{m,1},\ee
and ignoring the $O(1)$ term, we get the contribution
\be \label{C2}\tilde{\mathcal{I}}_p =4(-1)^{p/2} (2\pi)^{p+1}\frac{\G'(p+1)}{\G(p+1)\G(p+2)}.\ee

Thus \C{1C} and \C{C2} yield non--vanishing contribution to $\tilde{\mathcal{I}}_p$ leading to
\be \mathcal{I}_2 = -\frac{16\pi^5}{15} \Big({\rm ln} \pi +\gamma - \frac{137}{60} -\frac{2\zeta'(6)}{\zeta(6)}\Big).\ee

Next let us consider the integral
\be \label{later}\mathcal{I}_3   =\int_{\mathcal{F}_L}\frac{d^2\tau}{\tau_2^2}\frac{\bar{E}_4^2 \overline{\hat{E}}_2}{\bar{\eta}^{24}} R_{3,2}.\ee

To evaluate it, we start with the integral
\be \tilde{\mathcal{J}}_{s,p} =\int_{\mathcal{F}_L}\frac{d^2\tau}{\tau_2^2} \frac{\bar{E}_{10-p} \overline{\hat{E}}_2}{\bar{\eta}^{24}} R_{s,p} \ee
where $s \neq p+1$. Using \C{LaP} we have that
\be \label{epsilon2}s(s-p-1) \tilde{\mathcal{J}}_{s,p} = 2\int_{-1/2}^{1/2} d\tau_1 \frac{\bar{E}_{10-p}\overline{\hat{E}}_2}{\bar{\eta}^{24}} \bar{D}_p R_{s,p}\big\vert_{\tau_2 = L \rightarrow \infty} - \frac{6}{\pi} \int_{\mathcal{F}_L}\frac{d^2\tau}{\tau_2^2} \frac{\bar{E}_{10-p}}{\bar{\eta}^{24}} \bar{D}_p R_{s,p}.\ee
Expanding \C{epsilon2} for small $\epsilon = s -(p+1)$, we see that the left hand side gives $(p+1)\epsilon \tilde{\mathcal{J}}_p$ where 
\be \label{defJp}\tilde{\mathcal{J}}_p = \int_{\mathcal{F}_L}\frac{d^2\tau}{\tau_2^2} \frac{\bar{E}_{10-p} \overline{\hat{E}}_2}{\bar{\eta}^{24}} R_{p+1,p},\ee
and hence $\tilde{\mathcal{J}}_2=\mathcal{I}_3$, which is the integral we want. We now expand the right hand side of \C{epsilon2} for small $\epsilon$. The $O(1)$ term is proportional to
\be \frac{2\bar{E}_{10-p} \bar{E}_{p+2}\overline{\hat{E}}_2}{\bar{\eta}^{24}}\Big\vert_{\bar{q}^0}- \frac{6}{\pi} \int_{\mathcal{F}_L}\frac{d^2\tau}{\tau_2^2}\frac{\bar{E}_{10-p} \bar{E}_{p+2}}{\bar{\eta}^{24}}\ee
which vanishes using \C{I1}, and hence we consider the $O(\epsilon)$ term.

We first consider the contribution coming from the first term on the right hand side of \C{epsilon2}, which is a boundary term. The analysis is very similar to the analysis for $\tilde{\mathcal{I}}_p$ and we get a non--vanishing contribution
\be \label{first}\tilde{\mathcal{J}}_p = \frac{2c_{10-p}\zeta(p+2)}{\pi} \Big[ \frac{1}{p+1} + \frac{2\zeta'(p+2)}{\zeta(p+2)}-{\rm ln}\pi \Big] +4(-1)^{p/2} (2\pi)^{p+1}\frac{\G'(p+1)}{\G(p+1)\G(p+2)}.\ee  
We now consider the contribution from the second term on the right hand side of \C{epsilon2} at $O(\epsilon)$ for which we have to evaluate the integral
\be \tilde{\mathcal{J}}_0  = \int_{\mathcal{F}_L}\frac{d^2\tau}{\tau_2^2}\frac{\bar{E}_{10-p}}{\bar{\eta}^{24}}\bar{D}_p R_{s,p}\ee
to $O(\epsilon)$, while ignoring the $O(1)$ term.
Now using the identity
\be 4\tau_2^2 D_0\bar{D}_{p+2} R_{s+1,p+2} = [s(s-p-1)-(p+2)] R_{s+1,p+2},\ee
and noting that $\bar{D}_p R_{s,p} \sim R_{s+1,p+2}$, we have that
\be [s(s-p-1)-(p+2)]  \tilde{\mathcal{J}}_0  = 2\int_{-1/2}^{1/2} d\tau_1 \frac{\bar{E}_{10-p}}{\bar{\eta}^{24}} \bar{D}_{p+2}\bar{D}_p R_{s,p}\big\vert_{\tau_2 = L \rightarrow \infty}\ee
and hence the integral has been reduced purely to a boundary term, which we now evaluate. Proceeding as before, we see that the only finite contribution as $L\rightarrow \infty$ comes from the $\tau_1$ dependent part of $\bar{D}_{p+2}\bar{D}_pR_{s,p}$ arising from \C{np} leading to
\be \label{second}\tilde{\mathcal{J}}_0  =  4(-1)^{p/2} (2\pi)^{p+2}\epsilon\frac{\G'(p+1)}{(p+2)\G(p+1)^2}.\ee
Thus from \C{first} and \C{second} we get that
\be \mathcal{I}_3 = -\frac{32\pi^3}{3} \Big({\rm ln}\pi +\gamma -\frac{11}{6} - \frac{2\zeta'(4)}{\zeta(4)}\Big).\ee

As an aside, let us also consider the integral
\be \label{K1}\mathcal{K}_1 = \int_{\mathcal{F}_L}\frac{d^2\tau}{\tau_2^2}\frac{\bar{E}_4^2}{\bar{\eta}^{24}}(\bar{E}_4 - \overline{\hat{E}}_2^2)\mathcal{E}_2.\ee
This does not arise in our analysis given the very simple equations satisfied by the various graphs, but we expect integrals of this type to be relevant for higher point amplitudes where the relevant graphs need not satisfy such simple equations. In fact, as discussed before, in the cases we are considering, the Poisson equations could be always solved to obtain elementary relations between the graphs. This is not going to persist for generic interactions, and we shall have to directly deal with the Poisson equations which will involve integrating source term contributions over $\mathcal{F}_L$. The integral \C{K1} is a simple example of such a case.    
    
Now using
\be 4\tau_2^2 D_0 \bar{D}_0 \mathcal{E}_2 =  2\mathcal{E}_2,\ee
we easily get that
\be \mathcal{K}_1 = \int_{-1/2}^{1/2} d\tau_1 \frac{\bar{E}_4^2}{\bar{\eta}^{24}}(\bar{E}_4 - \overline{\hat{E}}_2^2) \bar{D}_0 \mathcal{E}_2 \Big\vert_{\tau_2 = L\rightarrow \infty} +\frac{6\mathcal{I}_3}{\pi^2} .\ee
Thus while the first term is a boundary term, the second term has already been evaluated above. 
 
Hence let us evaluate the boundary term
\be \int_{-1/2}^{1/2} d\tau_1 \frac{\bar{E}_4^2}{\bar{\eta}^{24}}(\bar{E}_4 - \overline{\hat{E}}_2^2) \bar{D}_0 \mathcal{E}_2 \Big\vert_{\tau_2 = L\rightarrow \infty}\ee 
that arises in $\mathcal{K}_1$. Considering the power behaved contribution to $\bar{D}_0 \mathcal{E}_2$, we see that the $O(\tau_2)$ term given by
\be \frac{2\zeta(4)}{\pi^2} \tau_2\ee
combines with the $O(1/\tau_2)$ term in $\overline{\hat{E}}_2^2$ given by
\be -\frac{6}{\pi\tau_2}\bar{E}_2\ee
to give a finite contribution. There is no contribution from the $\tau_1$ dependent terms in $\bar{D}_0 \mathcal{E}_2$, leading to 
\be \int_{-1/2}^{1/2} d\tau_1 \frac{\bar{E}_4^2}{\bar{\eta}^{24}}(\bar{E}_4 - \overline{\hat{E}}_2^2) \bar{D}_0 \mathcal{E}_2 \Big\vert_{\tau_2 = L\rightarrow \infty}=\frac{12\zeta(4)}{\pi^3}  \frac{\bar{E}_4^2\bar{E}_2}{\bar{\eta}^{24}}\Big\vert_{\bar{q}^0} = 64\pi,\ee
hence leading to
\be \mathcal{K}_1 = 64\pi +\frac{6\mathcal{I}_3}{\pi^2} .\ee

Thus adding the various contributions, we see that the contribution to the $D^2\mathcal{R}^4$ term from $\mathcal{F}_L$ is given by
\bea A^{1-loop}_{D^2 \mathcal{R}^4} = \frac{2}{(4\pi)^4}\prod_{i=1}^4 \epsilon^{(i)}_{\mu_i\nu_i} K^{\mu_1 \mu_2 \mu_3 \mu_4}  \Big[ A Z_2 + B Z_3 + C Z_4\Big]^{\n_1\n_2\n_3\n_4},\eea
where
\bea \label{ABC}A &=& \frac{4888\pi^5}{675} - \frac{416\pi^5}{45} ({\rm ln} \pi+\gamma) +\frac{64\pi^5}{3}\Big(\frac{2\zeta'(4)}{3\zeta(4)} +\frac{\zeta'(6)}{5\zeta(6)}\Big), \non \\ B &=&  -\frac{2408\pi^5}{675}- \frac{224\pi^5}{45} ({\rm ln} \pi+\gamma) -64\pi^5\Big(\frac{4\zeta'(4)}{9\zeta(4)} -\frac{3\zeta'(6)}{5\zeta(6)}\Big)  , \non \\ C &=&\frac{3068\pi^5}{675} - \frac{16\pi^5}{45} ({\rm ln} \pi+\gamma) +32\pi^5\Big(\frac{2\zeta'(4)}{9\zeta(4)} -\frac{\zeta'(6)}{5\zeta(6)}\Big).\eea

\subsection{The two graviton--two gluon amplitude}

Next we consider the low momentum expansion of the  two graviton--two gluon amplitude. We define the tensor
\be \mathcal{T}_{\m_1\m_2} = \epsilon^{(1)}_{\m_1\n_1} \epsilon^{(2)}_{\m_2\n_2} e^{(3)}_{a\n_3} e^{(4)}_{b\n_4}Tr(T^aT^b) K^{\nu_1 \nu_2 \nu_3 \nu_4} \ee
to simplify the notation.

Now from \C{2C}, we obtain the expression for the $\mathcal{R}^2\mathcal{F}^2$ term given by
\be \label{R2F2}A^{1-loop}_{\mathcal{R}^2\mathcal{F}^2} = -\frac{\zeta(2)}{12}\mathcal{T}_{\m_1\m_2} J_1^{\m_1\m_2}\int_{\mathcal{F}_L} \frac{d^2\tau}{\tau_2^2} \frac{\overline{\hat{E}}_2 \bar{E}_4}{{\bar{\eta}}^{24}}(\bar{E}_6 - \overline{\hat{E}}_2 \bar{E}_4) ,\ee
while the $D^2\mathcal{R}^2\mathcal{F}^2$ term is given by
\bea \label{D2R2F2}A^{1-loop}_{D^2\mathcal{R}^2\mathcal{F}^2} &=& -\frac{\alpha'}{32}\mathcal{T}_{\m_1\m_2}\int_{\mathcal{F}_L} \frac{d^2\tau}{\tau_2^2} \frac{\bar{E}_4}{{\bar{\eta}}^{24}}\Big[ \Big(\frac{Q_3}{3} (\bar{E}_6 - \overline{\hat{E}}_2 \bar{E}_4) + \frac{\bar{E}_4}{\pi^2}(Q_1 + Q_2^2)\Big)sJ_1^{\m_1\m_2} \non \\ &&- 2 \Big(\frac{Q_4}{3} (\bar{E}_6 - \overline{\hat{E}}_2 \bar{E}_4)-\frac{\bar{E}_4}{\pi^2}Q_1\Big)J_2^{\mu_1\m_2}\Big] .\eea
Finally the $D^4\mathcal{R}^2\mathcal{F}^2$ term given by
\be  \label{long}  A^{1-loop}_{D^4\mathcal{R}^2\mathcal{F}^2} =  A^{1-loop,\mathcal{Y}_1}_{D^4\mathcal{R}^2\mathcal{F}^2} +  A^{1-loop,\mathcal{Y}_2}_{D^4\mathcal{R}^2\mathcal{F}^2},\ee
where
\bea A^{1-loop,\mathcal{Y}_1}_{D^4\mathcal{R}^2\mathcal{F}^2} &=& -\frac{\alpha'^2}{3\cdot 256}\mathcal{T}_{\m_1\m_2} \int_{\mathcal{F}_L} \frac{d^2\tau}{\tau_2^2} \frac{\bar{E}_4(\bar{E}_6 - \overline{\hat{E}}_2 \bar{E}_4)}{{\bar{\eta}}^{24}} \Big[\Big(s^2(4 Q_{11} - 4 Q_4 +4Q_2\mathcal{E}_2 + 2Q_2) \non \\ &&-2ut(2Q_4 -3Q_{11} + 2Q_2\mathcal{E}_2)\Big)J_1^{\m_1\m_2} + 2sQ_{11} J_2^{\m_1\m_2}\Big], \non \\ A^{1-loop,\mathcal{Y}_2}_{D^4\mathcal{R}^2\mathcal{F}^2} &=& -\frac{\alpha'^2}{64\pi^2} \mathcal{T}_{\m_1\m_2} \int_{\mathcal{F}_L} \frac{d^2\tau}{\tau_2^2} \frac{\bar{E}_4^2}{{\bar{\eta}}^{24}}\Big[ \Big(s^2(Q_2 Q_3 - Q_5)\non \\ &&+ut(Q_2 Q_4 +Q_5 +2Q_6 - 3 Q_7 -2 Q_8)\Big)J_1^{\m_1\m_2} \non \\ &&-s(Q_2 Q_4 + Q_5 - Q_6 + 3 Q_7 + Q_8)J_2^{\m_1\m_2}\Big],\eea
where we have also used the expressions for $Q_9$ and $Q_{11}$ using \C{rel}. 

The coefficient of the $\mathcal{R}^2 \mathcal{F}^2$ term in \C{R2F2} is obtained using \C{I1}, leading to~\cite{Ellis:1987dc}
\be A^{1-loop}_{\mathcal{R}^2\mathcal{F}^2}  = \frac{4\pi^3}{3} \mathcal{T}_{\m_1\m_2} J_1^{\m_1\m_2}.\ee

Again on using \C{I1}, the relations between the graphs \C{rel}, and the definitions \C{tilde} and \C{later}, we obtain the expression for the $D^2\mathcal{R}^2\mathcal{F}^2$ term in \C{D2R2F2} which is given by
\bea A^{1-loop}_{D^2\mathcal{R}^2\mathcal{F}^2} &=& -\frac{\alpha'}{32}\mathcal{T}_{\m_1\m_2} \Big[ \Big(\frac{1920\zeta(4)}{\pi} +\frac{2}{3} (\tilde{\mathcal{I}}_2 -\mathcal{I}_3) \Big)sJ_1^{\m_1\m_2} \non \\ &&+ \Big(\frac{960\zeta(4)}{\pi} -\frac{2}{3} (\tilde{\mathcal{I}}_2 -\mathcal{I}_3) \Big)J_2^{\m_1\m_2}\Big].\eea
Now using $2\tilde{\mathcal{I}_2} = -\mathcal{I}_3$, we get that
\bea  A^{1-loop}_{D^2\mathcal{R}^2\mathcal{F}^2} &=& -\frac{\pi^3\alpha'}{3}\mathcal{T}_{\m_1\m_2}\Big[ \Big({\rm ln}\pi+\gamma +\frac{1}{6} - \frac{2\zeta'(4)}{\zeta(4)}\Big)sJ_1^{\m_1\m_2} \non \\ &&-\Big({\rm ln}\pi+\gamma - \frac{17}{6} - \frac{2\zeta'(4)}{\zeta(4)}\Big) J_2^{\m_1\m_2}\Big].\eea

We now consider the contribution to the $D^4\mathcal{R}^2 \mathcal{F}^2$ term in \C{long}. To start with, we consider the contribution coming from $A^{1-loop,\mathcal{Y}_1}_{D^4\mathcal{R}^2\mathcal{F}^2}$, where we calculate the various integrals that have already not been calculated.

To begin, we consider the integral
\be \mathcal{I}_4 = \int_{\mathcal{F}_L} \frac{d^2\tau}{\tau_2^2} \frac{\bar{E}_4(\bar{E}_6 - \overline{\hat{E}}_2 \bar{E}_4)}{{\bar{\eta}}^{24}}R_{4,2}\ee
since $Q_{11} = R_{4,2}$.
Using the relation
\be \bar{D}_4 \bar{E}_4 = \frac{2\pi}{3}(\bar{E}_6 - \overline{\hat{E}}_2 \bar{E}_4)\ee
and noting that $\bar{D}_8 \bar{E}_8 = 2\bar{E}_4 \bar{D}_4 \bar{E}_4$, we have that
\be \mathcal{I}_4 = \frac{3}{4\pi}\int_{\mathcal{F}_L} d^2\tau \frac{\bar{D}_8\bar{E}_8}{{\bar{\eta}}^{24}} D_0\bar{D}_2R_{4,2}\ee 
where we have also used the eigenvalue equation for $R_{4,2}$. Thus we have that
\be \label{4I}\mathcal{I}_4 = \frac{3}{8\pi} \int_{-1/2}^{1/2} d\tau_1 \frac{\bar{D}_8\bar{E}_8}{\bar{\eta}^{24}} \bar{D}_2R_{4,2} \big\vert_{\tau_2 = L \rightarrow \infty}+\frac{3}{\pi^2} \mathcal{I}_2, \ee
where we have used
\be\tau_2^2 D_0\bar{D}_8 \bar{E}_8   = -2\bar{E}_8\ee
and 
\be R_{5,4} = \frac{\pi}{2} \bar{D}_2 R_{4,2}.\ee

We next consider the integral
\be \label{I5}\mathcal{I}_5 = \int_{\mathcal{F}_L} \frac{d^2\tau}{\tau_2^2} \frac{\bar{E}_4\overline{\hat{E}}_2(\bar{E}_6 - \overline{\hat{E}}_2 \bar{E}_4)}{{\bar{\eta}}^{24}} \mathcal{E}_2.\ee
Proceeding as above and using the eigenvalue equation for $\mathcal{E}_2$, we get that
\be \label{5I}\mathcal{I}_5 = \frac{3}{4\pi}  \int_{-1/2}^{1/2} d\tau_1 \frac{(\bar{D}_8\bar{E}_8) \overline{\hat{E}}_2 }{\bar{\eta}^{24}} \bar{D}_0 \mathcal{E}_2 \big\vert_{\tau_2 = L \rightarrow \infty} -\frac{3}{\pi^2} (\tilde{\mathcal{I}}_2-2{\mathcal{I}}_3), \ee
where we have used the definitions \C{tilde} and \C{later}.

Again, as an aside we also consider the integral
\be \mathcal{K}_2 = \int_{\mathcal{F}_L} d^2\tau \frac{\bar{E}_4 (\bar{E}_6 - \overline{\hat{E}}_2 \bar{E}_4)}{{\bar{\eta}}^{24}} (\bar{E}_4 - \overline{\hat{E}}_2^2)D_0\mathcal{E}_2  = \frac{9}{2\pi^2}\int_{\mathcal{F}_L} d^2\tau \frac{(\bar{D}_8\bar{E}_8)(\bar{D}_2 \bar{E}_2)}{{\bar{\eta}}^{24}} D_0\mathcal{E}_2,\ee
which is not needed for our analysis, but is useful to analyze keeping generic amplitudes in mind. Proceeding as before, we get that
\be  \label{K2}\mathcal{K}_2 = \frac{9}{4\pi^2}\int_{-1/2}^{1/2} d\tau_1 \frac{(\bar{D}_8\bar{E}_8) (\bar{D}_2\overline{\hat{E}}_2) }{\bar{\eta}^{24}} \mathcal{E}_2 \big\vert_{\tau_2 = L \rightarrow \infty} +\frac{3}{2\pi} (\mathcal{K}_1+2 \mathcal{I}_5)\ee
on using \C{K1} and \C{I5}.

Hence to obtain the expression for $A^{1-loop,\mathcal{Y}_1}_{D^4\mathcal{R}^2F^2}$ and $\mathcal{K}_2$, we have to calculate the remaining boundary contributions in \C{4I}, \C{5I} and \C{K2} using the asymptotic expansions. Proceeding as before, we get that
\bea &&\int_{-1/2}^{1/2} d\tau_1 \frac{\bar{D}_8\bar{E}_8}{\bar{\eta}^{24}}  \bar{D}_2 R_{4,2}\big\vert_{\tau_2 = L \rightarrow \infty} =\frac{16\zeta(6)}{\pi^2}\frac{\bar{E}_8}{\bar\eta^{24}}\Big\vert_{\bar{q}^0} = \frac{128\pi^4}{15}, \non \\&&\int_{-1/2}^{1/2} d\tau_1 \frac{(\bar{D}_8\bar{E}_8)\overline{\hat{E}}_2 }{\bar{\eta}^{24}}   \bar{D}_0\mathcal{E}_2\big\vert_{\tau_2 = L \rightarrow \infty} = \frac{8\zeta(4)}{\pi^2} \frac{\bar{E}_4(2\bar{E}_2\bar{E}_4 - \bar{E}_6)}{\bar\eta^{24}}\Big\vert_{\bar{q}^0}= \frac{320\pi^2}{3}, \non \\ &&\int_{-1/2}^{1/2} d\tau_1 \frac{(\bar{D}_8\bar{E}_8)( \bar{D}_2 \overline{\hat{E}}_2)}{\bar{\eta}^{24}}  \mathcal{E}_2\big\vert_{\tau_2 = L \rightarrow \infty} = \frac{4\zeta(4)}{\pi^2} \frac{\bar{E}_4(3\bar{E}_2\bar{E}_4 - \bar{E}_6)}{\bar\eta^{24}}\Big\vert_{\bar{q}^0}= \frac{224\pi^2}{3}, \eea
on only keeping finite terms as $L\rightarrow \infty$.  

These integrals involving $\bar{D}_2 R_{4,2}, \bar{D}_0 \mathcal{E}_2$ and $\mathcal{E}_2$ do not receive contributions from the $\tau_1$ dependent part of these modular forms, given the structure of the other factor in the integrands. Thus using the various relations, we have that
\bea \mathcal{I}_4 &=& -\frac{16\pi^3}{5} \Big({\rm ln} \pi+\g -\frac{197}{60}-\frac{2\zeta'(6)}{\zeta(6)}\Big), \non\\ \mathcal{I}_5 &=& -80\pi \Big({\rm ln} \pi+\g -\frac{17}{6}-\frac{2\zeta'(4)}{\zeta(4)}\Big)\eea
which are relevant for calculating $A^{1-loop,\mathcal{Y}_1}_{D^4\mathcal{R}^2F^2}$. Thus we get that
\be A^{1-loop,\mathcal{Y}_1}_{D^4\mathcal{R}^2F^2} = -\frac{\alpha'^2}{3\cdot 256} \mathcal{T}_{\m_1\m_2}\Big[ (A' s^2 + B' ut) J_1^{\m_1\m_2} + C' sJ_2^{\m_1\m_2} \Big],\ee
where
\bea A' &=& \frac{118256\pi^3}{225} - \frac{2752\pi^3}{15} ({\rm ln}\pi+\g) +\frac{1024\pi^3}{3} \Big(\frac{\zeta'(4)}{\zeta(4)} + \frac{3\zeta'(6)}{40\zeta(6)}\Big), \non \\ B' &=& -\frac{27416\pi^3}{225} + \frac{352\pi^3}{15} ({\rm ln}\pi+\g) -\frac{256\pi^3}{3} \Big(\frac{\zeta'(4)}{\zeta(4)} - \frac{9\zeta'(6)}{20\zeta(6)}\Big), \non \\ C' &=&  -\frac{32\pi^3}{5} \Big({\rm ln} \pi+\g -\frac{197}{60}-\frac{2\zeta'(6)}{\zeta(6)}\Big). \eea

Finally, we need to calculate $A^{1-loop,\mathcal{Y}_2}_{D^4\mathcal{R}^2F^2}$ in \C{long}. This is obtained in a straightforward manner following the discussion in section 5.2, given the equality of the various integrals that arise in this case and the $D^2\mathcal{R}^4$ amplitude, leading to
\be A^{1-loop,\mathcal{Y}_2}_{D^4\mathcal{R}^2F^2} = -\frac{\alpha'^2}{128\pi^2} \mathcal{T}_{\m_1\m_2}\Big[ (2A s^2 + B ut) J_1^{\m_1\m_2} -2 C sJ_2^{\m_1\m_2} \Big],\ee 
where $A, B$ and $C$ are given by \C{ABC}.

\section{Some modular graph functions at higher order in the momentum expansion}

This method of analyzing modular graph functions generalizes to all orders in the $\alpha'$ expansion. However, the technical details get more involved as the graphs get more complicated. Though we have not done a detailed analysis for the graphs that arise for interactions with twelve derivatives, we present some elementary results for illustrative purposes.  

\begin{figure}[ht]
\begin{center}
\[
\mbox{\begin{picture}(350,110)(0,0)
\includegraphics[scale=.7]{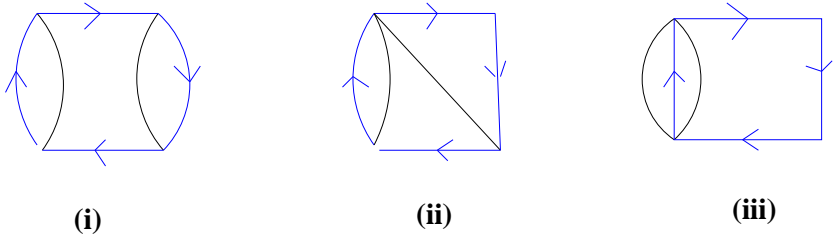}
\end{picture}}
\]
\caption{ (i) $Q_{12}$, (ii) $Q_{13}$ and (iii) $Q_{14}$}
\end{center}
\end{figure}

Among the several graphs that arise at this order in the derivative expansion, consider the graphs $Q_{12}$, $Q_{13}$ and $Q_{14}$ given in figure 5. As usual, the vertices are integrated with measure $d^2z/\tau_2$. For the sake of brevity, we write down the equations satisfied by the action of $\delta_{\m}$ on these graphs pictorially, where the analysis is done along the lines of that done before, and then the action of $\delta_{\bar\m} \delta_{\m}$ on every graph can be easily obtained. None of these graphs have a factor of $(\bar\p_1 G_{12})^2$ in the integrand, and hence we do not expect any non--vanishing contributions from $\delta_{\mu} \bar\p G$.

\begin{figure}[ht]
\begin{center}
\[
\mbox{\begin{picture}(410,60)(0,0)
\includegraphics[scale=.65]{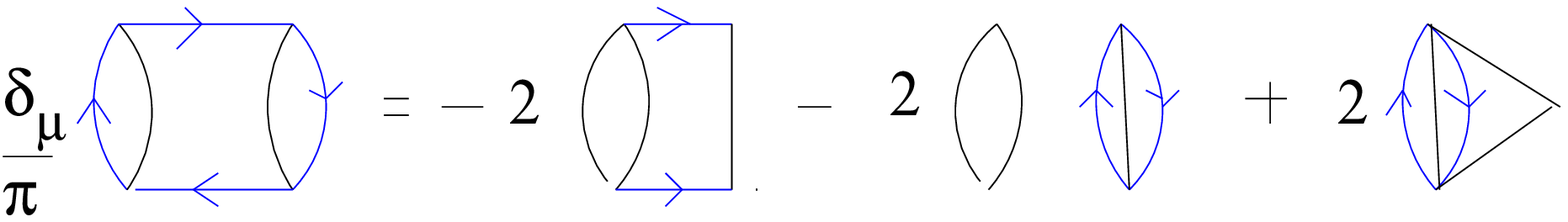}
\end{picture}}
\]
\caption{ The equation for $\delta_{\m} Q_{12}$}
\end{center}
\end{figure} 

\begin{figure}[ht]
\begin{center}
\[
\mbox{\begin{picture}(320,120)(0,0)
\includegraphics[scale=.65]{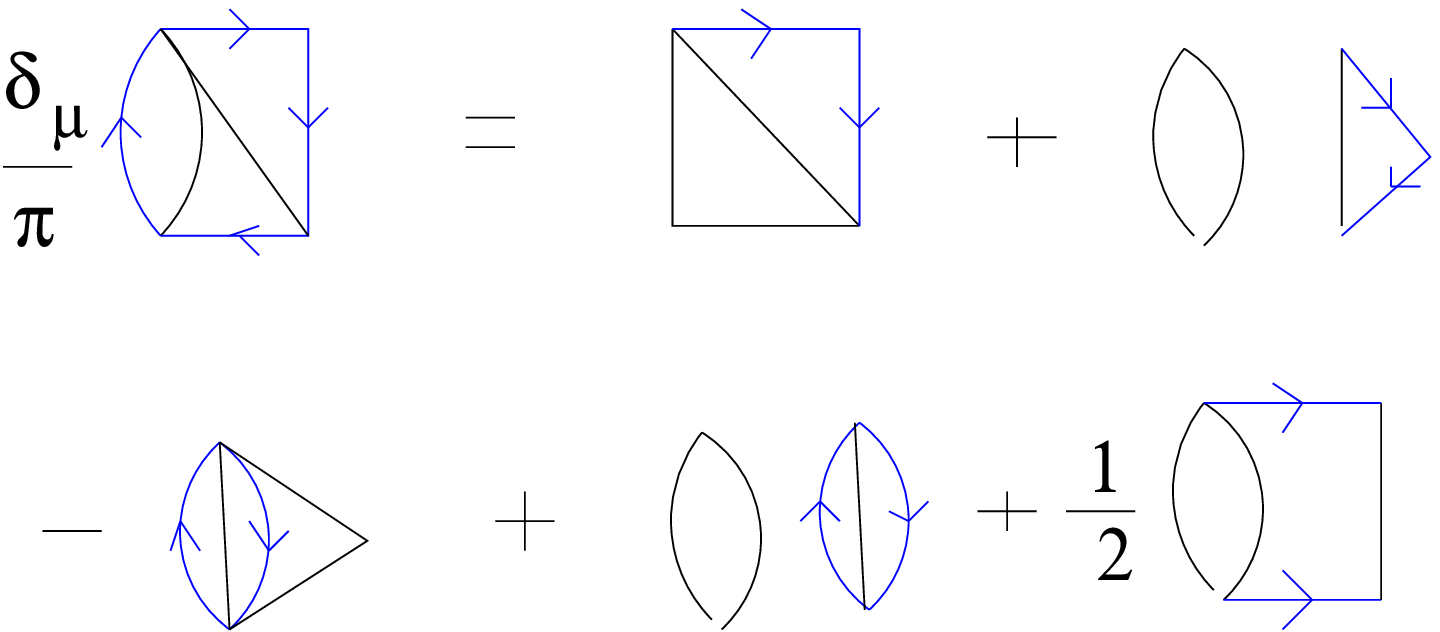}
\end{picture}}
\]
\caption{ The equation for $\delta_{\m} Q_{13}$}
\end{center}
\end{figure}

\begin{figure}[ht]
\begin{center}
\[
\mbox{\begin{picture}(410,240)(0,0)
\includegraphics[scale=.65]{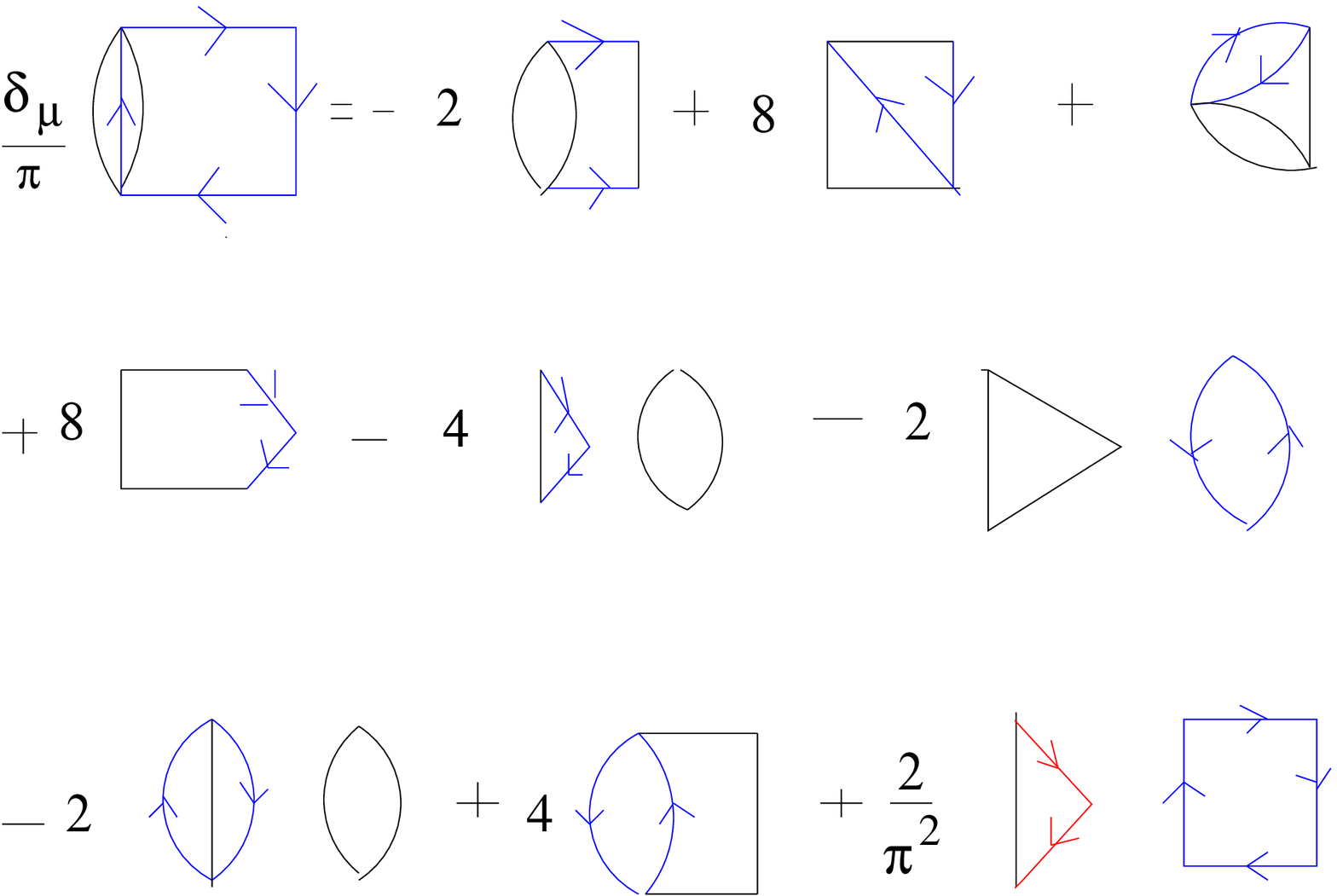}
\end{picture}}
\]
\caption{ The equation for $\delta_{\m} Q_{14}$}
\end{center}
\end{figure} 

\begin{figure}[ht]
\begin{center}
\[
\mbox{\begin{picture}(230,100)(0,0)
\includegraphics[scale=.6]{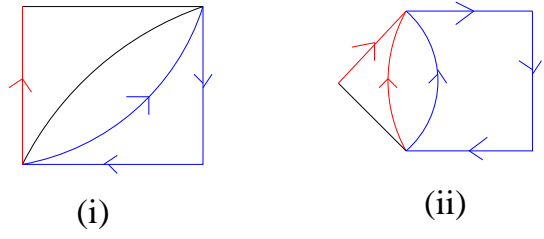}
\end{picture}}
\]
\caption{ The graphs $(i) S_1$ and  $(ii) S_2$}
\end{center}
\end{figure} 

The equations for $\delta_{\m} Q_{12}$, $\delta_{\m} Q_{13}$ and $\delta_{\m} Q_{14}$ are given by figures 6, 7 and 8 respectively, where we have suppressed a factor of $1/\tau_2$ for each term on the right hand side for brevity\footnote{In the last term on the right hand side of $\delta_{\m} Q_{14}$, we have suppressed a factor of $\tau_2$ instead.}.  While the equations for $\delta_{\m} Q_{12}$ and $\delta_{\m} Q_{13}$  are straightforward, the equation for $\delta_{\m} Q_{14}$ involves simplifying the graphs $S_1$ and $S_2$ given in figure 9. To do so, we introduce the auxiliary graphs $S_3$ and $S_4$ respectively given in figure 10.   

\begin{figure}[ht]
\begin{center}
\[
\mbox{\begin{picture}(250,120)(0,0)
\includegraphics[scale=.7]{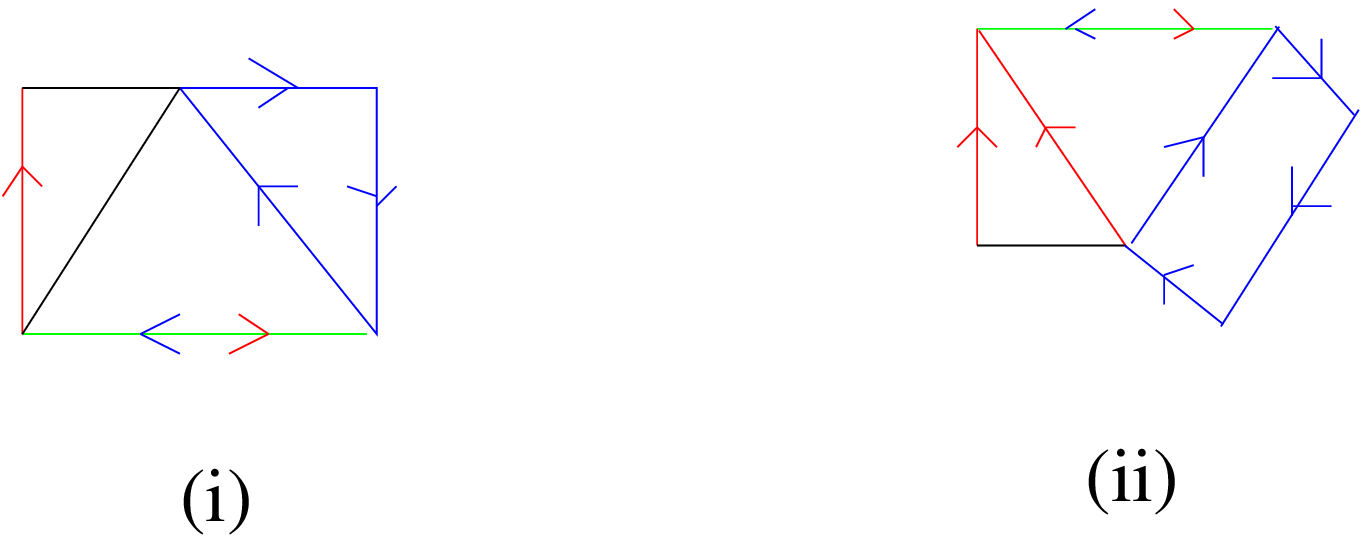}
\end{picture}}
\]
\caption{ The graphs $(i) S_3$ and  $(ii) S_4$}
\end{center}
\end{figure} 

In fact, there are also some simple relations between graphs of distinct topologies. They are obtained by starting with appropriate auxiliary graphs and generalize the analysis of~\cite{Basu:2016xrt} for modular invariant graphs. One such relation is given by $(i)$ in figure 11, which is obtained by starting with the auxiliary graph $(ii)$, and evaluating it in two different ways. Another such relation is given by $(i)$ in figure 12, which is obtained by starting with the auxiliary graph $(ii)$. Note that in these auxiliary graphs, there is a link of the form $\bar\p^2 G$.       

\begin{figure}[ht]
\begin{center}
\[
\mbox{\begin{picture}(250,200)(0,0)
\includegraphics[scale=.7]{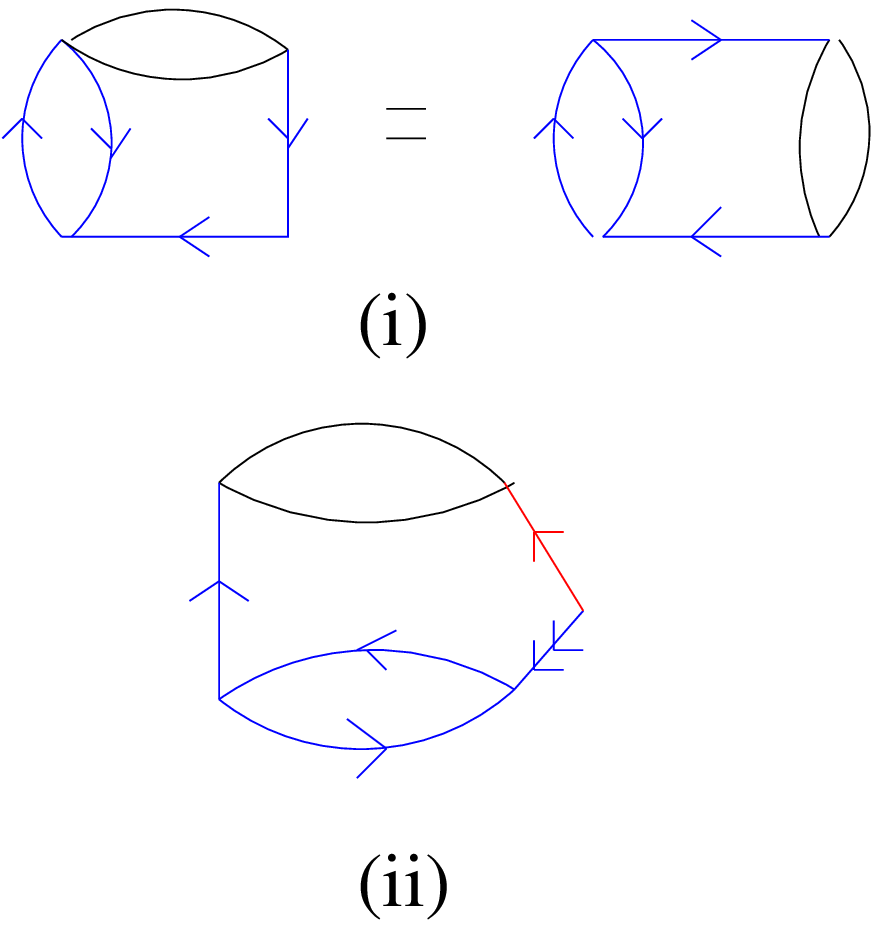}
\end{picture}}
\]
\caption{ $(i)$ the equal graphs, and $(ii)$ the auxiliary graph}
\end{center}
\end{figure} 

\begin{figure}[ht]
\begin{center}
\[
\mbox{\begin{picture}(250,210)(0,0)
\includegraphics[scale=.7]{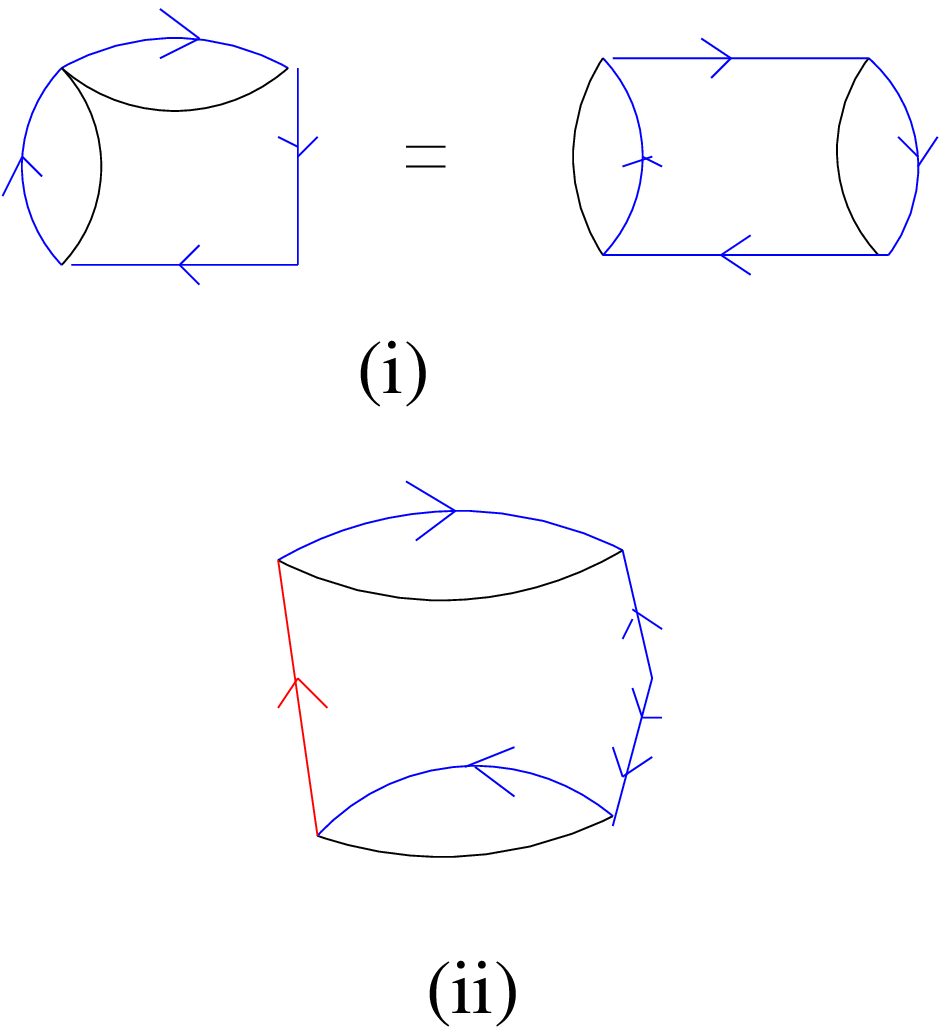}
\end{picture}}
\]
\caption{$(i)$ the equal graphs, and $(ii)$ the auxiliary graph}
\end{center}
\end{figure}

\vspace{.5cm}

{\bf{Acknowledgements:}} I am thankful to Dileep Jatkar and Ashoke Sen for useful comments.

\section{Appendix}

\appendix

\section{Various crossing symmetric combinations}

To denote the various spacetime structures that arise in our analysis of the four graviton amplitude at tree level and at one loop, we introduce the manifestly crossing symmetric combinations given below. 

The tensor involving two factors of $\eta^{\m\n}$ is given by
\bea I_{1;m,n}^{\m_1\m_2\m_3\m_4} &=& \Big(\frac{\alpha's}{4}\Big)^m \Big(\frac{\alpha'^2 ut}{16}\Big)^n \eta^{\m_1\m_2} \eta^{\m_3\m_4} + \Big(\frac{\alpha't}{4}\Big)^m \Big(\frac{\alpha'^2 su}{16}\Big)^n \eta^{\m_1\m_4}\eta^{\m_2\m_3} \non \\ &&+\Big(\frac{\alpha'u}{4}\Big)^m \Big(\frac{\alpha'^2 st}{16}\Big)^n\eta^{\m_1\m_3}\eta^{\m_2\m_4}. \eea 

The two tensors involving one factor of $\eta^{\m\n}$ are
\bea  I_{2;m,n}^{\m_1\m_2\m_3\m_4} &=& \Big(\frac{\alpha's}{4}\Big)^m \Big(\frac{\alpha'^2 ut}{16}\Big)^n (\eta^{\m_1\m_2} k_3^{\m_4} k_4^{\m_3} + \eta^{\m_3\m_4} k_1^{\m_2} k_2^{\m_1})  \non \\ && +  \Big(\frac{\alpha't}{4}\Big)^m \Big(\frac{\alpha'^2 su}{16}\Big)^n(\eta^{\m_1\m_4} k_2^{\m_3} k_3^{\m_2} + \eta^{\m_2\m_3} k_1^{\m_4} k_4^{\m_1})\non \\ &&+   \Big(\frac{\alpha'u}{4}\Big)^m \Big(\frac{\alpha'^2 st}{16}\Big)^n(\eta^{\m_1\m_3} k_2^{\m_4} k_4^{\m_2} + \eta^{\m_2\m_4} k_1^{\m_3} k_3^{\m_1}), \non \\ I_{3,n}^{\m_1\m_2\m_3\m_4} &=& \Big(\frac{\alpha's}{4}\Big)^n  \Big[\eta^{\m_1\m_2}(tk_1^{\m_3} k_2^{\m_4} + u k_1^{\m_4} k_2^{\m_3}) + \eta^{\m_3\m_4} (t k_3^{\m_1} k_4^{\m_2} + u k_3^{\m_3} k_4^{\m_1})\Big]\non \\ && +\Big(\frac{\alpha't}{4}\Big)^n  \Big[\eta^{\m_1\m_4}(sk_1^{\m_3} k_4^{\m_2} + u k_1^{\m_2} k_4^{\m_3}) + \eta^{\m_2\m_3} (s k_2^{\m_4} k_3^{\m_1} + u k_2^{\m_1} k_3^{\m_4})\Big]\non \\ && +\Big(\frac{\alpha'u}{4}\Big)^n  \Big[\eta^{\m_1\m_3}(sk_1^{\m_4} k_3^{\m_2} + t k_1^{\m_2} k_3^{\m_4}) + \eta^{\m_2\m_4} (s k_2^{\m_3} k_4^{\m_1} + t k_2^{\m_1} k_4^{\m_3})\Big].\non \\\eea 

Finally, the three tensors involving no factors of $\eta^{\m\n}$ are
\bea
I_{4,n}^{\m_1\m_2\m_3\m_4} &=& \Big(\frac{\alpha's}{4}\Big)^n k_1^{\m_2} k_2^{\m_1} k_3^{\m_4} k_4^{\m_3} + \Big(\frac{\alpha't}{4}\Big)^n k_1^{\m_4} k_4^{\m_1} k_2^{\m_3} k_3^{\m_2} + \Big(\frac{\alpha'u}{4}\Big)^n k_1^{\m_3} k_3^{\m_1} k_2^{\m_4} k_4^{\m_2}, \non \\ 
I_{5,n}^{\m_1\m_2\m_3\m_4} &=& \Big(\frac{\alpha's}{4}\Big)^n (k_1^{\n_3} k_2^{\n_4} k_4^{\n_1} k_3^{\n_2} + k_1^{\n_4} k_2^{\n_3} k_3^{\n_1} k_4^{\n_2})  +\Big(\frac{\alpha't}{4}\Big)^n (k_1^{\n_3} k_4^{\n_2} k_2^{\n_1} k_3^{\n_4} + k_1^{\n_2} k_4^{\n_3} k_3^{\n_1} k_2^{\n_4})\non \\ &&+ \Big(\frac{\alpha'u}{4} \Big)^n(k_1^{\n_2} k_3^{\n_4} k_4^{\n_1} k_2^{\n_3} + k_1^{\n_4} k_3^{\n_2} k_2^{\n_1} k_4^{\n_3}),
\non \\ 
I_{6,n}^{\m_1\m_2\m_3\m_4} &=& \Big(\frac{\alpha's}{4}\Big)^n (k_3^{\n_1} k_3^{\n_2} + k_4^{\n_1} k_4^{\n_2})(k_1^{\n_3}k_1^{\n_4} + k_2^{\n_3} k_2^{\n_4})+ \Big(\frac{\alpha't}{4}\Big)^n (k_2^{\n_1} k_2^{\n_4} + k_3^{\n_1} k_3^{\n_4}) \times \non \\ && (k_1^{\n_2}k_1^{\n_3} + k_4^{\n_2} k_4^{\n_3})+\Big(\frac{\alpha'u}{4}\Big)^n (k_1^{\n_2} k_1^{\n_4} + k_3^{\n_2} k_3^{\n_4})(k_2^{\n_1}k_2^{\n_3} + k_4^{\n_1} k_4^{\n_3}), 
  \eea
which repeatedly arise in our analysis.

Thus note that
\be  K^{\m_1\m_2\m_3\m_4} = \frac{4}{\alpha'^2} I_{1;0,1}^{\m_1\m_2\m_3\m_4} - \frac{1}{2} I_{3,0}^{\m_1\m_2\m_3\m_4}.\ee

\section{The various modular graph functions}

In the main text, various modular graph functions arise which we summarize below. They are $SL(2,\mathbb{Z})$ covariant expressions which are functions of the complex structure of the torus. The vertices of these graphs are the locations of the vertex operators on the toroidal worldsheet which are integrated with measure $d^2z/\tau_2$, while the links are either given by Green functions or their derivatives. These arise while performing the derivative expansion of the one loop amplitude. The graphs we need to consider for our analysis have either two of four factors of $\bar\p G$,  and hence are modular forms of weight $(0,2)$ of $(0,4)$ respectively.

We find it very convenient for our analysis to denote the graphs diagrammatically. In the various graphs, black links stand for the Green function.  
On the other hand, the notations for links having holomorphic and antiholomorphic derivatives acting on the Green function are given in figure 13, along with the notation for a link having a single Green function having both these derivatives.  

\begin{figure}[ht]
\begin{center}
\[
\mbox{\begin{picture}(280,60)(0,0)
\includegraphics[scale=.75]{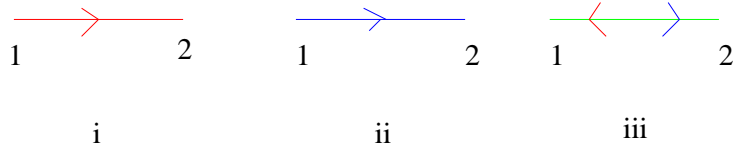}
\end{picture}}
\]
\caption{ (i) $\p_2 G_{12} = -\p_1 G_{12}$, (ii) $\bar\p_2 G_{12} = -\bar\p_1 G_{12}$ and (iii) $\p_1 \bar\p_2 G_{12}$}
\end{center}
\end{figure}

\begin{figure}[ht]
\begin{center}
\[
\mbox{\begin{picture}(280,325)(0,0)
\includegraphics[scale=.65]{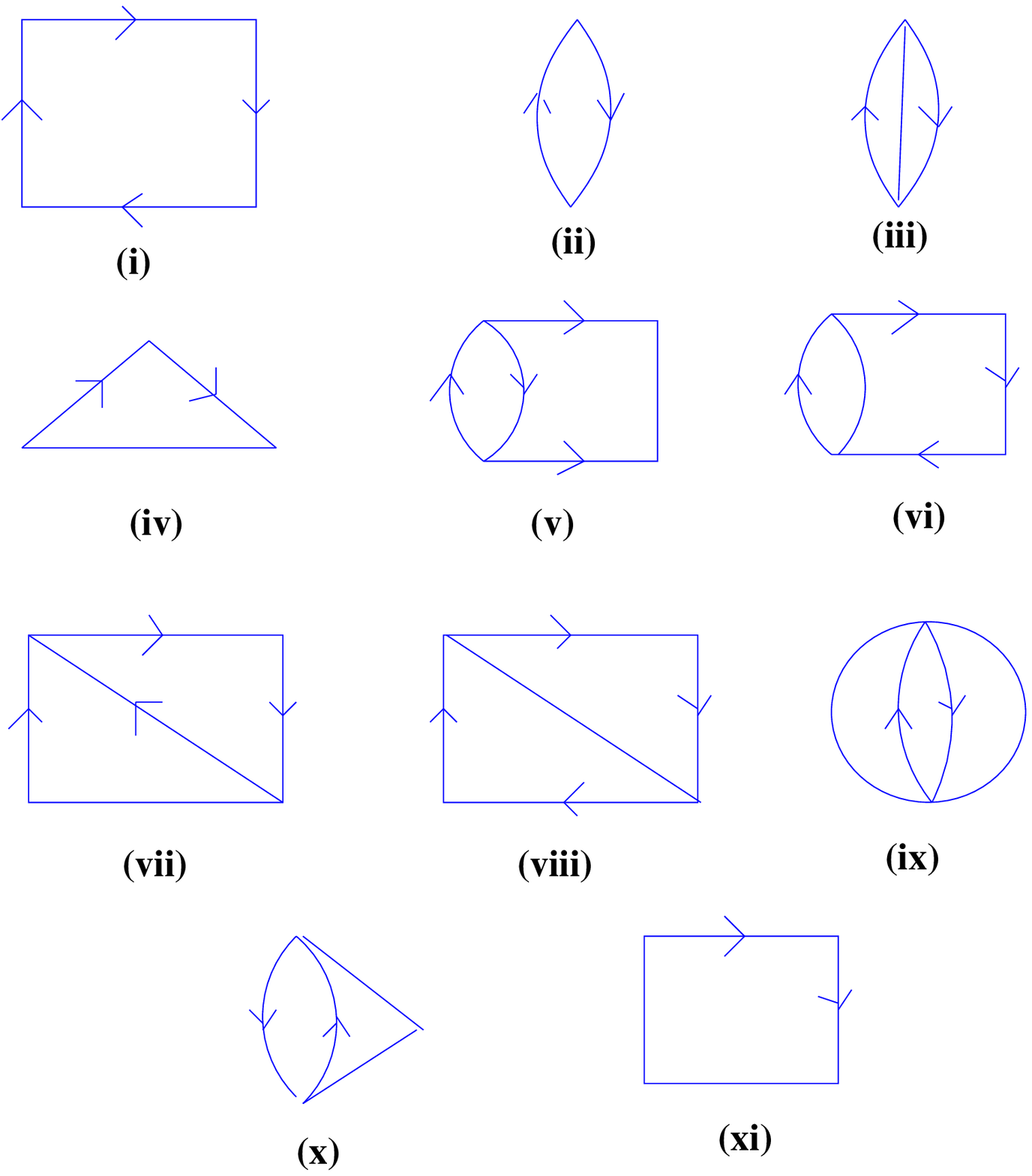}
\end{picture}}
\]
\caption{The graphs (i) $Q_1$, (ii) $Q_2$, (iii) $Q_3$, (iv) $Q_4$, (v) $Q_5$, (vi) $Q_6$, (vii) $Q_7$, (viii) $Q_8$, (ix) $Q_9$, (x) $Q_{10}$ and (xi) $Q_{11}$}
\end{center}
\end{figure}

For our purposes, the relevant graphs are those that appear in the amplitude that yield terms with upto ten derivatives in the effective action, and are given in figure 14.  Of them, the graphs $Q_2$, $Q_3$, $Q_4$, $Q_9$, $Q_{10}$ and $Q_{11}$ have two links involving $\bar\p G$ while the rest involve $G$, and hence are weight $(0,2)$ modular forms. The remaining graphs $Q_1$, $Q_5$, $Q_6$, $Q_7$ and $Q_8$ have four links involving $\bar\p G$ while the rest involve $G$, and hence are weight $(0,4)$ modular forms. These combine with other modular covariant expressions to yield modular invariant integrands which have to be integrated over the truncated fundamental domain of $SL(2,\mathbb{Z})$ to give us coefficients of appropriate terms in the effective action.

The relevant graphs are
\bea \label{graphs}Q_1 &=& \prod_{i=1}^4 \int_\S  \frac{d^2 z^i}{\tau_2} \bar\p_2 G_{12} \bar\p_3 G_{23} \bar\p_4 G_{34} \bar\p_1 G_{14}, \quad Q_2 = \prod_{i=1}^2 \int_\S  \frac{d^2 z^i}{\tau_2} \bar\p_1 G_{12} \bar\p_2 G_{12}, \non \\ 
Q_3 &=& \prod_{i=1}^2 \int_\S  \frac{d^2 z^i}{\tau_2} G_{12} \bar\p_1 G_{12} \bar\p_2 G_{12}, \quad Q_4 = \prod_{i=1}^3 \int_\S  \frac{d^2 z^i}{\tau_2} \bar\p_2 G_{12} \bar\p_3 G_{23} G_{13}, \non \\ Q_5 &=& \prod_{i=1}^4 \int_\S  \frac{d^2 z^i}{\tau_2} \bar\p_1 G_{12} \bar\p_2 G_{12} \bar\p_3 G_{23} G_{34} \bar\p_4 G_{14}, \non \\ Q_6 &=& \prod_{i=1}^4 \int_\S  \frac{d^2 z^i}{\tau_2} G_{12} \bar\p_2 G_{12} \bar\p_3 G_{23} \bar\p_4 G_{34} \bar\p_1 G_{14} , \non \eea
 
as well as
\bea Q_7 &=& \prod_{i=1}^4 \int_\S  \frac{d^2 z^i}{\tau_2} \bar\p_2 G_{12} \bar\p_3 G_{23} \bar\p_4 G_{34} G_{14} \bar\p_2 G_{24}, \non \\ Q_8 &=& \prod_{i=1}^4 \int_\S  \frac{d^2 z^i}{\tau_2} \bar\p_2 G_{12} \bar\p_3G_{23} \bar\p_4G_{34} \bar\p_1G_{14} G_{24}, \quad
Q_9 = \prod_{i=1}^2 \int_\S  \frac{d^2 z^i}{\tau_2} G_{12}^2 \bar\p_1 G_{12} \bar\p_2 G_{12}, \non \\ Q_{10} &=& \prod_{i=1}^3 \int_\S  \frac{d^2 z^i}{\tau_2} G_{13} G_{23} \bar\p_1 G_{12} \bar\p_2 G_{12}, \quad Q_{11} =\prod_{i=1}^4 \int_\S  \frac{d^2 z^i}{\tau_2} \bar\p_2 G_{12} \bar\p_3 G_{23} G_{34} G_{14},\eea
as depicted in figure 14.

\section{Asymptotic expansions of the various modular graph functions}

In the main text, the asymptotic expansions of the various graphs provide a useful check not only of the structure of the differential equations but they also provide data to determine the boundary terms. The asymptotic expansion as well provides necessary boundary data in order to calculate the integrals. In our analysis, we shall consider the terms that are independent of $\tau_1$ and power behaved in $\tau_2$ in the large $\tau_2$ expansion. In general, such expansions lead to powerful constraints in analyzing the graphs~\cite{Green:2008uj,D'Hoker:2015foa,Zerbini:2015rss} using various techniques. 

For our analysis, we shall obtain the terms in the asymptotic expansion by setting various subsets of the lattice momenta to zero, and performing Poisson resummation over some of the appropriate remaining lattice momenta. On setting the dual winding momenta to zero, we shall obtain the necessary terms in the series. The various powers of $\tau_2$ that arise are determined by the set of lattice momenta that are set to zero. This method (which has been used in the type II theory~\cite{Green:2008uj}) gives a clear intuitive picture of the source of origin of the various terms, apart from being tractable enough to implement for the graphs we are interested in.         

In our analysis, we shall make use of the asymptotic expansions
\bea \mathcal{E}_s (\tau,\bar\tau) &=& \frac{2\zeta(2s)}{\pi^s}  \tau_2^s + \frac{2\zeta(2s-1) \Gamma(s-1/2)}{\pi^{s-1/2}\Gamma(s)} \tau_2^{1-s}+\dots,\non \\ G_{2k} (\tau) &=& 2\zeta(2s) +\ldots,\eea
for $k \geq 2$ in the large $\tau_2$ limit, and
\be \hat{G}_2 (\tau,\bar\tau) = 2\zeta(2) - \frac{\pi}{\tau_2} +\ldots,\ee
where the terms we have neglected are exponentially suppressed in $\tau_2$ and also depend on $\tau_1$. We now consider the asymptotic expansions of the various graphs needed in the main text, focussing on only those terms that are power behaved in the large $\tau_2$ expansion.

\subsection{Asymptotic expansions of $Q_3$ and $Q_{10}$}

From \C{graphs}, we have that
\bea \label{DeF}Q_3 = - \sum' \frac{\tau_2}{\pi (m_1 + n_1\bar\tau)(m_2 + n_2 \bar\tau)\vert m_3 + n_3\tau\vert^2}, \non \\ Q_{10} = - \sum' \frac{\tau_2^2}{\pi^2 (m_1 + n_1\bar\tau)(m_2 + n_2 \bar\tau)\vert m_3 + n_3\tau\vert^4}, \eea
where the sum is over the integers $m_i, n_i$ satisfying the constraints
\be \label{sum1}(m_i,n_i) \neq (0,0), ~i=1,2,3, \quad \sum_i m_i = \sum_i n_i =0.\ee

In our analysis, we extract the contributions that are power behaved in $\tau_2$ by setting $n_i =0$ in the sums for appropriate $i$, which determines the $\tau_2$ dependence. In the discussions below, we mention which of the $n_i$ are set to zero, while it is implied that the others take only non-zero values.

For all the expressions involving $Q_3,\ldots,Q_9$ the leading contribution at large $\tau_2$ is given by setting all $n_i=0$. For $Q_3$ we get
\be \label{A1}-\frac{\pi}{\tau_2} Q_3^{asymp} = \sum' \frac{1}{ m_1 m_2 (m_1+m_2)^2} = 2 W(1,1,2) - 4 W(2,1,1),\ee
while for $Q_{10}$ we get
\be  \label{A2}-\frac{\pi^2}{\tau_2^2} Q_{10}^{asymp} = \sum' \frac{1}{ m_1 m_2 (m_1+m_2)^4} = 2 W(1,1,4) - 4 W(4,1,1).\ee
In \C{A1} and \C{A2}, the sum is over integers $m_1, m_2$ satisfying the constraints
\be \label{sum2}m_i \neq 0, \quad m_1 + m_2 \neq 0.\ee
Here we make use of the expression for the Tornheim sum or the Witten zeta function $W(\alpha_1,\alpha_2,\beta)$ defined by~\footnote{The multi zeta value is defined by
\be \zeta(s_1,\ldots,s_r) = \sum_{n_1 > n_2 > \ldots > n_r \geq 1 } \frac{1}{n_1^{s_1}n_2^{s_2}\ldots n_r^{s_r}}.\ee}
\bea W(\alpha_1,\alpha_2,\beta) &=& \sum_{m,n=1}^\infty \frac{1}{m^{\alpha_1} n^{\alpha_2} (m+n)^{\beta}} \non \\ &=& \sum_{r+s = \alpha_1+\alpha_2; r,s>0} \Big[ \binom{r-1}{\alpha_1 -1}+\binom{r-1}{\alpha_2 -1}\Big] \zeta(\beta+r,s).\eea
Thus 
\bea \label{match}Q_3^{asymp} &=&  \frac{4\tau_2}{\pi}\Big(\zeta(3,1) + \zeta(2,2)\Big) = \frac{4\zeta(4)}{\pi}\tau_2, \non \\ Q_{10}^{asymp} &=& \frac{4\tau_2^2}{\pi^2}\Big(\zeta(5,1)+\zeta(2,4)+\zeta(4,2) +\zeta(3,3)\Big) = \frac{4\zeta(6)}{\pi^2}\tau_2^2,\eea
on using the relations
\bea \label{mzv}&& 4 \zeta(3,1) = \frac{4}{3}\zeta(2,2) =\zeta(4), \quad 4\zeta(5,1) = 3\zeta(6) - 2 \zeta(3)^2, \non \\ && \zeta(2,4)= -\zeta(3)^2 +\frac{25}{12} \zeta(6), \quad \zeta(4,2) = \zeta(3)^2 -\frac{4}{3} \zeta(6),\quad 2\zeta(3,3) = \zeta(3)^2 - \zeta(6)\non \\ \eea
between multi zeta values and Riemann zeta functions.

We would now like to obtain the subleading contributions in the large $\tau_2$ expansion in a systematic way. 
To perform these asymptotic expansions, we consider the general sum given by
\be \label{gen}Q = -\sum' \frac{(m_1 + n_1\tau) (m_2 +n_2\tau)\tau_2^{s_3}}{\pi^{s_3}\vert m_1 + n_1\tau \vert^{2s_1}\vert m_2 + n_2 \tau \vert^{2s_2}\vert m_3 + n_3\tau\vert^{2s_3}}.\ee
Thus $Q_3$ corresponds to the case where $s_1 = s_2 = s_3 =1$, while $Q_{10}$ corresponds to the case where $s_1 = s_2 = 1, s_3 =2$. While the leading contributions are given by \C{match}, for our cases the first subleading contribution is obtained from the sector $n_3 =0$. On relabelling the indices, this is given by
\be \label{Expmore}\pi^{s_3} Q^{asymp} = \sum_{m_1 \neq 0, n_2 \neq 0, m_2 } \frac{\tau_2^{s_3} (m_2 + n_2\tau) ((m_1 +m_2) +n_2\tau)}{ m_1^{2s_3}\vert m_2 + n_2 \tau \vert^{2s_1}\vert (m_1 + m_2) + n_2\tau\vert^{2s_2}}.\ee 
To extract the asymptotic behaviour, we perform a Poisson resummation on $m_2$ on using the integral representation
\be \label{repint}\frac{1}{\vert z \vert^{2s}} = \frac{1}{\G(s)} \int_0^\infty d\l \l^{s-1} e^{-\vert z \vert^2 \l}\ee
for the two denominators. Thus $\l$ is the Schwinger parameter denoting the proper time for the propagator involving the lattice momenta. Naming the integer that is summed over after Poisson resummation as $\hat{m}_2$ (thus is it the winding momentum label), we see that the only $\tau_1$ dependence appears as a phase $e^{2\pi i \hat{m}_2 n_2 \tau_1}$. Hence to consider the $\tau_1$ independent terms, we set $\hat{m}_2=0$, leading to
\bea \label{zeroterm}\pi^{s_3} Q^{asymp} &=& \sum_{m_1 \neq 0, n_2 \neq 0} \frac{\pi^{1/2}\tau_2^{s_3}}{m_1^{2s_3}\G(s_1)\G(s_2)}\int_0^\infty d\l d\r \frac{\l^{s_1 -1} \r^{s_2 -1}}{\sqrt{\l+\r}}\non \\ &&\times  \Big[ \frac{1}{2(\l+\r)} - \frac{\l\r m_1^2}{(\l+\r)^2} - n_2^2\tau_2^2\Big] e^{-n_2^2\tau_2^2 (\l+\r)- m_1^2\l\r/(\l+\r).}\eea 
We have dropped the purely imaginary term $i\tau_2m_1 n_2 (\l-\r)/(\l+\r)$ in the bracket in the integrand as it is odd in $m_1$ as well as $n_2$ (it also vanishes on setting $s_1 = s_2$ which we shall eventually take) and hence it vanishes. Thus the expression is manifestly real.  

To obtain a systematic asymptotic expansion, we now define
\be \s = \l +\r, \quad \omega = \frac{\l}{\s}, \quad 0 \leq \omega \leq 1.\ee 
On further substituting $x = n_2^2\tau_2^2\s$, we get that 
\bea \label{Exp}\pi^{s_3} Q^{asymp} = \frac{4\pi^{1/2}}{\G(s_1)\G(s_2)}\tau_2^{1+s_3 -2 s_1 -2 s_2}\sum_{m_1 > 0, n_2 > 0} \frac{1}{m_1^{2s_3}n_2^{2s_1 + 2 s_2 -1}}\int_0^\infty dx x^{s_1 + s_2 -3/2}\non \\ \times  \int_0^1 d\omega \omega^{s_1 -1}(1-\omega)^{s_2 -1}\Big[ n_2^2\tau_2^2\Big(\frac{1}{2x}- 1\Big) -\omega(1-\omega)m_1^2\Big] e^{-x- \omega(1-\omega)m_1^2 x/n_2^2\tau_2^2.}\eea
Let us now consider the leading term in the asymptotic expansion. To obtain this, we set the term in the exponential $e^{-\omega(1-\omega)m_1^2 x/n_2^2\tau_2^2}$  equal to one. Thus to leading order we have that 
\bea\label{zeta} \pi^{s_3} Q^{asymp} = \frac{4\pi^{1/2}}{\G(s_1 + s_2)}\tau_2^{3+s_3 -2 s_1 -2 s_2}\zeta(2s_3)\zeta(2s_1 + 2s_2 -3)\int_0^\infty dx x^{s_1 + s_2-3/2}\Big(\frac{1}{2x}-1\Big)e^{-x}.\non \\ \eea
The $x$ integral yields
\be \label{g} (2-s_1 -s_2) \G(s_1 + s_2 -3/2).\ee
We now see why it is very useful and important for us to start with arbitrary $s_1$ and $s_2$. For $s_1 =s_2 =1$, the second zeta function in \C{zeta} diverges, while the multiplicative factor in \C{g} vanishes. Thus we take $s_1 = s_2 = 1+\epsilon$, and take the limit $\epsilon \rightarrow 0$~\footnote{Alternatively, we use the identity
\be \zeta(2s_1 + 2s_2 -3)(2-s_1 -s_2) \G(s_1 + s_2 -3/2) =\pi^{2s_1 + 2s_2 -7/2}\zeta(4-2s_1-2s_2)\G(3-s_1-s_2)\ee
which gives a finite value.}. In this limit
\be (2-s_1 -s_2) \zeta(2s_1 + 2s_2 -3)  \rightarrow -1/2\ee
giving us a finite answer~\footnote{In fact, this is exactly how $\hat{G}_2$ is defined by
\be \hat{G}_2 = {\rm lim}_{s\rightarrow 0}\sum_{(m,n)\neq (0,0)} \frac{1}{(m+n\tau)^2 \vert m+n\tau\vert^{2s}}.\ee
The regularization introduces the non--holomorphicity.}. Thus at leading order in this sector we get that
\be  \label{vv}Q^{asymp}_3 = -2\zeta(2) , \quad Q^{asymp}_{10} = -\frac{2\zeta(4)}{\pi} \tau_2 .\ee
Note that unlike the leading order expressions in \C{match}, this contribution arises from a regularized expression, and has transcendentality one less than what would be naively expected. This is because $0 \zeta(1)$ yields a constant of vanishing transcendentality on taking the limit, while $\zeta(1)$ formally has transcendentality one. We interpret such contributions as coming from the boundary of moduli space in the differential equation, as discussed in the main text.

It is very useful for our purposes to have an alternative representation for $Q^{asymp}$. 
To do so, we use a different representation for \C{zeroterm} when $\hat{m}_2= 0$ given by (generalizing the analysis in~\cite{Green:2008uj})
\be \label{equal}\pi^{s_3} Q^{asymp} = \sum_{m_1 \neq 0, n_2 \neq 0} \frac{\tau_2^{s_3}}{m_1^{2s_3}}\int_{-\infty}^\infty d\m \frac{(\m +i n_2 \tau_2)(\m + m_1 + in_2\tau_2)}{\vert \m +i n_2 \tau_2\vert^{2s_1} \vert \m + m_1 +i n_2 \tau_2\vert^{2s_2}}.\ee 
The equality easily follows by introducing Schwinger parameters $\l$ and $\r$ for the two propagators in \C{equal} with powers $2s_1$ and $2s_2$ respectively, using \C{repint} and integrating over $\m$. Let us set $s_1 = s_2 = s$ and consider the limit $s\rightarrow 1$. Thus we have that
\be \label{D}\pi^{s_3} Q^{asymp} = 4\sum_{m_1 > 0, n_2 > 0} \frac{\tau_2^{3+s_3-4s}}{m_1^{2s_3} n_2^{4s-3}}I(s;a),\ee 
where
\be \label{Integral}I(s;a) =  I(s;-a) = \int_{-\infty}^\infty d\n \frac{\n(\n + a)-1}{(\n^2 +1)^{s} [(\n+a)^2 + 1]^{s}},\ee
$a= m_1/n_2\tau_2$. We note that $I(1;a) =0$ for all $a$, and thus we can write~\footnote{This can also be directly seen on using
\be \label{s-1}I(s;a) = 2\sqrt{\pi}(1-s) \frac{\G(2s-3/2)}{\G^2(s)}\int_0^1 dx \frac{x^{s-1}(1-x)^{s-1}}{[1+a^2 x(1-x)]^{2s-3/2}}\ee
on introducing the Feynman parameter $x$ and performing the $\n$ integral.}
\be I(s;a) = (s-1) \hat{I} (s;a).\ee
Now the leading contribution in the large $\tau_2$ limit is obtained by setting $I(s;a) = I(s;0)$ in \C{D} which leads to \C{vv} as before. This contribution arose from regularizing the product of a vanishing integral and a divergent sum over the integer $n_2$ to yield a finite number.

What about a possible contribution from a potential divergence from the sum over the integer $m_1$? We now heuristically argue that this can lead to a contribution that is subleading in the large $\tau_2$ expansion. 

For $Q_3$, this can yield a term of the form $1/\tau_2$. Thus in this sector
\be \label{der1}Q_3^{asymp} = -2\zeta(2) + \frac{c_0}{\tau_2}.\ee 
To show the possible existence of the term involving $c_0$, it is very convenient to consider
\be \label{leaD}\frac{\p Q_3^{asymp}}{\p\tau_2} = -\frac{c_0}{\tau_2^2}\ee 
simply because $c_0$ is obtained as the leading contribution in \C{leaD}. From \C{D}, we have that 
\be \label{LeaD}\pi \frac{\p Q_3^{asymp}}{\p\tau_2} = \frac{16(1-s)}{\tau_2}\sum_{m_1 > 0, n_2 > 0}\frac{\tau_2^{4(1-s)}}{m_1^2 n_2^{4s-3}} I(s;a) -\frac{4}{\tau_2^2} \sum_{m_1 > 0, n_2 > 0} \frac{\tau_2^{4(1-s)}}{m_1n_2^{4s-2}} \frac{\p  I(s;a)}{\p a}.\ee
The contribution from the first term vanishes as $s\rightarrow 1$, as the sum involves $Q_3^{asymp}$ which is finite on regularization. On the other hand, the second term which has a prefactor of $O(1/\tau_2^2)$ can yield a potentially non--vanishing contribution. This happens on evaluating $\p I(s;a)/\p a$ at $a=0$. This yields an integral over $\n$ which trivially vanishes for arbitrary $s$.
However, the sum over $m_1$ then formally yields $\zeta(1)$ which diverges. Thus only such a contribution can yield a non--vanishing result on regularizing, giving us \C{der1}. Rather than obtaining this coefficient by a direct computation, we shall later argue for its value using modular covariance. 

Note that unlike the constant term in \C{der1} whose regularization only depended on $s_1$ and $s_2$, the regularization of the second term also depends on $s_3$. While the vanishing of the integral involving $\p I(s;a)/\p a$ evaluated at $a=0$ is trivial, what is non--trivial is the divergence arising from the sum over $m_1$ which is dictated by the value of $s_3$. The product of these two contributions can lead to a finite answer.        

In fact, proceeding similarly we get that
\be Q_{10}^{asymp} = -\frac{2\zeta(4)}{\pi}\tau_2 +\frac{c_1}{\tau_2^2},\ee
where we start by considering
\be \frac{\p^2 Q_{10}^{asymp}}{\p\tau_2^2} = \frac{6c_1}{\tau_2^4}.\ee
Thus this heuristic argument shows that the potential non--vanishing subleading contribution is obtained simply by Taylor expanding $I(s;a)$ around $a=0$ to the appropriate power determined by $s_3$, and then by regularizing the product of a trivially vanishing integral and a divergent sum.

Next we consider the further subleading contribution where $n_1 =0$ in \C{gen} ($n_2 =0$ gives the same result since $s_1= s_2$). The total contribution from these two cases is given by
\bea \pi^{s_3} Q^{asymp} &=&  -2\sum_{m_1 \neq 0, n_2 \neq 0} \frac{\tau_2^{s_3}}{m_1^{2s_1-1}}\int_{-\infty}^\infty d\m \frac{(\m +i n_2 \tau_2)}{\vert \m +i n_2 \tau_2\vert^{2s_2} \vert \m + m_1 +i n_2 \tau_2\vert^{2s_3}}\non \\ &=&  -4\sum_{m_1 \neq 0, n_2 > 0} \frac{\tau_2^{s_3}}{m_1^{2s_1-1}(n_2\tau_2)^{2(s_2+s_3)-2}}\int_{-\infty}^\infty d\n \frac{\n}{(\n^2 +1)^{s_2} [(\n + a)^2 +1 ]^{s_3}},\non \\ \eea
where we have Poisson resummed over the momenta $m_2$ and set the winding mode $\hat{m}_2 =0$. 

Thus for $Q_3$ we have that
\be \label{V1}Q_3^{asymp} = 4\tau_2 \sum_{m \neq 0, n > 0} \frac{1}{n\tau_2(m^2 + 4 n^2\tau_2^2)},\ee
while for $Q_{10}$ we have that
\be \label{V2}Q_{10}^{asymp} = \frac{2\tau_2^2}{\pi} \sum_{m \neq 0, n > 0}\frac{m^2 + 8 n^2 \tau_2^2}{n^3\tau_2^3(m^2 + 4 n^2 \tau_2^2)^2}.\ee
To evaluate \C{V1} and \C{V2}, we first perform the sum over $m$ by simplifying the summands using partial fractions and using
\be \sum_{m\in \mathbb{Z}} \frac{1}{m+z} = \pi {\rm cot}\pi z.\ee
Thus, we have that
\be \label{Cancel1}Q_3^{asymp} = 4\tau_2 \sum_{n>0} \frac{1}{n\tau_2} \Big(-\frac{1}{4n^2\tau_2^2} +\frac{\pi}{2n\tau_2}{\rm coth}2\pi n\tau_2 \Big) = -\frac{\zeta(3)}{\tau_2^2} + \frac{2\pi\zeta(2)}{\tau_2}.\ee
where we have ignored terms that are exponentially suppressed at large $\tau_2$. Similarly,
\be \label{Cancel2}Q_{10}^{asymp} = -\frac{\zeta(5)}{\pi\tau_2^3}+\frac{3\zeta(4)}{2\tau_2^2}.\ee

Finally, the remaining contribution comes from $n_1 \neq 0, n_2 \neq 0, n_1 +n_2\neq 0$. Poisson resumming over $m_1$ and $m_2$, we set the winding modes $\hat{m}_1 =0$ and $\hat{m}_2=0$, leading to\footnote{For this choice of $n_i$, the sector with $\hat{m}_1 n_1 + \hat{m}_2 n_2 = 0$, with $\hat{m}_1 \neq 0$, $\hat{m}_2\neq 0$ is also independent of $\tau_1$. However these terms are exponentially suppressed in $\tau_2$.}
\be Q_3^{asymp} = -\frac{\tau_2}{\pi}\sum' \int_{-\infty}^\infty d\mu_1 \int_{-\infty}^\infty d\mu_2 \frac{(\m_1 + i n_1 \tau_2)(\m_2 + i n_2\tau_2)}{\vert \m_1 + i n_1 \tau_2 \vert^2 \vert \m_2 + i n_2\tau_2\vert^2\vert\m_1+\m_2 +i(n_1+n_2)\tau_2\vert^2},\ee 
where the sum is over integers $n_i$ satisfying
\be \label{non}n_1 \neq 0, n_2 \neq 0, n_1 +n_2\neq 0.\ee
Defining $\mu_i = \vert n_i \vert \n_i \tau_2$, we get that
\be Q_3^{asymp} = -\frac{1}{\pi\tau_2} \sum' \int_{-\infty}^\infty d\n_1 \int_{-\infty}^\infty d\n_2 \frac{(\n_1 +isgn(n_1))(\n_2 +isgn(n_2))}{(\n_1^2 +1 )(\n_2^2 +1)[(\vert n_1\vert \n_1 +\vert n_2\vert \n_2)^2 + (n_1+n_2)^2]},\ee
where the sign function $sgn(n) = \pm 1$ if $n \gtrless 0$. Performing the integrals, we get that
\bea \label{cancel1}Q_3^{asymp} &=& \frac{\pi}{\tau_2}\sum' \frac{1+sgn(n_1)sgn(n_2)}{\vert n_1+n_2\vert (\vert n_1\vert +\vert n_2\vert +\vert n_1+n_2\vert)} = \frac{2\pi}{\tau_2} \sum_{m=1}^\infty \sum_{n=1}^\infty \frac{1}{(m+n)^2}\non \\ &=& \frac{2\pi}{\tau_2}\sum_{n=1}^\infty \frac{n-1}{n^2} = -\frac{2\pi}{\tau_2} \zeta(2) +\frac{\tilde{c}_0}{\tau_2},\eea
where in the final expression we also have a formally divergent term of $O(1/\tau_2)$ which has transcendentality less than 3, which has to be appropriately regularized by starting with arbitrary $s_i$.

Proceeding similarly, we get that
\bea \label{cancel2}Q_{10}^{asymp} &=& \frac{1}{2\tau_2^2}\sum' \frac{(1+sgn(n_1)sgn(n_2))(\vert n_1\vert + \vert n_2\vert + 2 \vert n_1+n_2\vert)}{\vert n_1+n_2\vert^3 (\vert n_1\vert +\vert n_2\vert +\vert n_1+n_2\vert)^2} \non \\ &=& \frac{3}{2\tau_2^2} \sum_{m=1}^\infty \sum_{n=1}^\infty \frac{1}{(m+n)^4} = \frac{3}{2\tau_2^2} \Big( \zeta(3) -\zeta(4)\Big).\eea

Note that the terms involving $\zeta(2)$ and $\zeta(4)$ in \C{cancel1} and \C{cancel2} cancel terms in \C{Cancel1} and \C{Cancel2} respectively. This is crucial for the consistency of the differential equations deduced in the main text.

Thus adding all the contributions, we get that
\bea Q_3^{asymp} &=& \Big( \frac{4\zeta(4)}{\pi}\tau_2  -\frac{\zeta(3)}{\tau_2^2} \Big) -2\zeta(2) + \frac{c_0 +\tilde{c}_0}{\tau_2}, \non \\
 Q_{10}^{asymp} &=& \Big( \frac{4\zeta(6)}{\pi^2}\tau_2^2 -\frac{\zeta(5)}{\pi\tau_2^3}\Big) -\frac{2\zeta(4)}{\pi}\tau_2 +\frac{3\zeta(3) + 2c_1}{2\tau_2^2}.\eea
In each case, the terms in parentheses have uniform transcendentality, which is more than that for the the other terms that we have determined. We interpret these terms having lesser transcendentality, as boundary terms. For these contributions, given the $\tau_2$ dependence and the leading terms, we write down the exact power behaved $\tau_2$ dependent terms based on modular covariant expressions, which also fixes the undetermined constants. This gives us   
\bea \label{Match} Q_3^{asymp} &=& \Big( \frac{4\zeta(4)}{\pi}\tau_2  -\frac{\zeta(3)}{\tau_2^2} \Big)-2\zeta(2) \Big( 1 - \frac{3}{\pi\tau_2}\Big) , \non \\
 Q_{10}^{asymp} &=& \Big( \frac{4\zeta(6)}{\pi^2}\tau_2^2 -\frac{\zeta(5)}{\pi\tau_2^3}\Big)-\pi  \Big(\frac{2\zeta(4)}{\pi^2}\tau_2 -\frac{\zeta(3)}{2\pi\tau_2^2}\Big),\eea
which have $SL(2,\mathbb{Z})$ covariant completions given by
\bea \label{fin1}Q_3 &=& 2\pi \bar{D}_0 \mathcal{E}_2 -2\zeta(2) \overline{\hat{E}}_2, \non \\ Q_{10} &=& \frac{4\pi}{3}\bar{D}_0 \mathcal{E}_3 -\pi \bar{D}_0 \mathcal{E}_2.\eea

Thus we have that
\be \label{NeedLat} c_0+\tilde{c}_0 = \pi.\ee

Note that for the power behaved terms, we have obtained the answer for the boundary terms which need regularization based on the asymptotics and modular covariance, and have not explicitly determined all the coefficients. This is also true for the later graphs we analyze. A general understanding of such regularized contributions will be quite useful. 

Now there is a good reason why we separately covariantized the two sets of contributions in \C{Match} to obtain $SL(2,\mathbb{Z})$ covariant expressions. 
This is simply because \C{fin1} agrees with \C{f1} and \C{f2}. While for the first term on the right hand side of each equation this is a non--trivial consistency check of the eigenvalue equation, for the boundary contributions the asymptotic behavior is then used as an input in $\delta_{\m} Q$ which has been covariantized.     

We now consider the asymptotic expansions of the other graphs. A lot of the primary arguments follows the analysis done above, hence we give only the results in such cases and skip the details. Also in the various sectors,  the dual winding number shall always be set to zero for the momentum that is Poisson resummed, which is very similar to the above analysis.

\subsection{Asymptotic expansion of $Q_5$}

From \C{graphs}, we have that
\bea \label{Def5}Q_5 =  \sum' \frac{\tau_2}{\pi (m_1 + n_1\bar\tau)(m_2 + n_2 \bar\tau)(m_3 +n_3 \bar\tau)^2\vert m_3 + n_3\tau\vert^2},  \eea
where the sum is over integers $m_i$, satisfying \C{sum1}. 
The leading contribution, obtained by setting all $n_i =0$, is given by
\be \frac{\pi}{\tau_2} Q_5^{asymp} = \sum' \frac{1}{ m_1 m_2 (m_1+m_2)^4} = 2 W(1,1,4) - 4 W(4,1,1)\ee
where $m_i$ satisfies \C{sum2}, leading to
\be Q_5^{asymp} =  -\frac{4\zeta(6)}{\pi}\tau_2.\ee

To consider the subleading contributions, as before we start with 
\be \label{gen2}\hat{Q} = \sum' \frac{(m_1 + n_1\tau) (m_2 +n_2\tau)(m_3+n_3\tau)^2\tau_2^{s_3}}{\pi^{s_3}\vert m_1 + n_1\tau \vert^{2s_1}\vert m_2 + n_2 \tau \vert^{2s_2}\vert m_3 + n_3\tau\vert^{2s_3+4}}.\ee
The first subleading contribution is given by $n_3 =0$ and needs to be regularized as $s_1 = s_2 = s \rightarrow 1$. The analysis is exactly along the lines of what we did for $Q_3$ and $Q_{10}$ leading to
\be  Q^{asymp}_5 = 2\zeta(4) +\frac{c_2}{\tau_2^3}\ee
where $c_2$ shall be determined later using modular covariance.
Once again this is interpreted as a contribution coming from the boundary of moduli space. 

The contributions from $n_1=0$ and $n_2=0$ are the same, and they together give
\be Q_5^{asymp} = \tau_2 \sum_{m \neq 0, n > 0} \frac{1}{n^3\tau_2^3(m^2 + 4 n^2\tau_2^2)} = -\frac{\zeta(5)}{4\tau_2^4} +\frac{\pi\zeta(4)}{2\tau_2^3}\ee 
where we have dropped terms that are exponentially suppressed at large $\tau_2$ as before, which we continue to do later as well.

From the sector \C{non} we get that
\be Q_5^{asymp} = \frac{\pi}{4\tau_2^3}\sum' \frac{(1+sgn(n_1)sgn(n_2))\vert n_1\vert}{\vert n_1+n_2\vert^5 (\vert n_1\vert +\vert n_2\vert +\vert n_1+n_2 \vert)^3} f(\vert n_1\vert, \vert n_2\vert, \vert n_1+n_2\vert),\ee
where
\bea  f(\vert n_1\vert, \vert n_2\vert, \vert n_1+n_2\vert)  &=&-9 \vert n_1+n_2\vert^3 + 9\vert n_1+n_2\vert (\vert n_1 \vert +\vert n_2\vert)^2 \non \\ &&+ 5 (\vert n_1\vert +\vert n_2\vert) \vert n_1+n_2\vert^2 + 3 (\vert n_1\vert +\vert n_2\vert)^3.\eea
Thus
\be Q_5^{asymp} = \frac{\pi\zeta(3)}{2\tau_2^3} - \frac{\pi\zeta(4)}{2\tau_2^3}\ee
from this sector. 

Adding all the contributions,  we get that
\be Q_5^{asymp} = -\Big(\frac{4\zeta(6)}{\pi}\tau_2 +\frac{\zeta(5)}{4\tau_2^4} \Big) +2\zeta(4) + \frac{\pi\zeta(3) +2 c_2}{2\tau_2^3} .\ee
This naturally leads us to the asymptotic expansion
\be Q_5^{asymp} = -\Big(\frac{4\zeta(6)}{\pi}\tau_2 +\frac{\zeta(5)}{4\tau_2^4} \Big) +2\zeta(4),\ee
which has an $SL(2,\mathbb{Z})$ covariant completion given by
\be Q_5 = -\frac{2\pi^2}{3} \bar{D}_2 \bar{D}_0 \mathcal{E}_3 +\frac{2\pi^2}{3} \bar{D}_2 \bar{D}_0 \mathcal{E}_2,\ee
in agreement with the structure obtained in \C{f3}.

We next consider the graphs $Q_6$, $Q_7$ and $Q_8$ which do not require any regularization in performing their asymptotic expansions. For these graphs for the sake of brevity, we simply mention which $n_i$ is vanishing in a certain sector without mentioning which $m_i$ is Poisson resummed (and hence which $\hat{m}_i$ vanishes), as this is very much like the analysis before.

\subsection{Asymptotic expansion of $Q_6$}

From \C{graphs} we start with $Q_6$, which is given by
\bea \label{Def6}Q_6 =  -\sum' \frac{\tau_2}{\pi (m_1 + n_1\bar\tau)\vert m_2 + n_2 \tau\vert^2(m_3 +n_3 \bar\tau)^3},  \eea
where the sum is over integers $m_i$, satisfying \C{sum1}. The leading contribution, obtained by setting all $n_i =0$, is given by
\be \frac{\pi}{\tau_2} Q_6^{asymp} = \sum' \frac{1}{ m_1 m_2^2 (m_1+m_2)^3} = 2 \Big( W(1,2,3) + W(2,3,1)-  W(3,1,2)\Big)\ee
where $m_i$ satisfies \C{sum2}, leading to
\be Q_6^{asymp} = \frac{\zeta(6)}{\pi}\tau_2.\ee

When $n_3 =0$, we get that
\be Q_6^{asymp} = 2\tau_2\sum_{m\neq 0, n > 0}\frac{1}{n\tau_2 m^2(m^2 + 4 n^2\tau_2^2)} = \frac{\zeta(2)\zeta(3)}{\tau_2^2} -\frac{\pi\zeta(4)}{4\tau_2^3} + \frac{\zeta(5)}{8\tau_2^4},\ee
while when $n_2=0$, we have that
\be Q_6^{asymp} =0.\ee
Finally, when $n_1 =0$, we get that
\be Q_6^{asymp} = 2\tau_2\sum_{m\neq 0, n > 0}\frac{m^2 -12 n^2 \tau_2^2}{n\tau_2 (m^2 + 4 n^2\tau_2^2)^3} = \frac{3\zeta(5)}{8\tau_2^4} - \frac{\pi\zeta(4)}{4\tau_2^3}.\ee

Thus we get that
\be \label{Agree}Q_6^{asymp} =  \frac{\zeta(6)}{\pi}\tau_2 + \frac{\zeta(2)\zeta(3)}{\tau_2^2} -\frac{\pi\zeta(4)}{2\tau_2^3} + \frac{\zeta(5)}{2\tau_2^4},\ee
apart from the contribution from the \C{non} sector. Instead of performing this calculation, we can fix it simply by demanding consistency with \C{consis1}, which yields that this contribution is equal to
\be \label{agree}-\frac{\pi}{2\tau_2^3} \Big(\zeta(3)-\zeta(4)\Big),\ee
leading to~\footnote{In fact, direct calculation of this contribution yields
\be Q_6^{asymp} = -\frac{\pi}{\tau_2^3} \sum' \frac{(1+sgn(n_1)sgn(n_2))}{\vert n_1 +n_2\vert (\vert n_1\vert+\vert n_2\vert+\vert n_1+n_2\vert)^3}= -\frac{\pi}{2\tau_2^3} \Big(\zeta(3)-\zeta(4)\Big)\ee
in precise agreement with \C{agree}.}
\be \label{asympQ6}Q_6^{asymp} =  \frac{\zeta(6)}{\pi}\tau_2 + \frac{\zeta(2)\zeta(3)}{\tau_2^2} -\frac{\pi\zeta(3)}{2\tau_2^3} + \frac{\zeta(5)}{2\tau_2^4}.\ee
Note that the term $\pi\zeta(4)/2\tau_2^3$ cancels in \C{Agree} on adding it to \C{agree}, while the term  $-\pi\zeta(3)/2\tau_2^3$ remains in the sum, which has transcendentality less than the other terms. The origin of this boundary term is the presence of $Q_2 Q_4$ in \C{finQ6}. This also happens in the analysis for $Q_7$ and $Q_8$ below.

\subsection{Asymptotic expansion of $Q_7$}

From \C{graphs}, we have that
\bea \label{Def7}Q_7 =  \sum' \frac{\tau_2}{\pi (m_1 + n_1\bar\tau)\vert m_1 +n_1 \tau \vert^2 ( m_2 + n_2 \bar\tau) (m_3 +n_3 \bar\tau)^2},  \eea
where the sum is over integers $m_i$, satisfying \C{sum1}.  The leading contribution, obtained by setting all $n_i =0$, is given by
\be \frac{\pi}{\tau_2} Q_7^{asymp} = \sum' \frac{1}{ m_1^3 m_2 (m_1+m_2)^2} = -2 \Big( W(1,2,3) + W(2,3,1)-  W(3,1,2)\Big)\ee
where $m_i$ satisfies \C{sum2}, leading to
\be Q_7^{asymp} = -\frac{\zeta(6)}{\pi}\tau_2.\ee

When $n_3 =0$, we get that
\be Q_7^{asymp} = 2\tau_2\sum_{m\neq 0, n > 0}\frac{1}{n\tau_2 m^2(m^2 + 4 n^2\tau_2^2)} = \frac{\zeta(2)\zeta(3)}{\tau_2^2} -\frac{\pi\zeta(4)}{4\tau_2^3} + \frac{\zeta(5)}{8\tau_2^4},\ee
and when $n_2 =0$, we have that
\be Q_7^{asymp} = -4\tau_2\sum_{m\neq 0, n > 0}\frac{1}{n\tau_2 (m^2 + 4 n^2\tau_2^2)^2} = \frac{\zeta(5)}{4\tau_2^4} - \frac{\pi\zeta(4)}{4\tau_2^3}.\ee
Finally when $n_1 =0$, we have that
\be Q_7^{asymp} =0.\ee

Thus we get that
\be Q_7^{asymp} = -\frac{\zeta(6)}{\pi}\tau_2 +\frac{\zeta(2)\zeta(3)}{\tau_2^2} -\frac{\pi\zeta(4)}{2\tau_2^3} + \frac{3\zeta(5)}{8\tau_2^4},\ee
apart from the contribution from the \C{non} sector. Instead of performing this calculation, once again we fix it simply by demanding consistency with \C{consis2}, which yields that this contribution is equal to
\be -\frac{\pi}{2\tau_2^3} \Big(\zeta(3)-\zeta(4)\Big),\ee
leading to
\be Q_7^{asymp} = -\frac{\zeta(6)}{\pi}\tau_2 +\frac{\zeta(2)\zeta(3)}{\tau_2^2} -\frac{\pi\zeta(3)}{2\tau_2^3} + \frac{3\zeta(5)}{8\tau_2^4}.\ee

\subsection{Asymptotic expansion of $Q_8$}

From \C{graphs}, we have that
\bea \label{Def8}Q_8 =  \sum' \frac{\tau_2}{\pi (m_1 + n_1\bar\tau)^2 \vert m_2 +n_2 \tau \vert^2  (m_3 +n_3 \bar\tau)^2},  \eea
where the sum is over integers $m_i$, satisfying \C{sum1}.  The leading contribution, obtained by setting all $n_i =0$, is given by
\be \frac{\pi}{\tau_2} Q_8^{asymp} = \sum' \frac{1}{ m_1^2 m_2^2 (m_1+m_2)^2} = 6 \ W(2,2,2) \ee
where $m_i$ satisfies \C{sum2}, leading to
\be  Q_8^{asymp} = \frac{2\zeta(6)}{\pi}\tau_2,\ee
while when $n_3 =0$ (as well as $n_1=0$), we get that
\be Q_8^{asymp} = 4\tau_2\sum_{m\neq 0, n > 0}\frac{m^2 - 4 n^2 \tau_2^2}{m^2 n\tau_2 (m^2 + 4 n^2\tau_2^2)^2} = -\frac{2\zeta(2)\zeta(3)}{\tau_2^2}-\frac{3\zeta(5)}{4\tau_2^4} + \frac{\pi\zeta(4)}{\tau_2^3}\ee
on adding the two contributions.
Finally when $n_2=0$, we have that
\be Q_8^{asymp} =0.\ee

Thus we have that
\be Q_8^{asymp}=  \frac{2\zeta(6)}{\pi}\tau_2  -\frac{2\zeta(2)\zeta(3)}{\tau_2^2}-\frac{3\zeta(5)}{4\tau_2^4} + \frac{\pi\zeta(4)}{\tau_2^3}, \ee
apart from the contribution from the \C{non} sector. This can be easily determined by using $Q_8 = -2 Q_7$ as deduced in the main text, hence leading to the contribution
\be \frac{\pi}{\tau_2^3} \Big(\zeta(3)-\zeta(4)\Big)\ee
from this sector, leading to
\be Q_8^{asymp}=  \frac{2\zeta(6)}{\pi}\tau_2  -\frac{2\zeta(2)\zeta(3)}{\tau_2^2}-\frac{3\zeta(5)}{4\tau_2^4} + \frac{\pi\zeta(3)}{\tau_2^3}.\ee

\subsection{Asymptotic expansion of $Q_9$}

Finally from \C{graphs} we consider the asymptotic expansion of $Q_9$, which is given by
\bea \label{Def9}Q_9 =  -\sum' \frac{\tau_2^2}{\pi^2 (m_1 + n_1\bar\tau) (m_2 +n_2 \bar\tau )  \vert m_3 +n_3 \tau \vert^2 \vert m_4 + n_4\tau\vert^2},  \eea
where the sum is over integers $m_i$ satisfying
\be (m_i,n_i) \neq (0,0), ~i=1,2,3,4, \quad \sum_i m_i = \sum_i n_i =0.\ee

The leading contribution at large $\tau_2$ is obtained by setting all $n_i =0$ in \C{Def9}. This gives us
\bea Q_9^{asymp} &=& \frac{4\tau_2^2}{\pi^2} \Big[ \zeta(2)\zeta(4) - 2\zeta(2)\Big( W(1,1,2) - 2W(2,1,1)\Big) +3\Big( W(1,1,4) -2 W(4,1,1)\Big)\Big] \non \\ &=& \frac{11\zeta(6)}{\pi^2}\tau_2^2.\eea

The contribution from the $n_3 = n_4 =0$ sector is given by 
\be \label{9one}Q_9^{asymp} = \frac{2\tau_2^2}{\pi^2} \sum_{n_1 > 0, m_3 \neq 0, m_4 \neq 0} \frac{\tau_2^{3-4s}}{m_3^2 m_4^2  n_1^{4s-3}} I(s,\tilde{a})\ee
in the limit $s\rightarrow 1$, where $I(s,\tilde{a})$ is defined by \C{Integral} and
\be \tilde{a} = \frac{m_3+m_4}{n_1 \tau_2}.\ee
In obtaining \C{9one} we have Poisson resummed over $m_1$ and set $\hat{m}_1=0$.
Thus
\be Q_9^{asymp} = -\frac{10\zeta(4)}{\pi}\tau_2\ee
at leading order, on regularizing the expression. The subleading contributions can be heuristically argued as before, and yield
\be \label{finreg}Q_9^{asymp} = -\frac{10\zeta(4)}{\pi}\tau_2 + c_3 + \frac{c_4}{\tau_2}\ee
in this sector. Essentially, $c_3$ arises from the divergence in the sum over $m_3$ (and also separately from $m_4$) obtained from a term linear in $\tilde{a}$ in expanding $I(s,\tilde{a})$, while $c_4/\tau_2$ is obtained from a simultaneous divergence in the sums over $m_3$ and $m_4$ obtained from $m_3m_4$ in expanding $I(s,\tilde{a})$ to quadratic order in $\tilde{a}$. As before, these potential contributions can arise by regularizing the product of a vanishing integral and a divergent sum. Now from \C{LeaD} we directly see that
\be c_3 = \frac{4\zeta(2)}{\pi} c_0\ee
which shall be useful later.  

Next we consider the contribution from the $n_1 = n_2 =0$ sector which is given by\footnote{We have Poisson resummed over $m_3$ and set $\hat{m}_3 =0$.}
\bea \label{91}Q_9^{asymp} &=& -\frac{4\tau_2}{\pi} \sum_{n_3 > 0, m_1 \neq 0 , m_2 \neq 0}\frac{1}{m_1 m_2n_3 [(m_1 + m_2)^2 + 4 n_3^2 \tau_2^2]} \non \\ &=& \frac{\pi\zeta(3)}{\tau_2} -\frac{\zeta(4)}{\tau_2^2} + \frac{\zeta(5)}{2\pi\tau_2^3}.\eea

Also there are four equal contributions from the $(n_1, n_3) = (0,0)$, $(n_1, n_4) = (0,0)$, $(n_2, n_3) = (0,0)$ and $(n_2,n_4) = (0,0)$ sectors. The total contribution is given by\footnote{For $(n_1,n_3)=(0,0)$, we have Poisson resummed over $m_2$ and set $\hat{m}_2 =0$.}
\bea \label{92}Q_9^{asymp} &=& \frac{8\tau_2}{\pi} \sum_{n_2 > 0, m_1 \neq 0, m_3 \neq 0} \frac{m_1 + m_3}{m_1 m_3^2 n_2 [(m_1 + m_3)^2 + 4 n_2^2 \tau_2^2]} \non \\ &=& 8\zeta(2)^2 -\frac{8\pi\zeta(3)}{3\tau_2} + \frac{3\zeta(4)}{\tau_2^2} - \frac{3\zeta(5)}{2\pi\tau_2^3}.\eea

In both the expressions \C{91} and \C{92}, we have dropped terms that are exponentially suppressed in the large $\tau_2$ expansion. 

The contributions from the $n_3 =0$ and $n_4 =0$ sectors are equal. Hence the total contribution is\footnote{For $n_3 =0$, we have Poisson resummed over $m_1$, $m_2$ and set $\hat{m}_1 = \hat{m}_2 =0$.}
\bea Q_9^{asymp} &=& 2\sum' \frac{(1+sgn(n_1)sgn(n_2))(\vert n_1\vert+\vert n_2\vert+\vert n_1+n_2\vert)}{m_3^2 \vert n_1 +n_2\vert [(\vert n_1\vert+\vert n_2\vert +\vert n_1+n_2\vert)^2 + m_3^2/\tau_2^2]}\non \\ &=& -8\zeta(2)^2 +\tilde{c}_3-\frac{2\pi}{\tau_2}\Big(\zeta(2)-\zeta(3)\Big) +\frac{1}{\tau_2^2}\Big(\zeta(3)-\zeta(4)\Big),\eea
where the sum in the first line is over the integers \C{non} and $m_3 \neq 0$, and we have dropped exponentially suppressed terms. We also have a formally divergent constant term of transcendentality less than 4 which needs regularization. Now from \C{cancel1} we again directly see that
\be \tilde{c}_3 = \frac{4\zeta(2)}{\pi} \tilde{c}_0\ee
which we shall use later.

The contributions from the $n_1 =0$ and $n_2 =0$ sectors are equal, and for $n_1 =0$ we Poisson resummed over $m_2$, $m_3$ and set $\hat{m}_2 = \hat{m}_3 =0$. On relabelling variables, the total contribution is
\bea Q_9^{asymp} &=& \frac{2\pi}{\tau_2} \sum' \frac{1}{\vert n_1\vert \vert n_1+n_2\vert(\vert n_1\vert+\vert n_2\vert +\vert n_1+n_2\vert)} \non \\ &&- \frac{2}{\tau_2^2}\sum' \frac{1}{\vert n_1\vert \vert n_1+n_2\vert(\vert n_1\vert+\vert n_2\vert +\vert n_1+n_2\vert)^2},\eea
where the sum is over the integers satisfying \C{non} and we have neglected exponentially suppressed contributions. This gives~\footnote{Apart from some of the relations in \C{mzv}, we also use the relation $\zeta(2,1) = \zeta(3)$.}
\be Q_9^{asymp} = \frac{8\pi\zeta(3)}{\tau_2} - \frac{\zeta(4)}{\tau_2^2}.\ee

Finally we consider the contribution from the sector where
\be \label{morenon}n_1 \neq 0, \quad n_2 \neq 0,\quad  n_3 \neq 0 , \quad n_1+n_2+n_3 \neq 0.\ee
Poisson resumming over $m_1, m_2, m_3$, and setting $\hat{m}_1 = \hat{m}_2 = \hat{m}_3 =0$, this contribution gives\footnote{The contribution from $\sum_i \hat{m}_i n_i =0$, $m_i \neq 0$ is also independent of $\tau_1$, however these terms are exponentially suppressed in $\tau_2$.}
\bea Q_9^{asymp} &=& \frac{\pi}{\tau_2} \sum' \frac{1+sgn(n_1)sgn(n_2)}{\vert n_3\vert \vert n_1+n_2+n_3\vert (\vert n_1\vert+\vert n_2\vert+\vert n_3\vert+\vert n_1+n_2+n_3\vert)}\non \\ &=& \frac{1}{\tau_2}[ \tilde{c}_4 - 8\pi\zeta(3)],\eea
where the sum is over \C{morenon}, and we also have a formally divergent contribution of order $1/\tau_2$ having transcendentality less than 4. This has to be regularized, and will add to the coefficient $c_4$ in \C{finreg}.

Thus adding the various contributions, we get that
\bea Q_9^{asymp} &=& \frac{11\zeta(6)}{\pi^2}\tau_2^2 +\frac{\pi\zeta(3)}{3\tau_2} -\frac{\zeta(5)}{\pi\tau_2^3}\non \\ &&-\frac{10\zeta(4)}{\pi}\tau_2 +\frac{\zeta(3)}{\tau_2^2} + c_3 +\tilde{c}_3 +\frac{c_4 +\tilde{c}_4}{\tau_2},\eea
where we have also absorbed a term of the form $\pi\zeta(2)/\tau_2$ in $(c_4+\tilde{c}_4)/\tau_2$.
We simply rewrite this expression as
\bea \label{9q}Q_9^{asymp} &=& \Big(\frac{11\zeta(6)}{\pi^2}\tau_2^2 +\frac{\pi\zeta(3)}{3\tau_2} -\frac{\zeta(5)}{\pi\tau_2^3} - \frac{4\zeta(4)}{\pi}\tau_2 -\frac{\zeta(3)}{2\tau_2^2}\Big)\non \\ &&-3\pi\Big(\frac{2\zeta(4)}{\pi^2}\tau_2 -\frac{\zeta(3)}{2\pi\tau_2^2}\Big) + 4\zeta(2) +\frac{c_4 +\tilde{c}_4}{\tau_2}\eea
on using \C{NeedLat}.
The terms in the first line of \C{9q} precisely agree with what we obtain from $\delta_{\m} Q_9$ in the main text, on ignoring all contributions arising from $\delta_{\mu} \bar\p G$. Note that while the first three terms have transcendentality 4, the last two have transcendentality 3. They arise from the term of the form $1/\tau_2$ in $Q_2$ in the factor involving $Q_2 \bar{Q}_4$ in \C{MatcH}. These boundary terms of reduced transcendentality are already included in our analysis, and contributions coming from $\delta_{\mu} \bar\p G$ add more such terms. These terms are given in the second line of \C{9q}. Thus based on the asymptotic expansion and modular covariance, we are led to the expansion  
\bea \label{asymp9}Q_9^{asymp} &=& \Big(\frac{11\zeta(6)}{\pi^2}\tau_2^2 +\frac{\pi\zeta(3)}{3\tau_2} -\frac{\zeta(5)}{\pi\tau_2^3} - \frac{4\zeta(4)}{\pi}\tau_2 -\frac{\zeta(3)}{2\tau_2^2}\Big)\non \\ && -3\pi\Big(\frac{2\zeta(4)}{\pi^2}\tau_2 -\frac{\zeta(3)}{2\pi\tau_2^2}\Big) +4
\zeta(2)\Big( 1 -\frac{3}{\pi\tau_2} \Big).\eea
Note that the terms in the first line can be completed $SL(2,\mathbb{Z})$ covariantly to give
\be \label{last}\frac{4\pi}{3}\bar{D}_0 \mathcal{E}_3 -\pi\bar{D}_0 \mathcal{E}_2 +\frac{\pi^2}{3} \overline{\hat{E}}_2 \mathcal{E}_2,\ee
while the terms in the second line give 
\be \label{verylast}-3\pi\bar{D}_0 \mathcal{E}_2 +4\zeta(2) \overline{\hat{E}}_2.\ee


\providecommand{\href}[2]{#2}\begingroup\raggedright\endgroup

\end{document}